%% file: main.tex
\documentclass[preprint,12pt,5p,times,twocolumn]{elsarticle}

\usepackage{lineno}
\usepackage[draft]{hyperref}
\modulolinenumbers[5]

\usepackage{booktabs}       
\usepackage{mathtools}
\usepackage{amssymb}
\usepackage{amsmath}
\usepackage{algorithm}
\usepackage{algorithmic}
\usepackage{color}
\usepackage{subfig}
\usepackage{multirow}
\usepackage[table]{xcolor}
\usepackage{array}
\usepackage{graphicx}

\usepackage{etoolbox}
\makeatletter
    \patchcmd{\tnotemark}{\ding{73}}{\dag}{}{\@latex@error{Failed to path \string\tnotemark\space for \string\ding{73}}}
    \patchcmd{\tnotemark}{\ding{73}\ding{73}}{\dag\dag}{}{\@latex@error{Failed to path \string\tnotemark\space for \string\ding{73}\string\ding{73}}}
    \patchcmd{\tnotetext}{\ding{73}}{\dag}{}{\@latex@error{Failed to path \string\tnotetext\space for \string\ding{73}}}
    \patchcmd{\tnotetext}{\ding{73}\ding{73}}{\dag\dag}{}{\@latex@error{Failed to path \string\tnotetext\space for \string\ding{73}\string\ding{73}}}
\makeatother

\journal{Medical Image Analysis}

\begin{document}

\begin{frontmatter}


\title{Dynamic Coronary Roadmapping via Catheter Tip Tracking in X-ray Fluoroscopy with Deep Learning Based Bayesian Filtering\tnoteref{f1}\tnoteref{license}}

\author[address1]{Hua Ma\corref{mycorrespondingauthor}}
\cortext[mycorrespondingauthor]{Corresponding author}
\ead{huama.research@gmail.com}
\author[address2]{Ihor Smal}
\author[address3]{Joost Daemen}
\author[address1]{Theo van Walsum}

\address[address1]{Biomedical Imaging Group Rotterdam, Erasmus MC, University Medical Center Rotterdam, Rotterdam, The Netherlands}
\address[address2]{Department of Geoscience and Remote Sensing, Delft University of Technology, Delft, The Netherlands}
\address[address3]{Department of Cardiology, Erasmus MC, University Medical Center Rotterdam, Rotterdam, The Netherlands}

\tnotetext[f1]{This manuscript has been accepted for publication in \textit{Medical Image Analysis}. For the published journal article, please refer to DOI: https://doi.org/10.1016/j.media.2020.101634.}
\tnotetext[license]{\copyright 2020. This manuscript version is made available under the CC-BY-NC-ND 4.0 license. For license details, please check via http://creativecommons.org/licenses/by-nc-nd/4.0/.}

%
%
\input{sections/abstract/abstract.tex}

\end{frontmatter}



%
%
\input{sections/introduction/introduction.tex}

%
%
\input{sections/methods/methods.tex}

%
%
\input{sections/setup/setup.tex}

%
%
\input{sections/experiments/experiments.tex}

%
%
\input{sections/discussion/discussion.tex}

%
%
\input{sections/acknowledgment/acknowledgment.tex}

%
%
\input{sections/references/references.tex}
%
%
\input{sections/appendix/appendix.tex}

\end{document}

%% file: sections/abstract/abstract.tex
\begin{abstract}

Percutaneous coronary intervention (PCI) is typically performed with image guidance using X-ray angiograms in which coronary arteries are opacified with X-ray opaque contrast agents. Interventional cardiologists typically navigate instruments using non-contrast-enhanced fluoroscopic images, since higher use of contrast agents increases the risk of kidney failure. When using fluoroscopic images, the interventional cardiologist needs to rely on a mental anatomical reconstruction. This paper reports on the development of a novel dynamic coronary roadmapping approach for improving visual feedback and reducing contrast use during PCI. The approach compensates cardiac and respiratory induced vessel motion by ECG alignment and catheter tip tracking in X-ray fluoroscopy, respectively. In particular, for accurate and robust tracking of the catheter tip, we proposed a new deep learning based Bayesian filtering method that integrates the detection outcome of a convolutional neural network and the motion estimation between frames using a particle filtering framework. The proposed roadmapping and tracking approaches were validated on clinical X-ray images, achieving accurate performance on both catheter tip tracking and dynamic coronary roadmapping experiments. In addition, our approach runs in real-time on a computer with a single GPU and has the potential to be integrated into the clinical workflow of PCI procedures, providing cardiologists with visual guidance during interventions without the need of extra use of contrast agent.

\end{abstract}


%% file: sections/introduction/introduction.tex
\section{Introduction}

\subsection{Clinical Background}
\label{intro-subsec-clinical}

Percutaneous coronary intervention (PCI) is a minimally invasive procedure for treating patients with coronary artery disease. During these procedures, medical instruments inserted through a guiding catheter are advanced to treat coronary stenoses. A guiding catheter is firstly positioned into the ostium of the coronary artery. Through the guiding catheter, a balloon catheter carrying a stent is introduced over a guidewire to the stenosed location. The balloon is then inflated and the stent is deployed to prevent the vessel from collapsing and restenosing.

PCI is typically performed with image-guidance using X-ray angiography (XA). Coronary arteries are visualized with X-ray opaque contrast agent. During the procedure, interventional cardiologists may repeatedly inject contrast agent to visualize the vessels, as the opacification of coronary arteries only lasts for a short period. The amount of periprocedural contrast use has been correlated to operator experience, procedural complexity, renal function and imaging setup (\cite{piayda2018dynamic}). Furthermore, the risk for contrast induced nephropathy has been associated to contrast volume (\cite{tehrani2013contrast}). Manoeuvring guidewires and material, however, typically occurs without continuous contrast injections. In these situations, the navigation of devices is guided with "vessel-free" fluoroscopic images. Cardiologists have to mentally reconstruct the position of vessels and stenosis based on previous angiograms.

\subsection{Dynamic Coronary Roadmapping}
\label{intro-subsec-dcr}


Dynamic coronary roadmapping (DCR) is a promising solution towards improving visual feedback and reducing usage of contrast medium during PCI (\cite{elion1989dynamic, zhu2010image, manhart2011self, kim2018registration}). This approach dynamically superimposes images or models of coronary arteries onto live X-ray fluoroscopic sequences. The dynamic overlay serves as a roadmap that provides immediate feedback to cardiologists during the intervention, so as to assist in navigating a guidewire into the appropriate coronary branch and proper placement of a stent at the stenotic site with reduced application of contrast agent. Studies with a phantom setup using research software (\cite{kim2018registration}) or on first cases of clinical interventions using commercially available systems (\cite{dannenberg2016road, yabe2018impact, takimura2018usefulness}) have investigated the usefulness of DCR in assisting PCI, reporting that DCR helps to reduce procedure time, radiation dose and contrast volume.

To develop a DCR system, it is important but yet a challenge to accurately overlay a roadmap of coronary arteries onto an X-ray fluoroscopic image, as limited information of vessels is present in the target fluoroscopic image for inferring the compensation of the vessel motion resulting from patient respiration and heartbeat. The methods that have been proposed on motion compensation for DCR can be generally grouped into two categories: direct roadmapping and model-based approaches.

Direct roadmapping methods use information from X-ray images and ECG signals to directly correct the motion caused by respiration and heartbeat. The first DCR system (\cite{elion1989dynamic}) used digital subtraction of a contrast sequence and a mask sequence to create a full cardiac cycle of coronary roadmaps. The roadmaps were stored and later synchronized with the live fluoroscopic sequence by aligning the R waves of their corresponding ECG signals. This system compensates the cardiac motion of vessels, yet does not correct the respiratory motion during interventions. Two later studies by \cite{zhu2010image} and \cite{manhart2011self} introduced image-based respiratory motion compensation methods for DCR. Their methods assumed an affine respiratory motion model in ECG-gated fluoroscopic frames and recovered the respiratory motion from soft tissues with special handling of static structures. The limitation of these approaches is that they require relevant tissue to be sufficiently visible in the field of view for reliable motion compensation which is not always the case. In addition, they require to be run on cardiac-gated frames. In a more recent work by \cite{kim2018registration}, binary vessel masks were created as the roadmaps from at least one cardiac cycle of angiographic images. Temporal alignment of the roadmaps and the fluoroscopic sequence, which compensated the cardiac motion of vessels, was performed by registering ECG signals using cross-correlation. Additionally, the respiratory motion was corrected by aligning the guidewire centerline in the fluoroscopy to the contour of vessels in the angiogram where the roadmaps were created. The system has been shown useful in a phantom-based study, nevertheless no accuracy evaluation of the spatiotemporal alignment was presented. Furthermore, the spatial registration relies on robust extraction of vessels and guidewires which is often challenging for X-ray images.

Unlike direct roadmapping, the model-based approaches build a model to predict motion in fluoroscopic frames. The motion models are often functions that relate the motion of roadmaps to surrogate signals derived from images or ECG, so that once the surrogates for fluoroscopic frames are obtained, the motion can be computed by the model. For cardiac interventions including PCI, the organ motion is mainly affected by respiratory and cardiac motion. Many previous works often built a motion model parameterized by a cardiac signal derived from ECG and a respiratory signal obtained from diaphragm tracking (\cite{shechter2005prospective, timinger2005motion, faranesh2013integration}) or automatic PCA-based surrogate (\cite{fischer2018mr}). Some other works model only the respiratory motion in cardiac-gated images (\cite{schneider2010model, king2009subject, peressutti2013novel}). For a complete review on respiratory motion modeling, we refer readers to the survey article by \cite{mcclelland2013respiratory}. One limitation of the model-based approaches is that the motion models are often patient-specific, which requires training the model every time for a new subject. Additionally, once the surrogate values during inference are out of the surrogate range for building the model, e.g. for abnormal motion, extrapolation is needed, which may hamper accurate motion compensation.

\subsection{Interventional / Surgical Tool Tracking}
\label{intro-subsec-tracking}

Tracking interventional tools is relevant for motion compensation (\cite{schneider2010model, brost2010respiratory, ma2012clinical, baka2015respiratory, ambrosini2017hidden}). In particular for PCI, the guiding catheter tip typically remains within the coronary ostium which is in the field of view during the largest part of the intervention, making it a suitable landmark for tracking. \cite{baka2015respiratory} have shown that catheter tip motion during PCI can be modeled as a combination of cardiac and respiratory motion. As using catheter tip displacement can only correct translational motion, \cite{baka2015respiratory} further showed that, compared to a rigid motion model for the respiratory motion, modeling only the translational part of the respiratory motion deteriorated the accuracy marginally, which confirms the observations by \cite{shechter2004respiratory} that the rotational part of respiratory motion is small. These findings motivate motion compensation for DCR through tracking the catheter tip in X-ray fluoroscopic sequences.


Many works have been proposed to address the problem of tracking interventional or surgical tools in medical images for various applications. The tracking methods from these works can be generally categorized into two kinds of approaches: tracking by detection, and temporal tracking.


The tracking by detection approaches treat tracking as a detection problem, which rely on features only from the current image without using information from previous frames. For example in electrophysiology procedure, as the catheters present specific features in shape or intensity, \textit{ad hoc} methods were proposed based on, e.g. blob detection, shape constrained searching and model-/template- based detection (\cite{ma2012clinical, ma2013real}). \cite{chang2016robust} modeled the catheter tracking problem by optimizing the posterior in a Bayesian framework, in which the catheter was represented as a B-spline tube model and was tracked by fitting the B-spline to measurements based on gray intensity and vesselness image. \cite{baur2016cathnets} proposed a convolutional neural network (CNN) to detect catheter electrodes in X-ray images, which treated catheter detection as a segmentation problem. The method used a weighted cross-entropy loss to cope with the class imbalancing problem due to the small size of the target. \cite{laina2017concurrent} and \cite{du2018articulated} tracked surgical instruments using a deep network having an encoder-decoder architecture. Their approaches combined instrument segmentation and detection in a multi-task learning problem to make the tool detection in a cluttered background more robust.


Different from tracking by detection, which relies solely on the current image, temporal tracking also uses information from previous frames. The temporal information can reduce the search space for detection, or put additional constraints in the model, making the tracking more robust. 

Temporal information has been used in various ways. Some methods mainly relied on a detection model, but incorporate temporal information in the preprocessing (\cite{brost2010respiratory}) or post-processing (\cite{garcia2016real}) step or in the input (\cite{rieke2016real, ambrosini2017fully}). Approaches based on background estimation have been used for catheter (\cite{yatziv2012toward}) or guidewire (\cite{petkovic2010guidewire}) tracking. In these approaches, the background was recursively updated for every frame, and was used for enhancing the foreground containing instruments. Apart from those, many works adopted a Bayesian framework for tracking instruments via a \textit{maximum a posteriori} (MAP) formulation. Representations based on key points (\cite{wu2015fast}), B-splines (\cite{wang2009robust, pauly2010machine, honnorat2011graph, heibel2013interventional}), or segment-like features (\cite{vandini2017robust}) have been used to model catheters or guidewires. Markov random field (MRF) was used to model the position or deformation of the control points in the B-spline (\cite{pauly2010machine, honnorat2011graph, heibel2013interventional, wu2015fast}). In the work by \cite{vandini2017robust}, temporal information was incorporated in the prior term using Kalman filter. Particularly, learning-based approaches were used in several works to obtain the likelihood for a more robust measurement using probabilistic boosting tree (\cite{wang2009robust, wu2012fast}) or support vector regression (\cite{pauly2010machine}). In addition, temporal tracking models based on Bayesian filtering were also a popular approach for instrument tracking. \cite{ambrosini2017hidden} used a hidden Markov model (HMM) to track catheter tip in a 3D vessel tree, for which the likelihood was obtained based on the 3D-2D registration outcome. \cite{speidel2006tracking} used particle filters to track surgical tools in medical images. They used a likelihood based on the segmentation of instruments, and a dynamic model that incorporates samples from two previous time steps. In a later work, \cite{speidel2014visual} used a multi-object particle filter to track multiple instrument regions simultaneously, in which a particle is the concatenation of the states of several objects.


Despite of many existing works on inverventional or surgical tool tracking in medical images, an automatic approach for tracking the tip of guiding catheter in X-ray fluoroscopy for PCI has not been investigated yet. The challenges of this task are: (1) different from the catheters for EP that can be viewed as blobs or a circle, the guiding catheter for PCI presents a dark tubular appearance which shows no prominent features; (2) the shape of the guiding catheter tip segment varies depending on the orientation of the C-arm, making feature-/model- based detection challenging; (3) the background may contain structures that have similar appearance to a catheter tip, such as vertebral structures or residual contrast agent, which makes robust tracking difficult.

\subsection{Contributions}
\label{intro-subsec-contribution}

We propose and evaluate a novel approach for dynamic coronary roadmapping. The approach compensates changes in vessel shapes and cardiac motion by selecting the roadmap of the same cardiac phase through ECG alignment, and corrects the respiratory induced motion via tracking the tip of the guiding catheter. Our contributions are:

\begin{enumerate}
	\item We develop a new way to perform dynamic coronary roadmapping on free breathing, non-cardiac-gated X-ray fluoroscopic sequences. Particularly, the respiratory-induced vessel motion is robustly compensated via the displacement of catheter tip.
	\item We proposed a deep learning based method within a Bayesian filtering framework for online detection and tracking of guiding catheter tip in X-ray fluoroscopic images. The method models the likelihood term of Bayesian filtering with a convolutional neural network, and integrates it with particle filtering in a comprehensive manner, leading to more robust tracking.
	\item We evaluate the proposed approach visually and quantitatively on clinical X-ray sequences, achieving low errors on both tracking and roadmapping tasks.
	\item The proposed DCR method runs in real-time with a modern GPU, thus can potentially be used during PCI in real clinical settings.
\end{enumerate}

%% file: sections/methods/methods.tex

\section{Scenario Setup and Method Overview}
\label{method-sec-overview}

\begin{figure*}[t]
	\centering
	\includegraphics[width=0.99\textwidth]{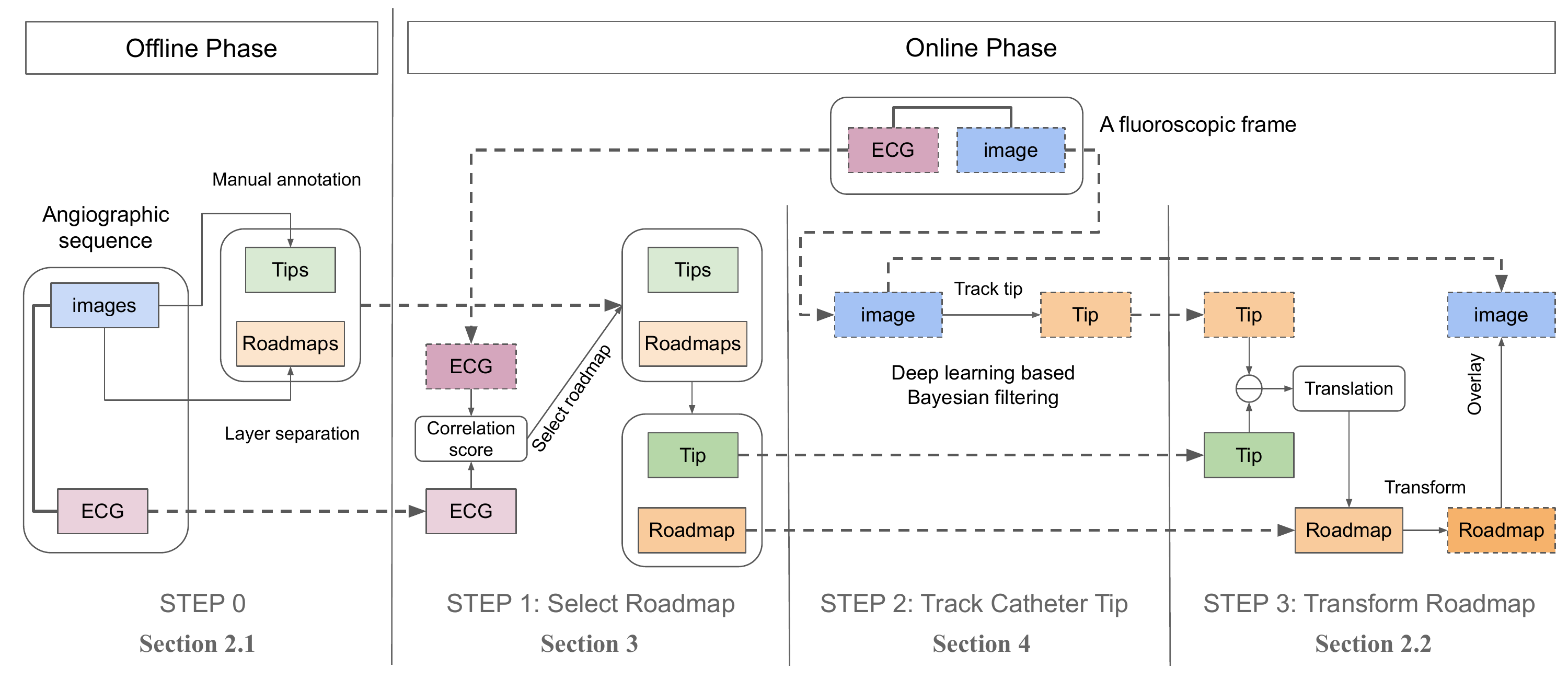}
	\caption{The overview of the proposed dynamic coronary roadmapping method. The colored blocks with a dash line border denote objects acquired in the online phase; the colored blocks with a solid line border are objects originated from the offline phase.}
	\label{method-figure-overview}
\end{figure*}

The proposed method assumes that the scenario of performing dynamic coronary roadmapping to guide a PCI procedure consists of an offline phase and an online phase. An overview of the method is shown in Figure \ref{method-figure-overview}.

\subsection{Offline Phase}
\label{method-subsubsec-roadmap-preparation}
This phase (Step 0 in Figure \ref{method-figure-overview}) is performed off-line before the actual roadmapping is conducted. In this stage, roadmaps of coronary arteries containing multiple cardiac phases are created from an X-ray angiography sequence acquired with injection of contrast agent. A roadmap can be a vessel model in the form of centerlines, contours, masks, etc. It may also contain information of clinical interest, e.g. stenosis. Since the main focus of this paper is on accurate overlay of a roadmap, we do not investigate how to create the most suitable roadmaps, but use the images containing only vessels and catheters that are created using the layer separation method by \cite{ma2015layer} as the roadmaps to show the concept of dynamic coronary roadmapping. Along with the XA sequence, ECG signals are also acquired and stored for later selecting a roadmap that has similar cardiac phase to a given X-ray fluoroscopic frame in the online phase (see details in Section \ref{method-sec-roadmap}). Once the image sequence and ECG signals are acquired, the catheter tip location in every frame is obtained to serve as a reference point for roadmap transformation. In this work we manually annotated the catheter tip in the offline XA sequence. In real clinical scenarios, the annotation work can be done by the clinician or a person who assists the intervention, such as a technician or a nurse.

\subsection{Online Phase}
\label{method-subsubsec-roadmap-transformation}
This is when the dynamic roadmapping is actually performed. In this phase, non-contrast X-ray fluoroscopic images with the same view angles as the roadmaps created during the offline phase are acquired sequentially. At the same time, ECG signals along with the roadmapping frames are also obtained and are compared with the stored ECG to select the most matched roadmap (Step 1 in Figure \ref{method-figure-overview}; see details in Section \ref{method-sec-roadmap}). This is to compensate the change of vessel shape and position between frames due to cardiac motion. Simultaneously, the catheter tip location in the acquired X-ray fluoroscopic images is tracked online using the proposed deep learning based Bayesian filtering method in Section \ref{method-sec-bayesian-filter} (Step 2 in Figure \ref{method-figure-overview}). The displacement of catheter tip between the current image and the selected roadmap image is then obtained and are applied to transform the roadmap. Finally, the transformed roadmap is overlaid on the current non-contrast frame to guide the procedure (Step 3 in Figure \ref{method-figure-overview}).

\section{ECG matching for Roadmap Selection}
\label{method-sec-roadmap}

Roadmap selection in this work is achieved by comparing the ECG signal associated with the fluoroscopic image and the ECG of the angiographic sequence, such that the most suitable candidate roadmap is selected where the best match of the ECG signals is found. The selected roadmap has the same (or very similar) cardiac phase with the X-ray fluoroscopic image, which compensates the difference of vessel shape and pose induced by cardiac motoin. An approach similar to the ECG matching method by \cite{kim2018registration} is used to accomplish this task.

To select roadmaps images based on ECG, a temporal mapping between X-ray images and ECG signal points needs to be built first. We assume that ECG signals and X-ray images are well synchronized during acqusition. In the offline phase, the beginning and the end of the image sequence are aligned with the start and end ECG signal points; the XA frames in between are then evenly distributed on the timeline of ECG. This way, a mapping between the stored sequence images and its ECG signal can be set up: for each image, the closest ECG signal point to the location of the image on the timeline can be found; for each ECG point, an image that is closest to this point on the timeline can be similarly located. Once the mapping is available, all images with good vessel contrast filling and the ECG points that are associated to these images are selected from the XA sequence for the pool of roadmaps. In this process, at least one heartbeat of frames should be acquired, which is generally the case in our data. In the online phase, similar to the approach of \cite{kim2018registration}, for acquisition of each image, a block of $N_{ECG}$ latest ECG signal points is constantly stored and updated in the history buffer. This is considered as the ECG signal corresponding to the fluoroscopic frame.


To compare the ECG signals associated with the angiographic sequence and the online fluoroscopic image, a temporal registration of the two signals using cross-correlation is applied (\cite{kim2018registration}). The two ECG signals are first cross-correlated for every possible position on the signals, resulting in a 1D vector of correlation scores. The candidate frame for dynamic overlay is then selected as the one associated with the point on the ECG of the angiographic sequence that is corresponding to the highest correlation score.

\section{Bayesian Filtering for Catheter Tip Tracking}
\label{method-sec-bayesian-filter}

Bayesian filtering is a state-space approach aiming at estimating the true state of a system that changes over time from a sequence of noisy measurement made on the system (\cite{arulampalam2002tutorial}). One popular application area of this approach is tracking objects in a series of images. 

\subsection{Theory of Bayesian Filtering}
\label{method-subsec-bayesian-filter}


Bayesian filtering typically includes the following components: hidden system states, a state transition model, observations and a observation model. Let $\textbf{x}_k \in \mathbb{R}^2$ $(k=\{0, 1, 2, ...\})$ denote the state, the location of guiding catheter tip in the $k$-th frame, a 2D vector representing the coordinates in the X-ray image space. The transition of the system from one state to the next state is given by the state transition model $\textbf{x}_k=f_k(\textbf{x}_{k-1}, \textbf{v}_{k-1})$, where $\textbf{v}_{k-1} \in \mathbb{R}^2$ is an independent and identically distributed (i.i.d.) process noise, $f_k: \mathbb{R}^2 \times \mathbb{R}^2 \rightarrow \mathbb{R}^2$ is a possibly nonlinear function that maps the previous state $\textbf{x}_{k-1}$ to the current state $\textbf{x}_k$ with noise $\textbf{v}_{k-1}$. The observation $\textbf{z}_k$ in this work is defined as the $k$-th X-ray image of a sequence, so that $\textbf{z}_k \in \mathbb{R}^{w \times h}$, where $w$ and $h$ are the width and height of an X-ray image. We further define the observation model as $\textbf{z}_k=h_k(\textbf{x}_k, \textbf{n}_k)$, where $\textbf{n}_k \in \mathbb{R}^{n_k}$ is an i.i.d~measurement noise ($n_k$ is the dimension of $\textbf{n}_k$), $h_k: \mathbb{R}^2 \times \mathbb{R}^{n_k} \rightarrow \mathbb{R}^{w \times h}$ is a highly nonlinear function that generates the observation $\textbf{z}_k$ from the state $\textbf{x}_k$ with noise $\textbf{n}_k$. The state transition model $f_k$ and the observation model $h_k$, respectively, can also be equivalently represented using probabilistic forms, i.e. the state transition prior $p(\textbf{x}_k|\textbf{x}_{k-1})$ and the likelihood $p(\textbf{z}_k|\textbf{x}_k)$ from which $\textbf{x}_k$ and $\textbf{z}_k$ can be obtained by sampling.

With these definitions and $p(\textbf{x}_0)$, the inital belief of $\textbf{x}_0$, Bayesian filtering seeks an estimation of $\textbf{x}_k\ (k\geq1)$ based on the set of all available observations $\textbf{z}_{0:k}=\{\textbf{z}_i, i=0,...,k\}$ up to time $k$ via recursively computing the posterior probability $p(\textbf{x}_k|\textbf{z}_{0:k})$ as Eq.(\ref{method-eq-bayes-filter}) (\cite{arulampalam2002tutorial}):

\begin{equation}
	\label{method-eq-bayes-filter}
	p(\textbf{x}_k|\textbf{z}_{0:k}) \propto p(\textbf{z}_k|\textbf{x}_k)\underbrace{\int p(\textbf{x}_k|\textbf{x}_{k-1})p(\textbf{x}_{k-1}|\textbf{z}_{0:k-1}) \mathrm{d}\textbf{x}_{k-1}}_\text{$p(\textbf{x}_k|\textbf{z}_{0:k-1})$} .
\end{equation}

\noindent Assuming the initial probability $p(\textbf{x}_0|\textbf{z}_0)=p(\textbf{x}_0)$ is known, based on Eq.(\ref{method-eq-bayes-filter}), Bayesian filtering runs in cycles of two steps: prediction and update. In the prediction step, the prior probability $p(\textbf{x}_k|\textbf{z}_{0:k-1})$, the initial belief of $\textbf{x}_k$ given previous observations, is estimated by computing the integral in Eq.(\ref{method-eq-bayes-filter}). In the update step, the prior probability is corrected by the current likelihood $p(\textbf{z}_k|\textbf{x}_k)$ to obtain the posterior $p(\textbf{x}_k|\textbf{z}_{0:k})$.

In Section \ref{method-subsec-likelihood}, we will firstly introduce how to model the likelihood. Then in Section \ref{method-subsec-posterior}, a way to represent and efficiently approximate the posterior will be discussed. Finally in Section \ref{method-subsec-summary}, a summary of the complete catheter tip tracking method will be given.

\subsection{A Deep Learning based Likelihood}
\label{method-subsec-likelihood}


Directly modeling the likelihood $p(\textbf{z}_k|\textbf{x}_k)$ is challenging due to (1) the complexity of the generation process $h_k$ and (2) the computational complexity of $p(\textbf{z}_k|\textbf{x}_k)$ for every value  $\textbf{x}_k \in \mathbb{R}^2$. In this work, we simplify the problem by only computing the likelihood $p(\textbf{z}_k|\textbf{x}_k)$ in the image pixel space, i.e. the integer pixel coordinate. For a subpixel $\textbf{x}_k$, the value of $p(\textbf{z}_k|\textbf{x}_k)$ can possibly be approximated by interpolation. To this end, we propose to use a deep neural network $\mathcal{D}$ to approximate $p(\textbf{z}_k|\textbf{x}_k)$ for integer pixel locations. The network takes an image $\textbf{z}_k$ as input and outputs a probability of observing the input $\textbf{z}_k$ for every pixel location $\textbf{x}_k$. Therefore, the approximated likelihood is a function of $\textbf{x}_k$, denoted as $\mathcal{D}_{\textbf{z}_k}(\textbf{x}_k)$. Since $\textbf{x}_k$ is defined within the scope of the image pixel space, $\mathcal{D}_{\textbf{z}_k}(\textbf{x}_k)$ is essentially a probability map having the same dimension and size with the input image $\textbf{z}_k$, in which the entry at each location $\textbf{x}_k^{j}$ ($j=1, 2, \dots, wh$) in the map represents the probability of observing $\textbf{z}_k$ given $\textbf{x}_k^{j}$. It is worth mentioning that the deep neural network is used for approximation of $p(\textbf{z}_k|\textbf{x}_k)$, which should be clearly distinguished from the generation model $h_k$ that maps an $\textbf{x}_k$ to $\textbf{z}_k$. The existence of $h_k$ is merely for the convenience of definition, its explicit form, however, is not required in the context of this work.

\begin{figure}[t]
	\centering
	\subfloat[]{
		\includegraphics[width=0.14\textwidth]{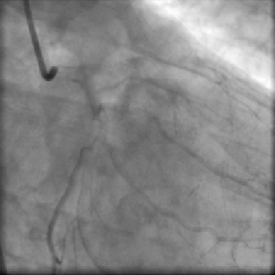}
		\label{method-subfigure-xa}
	}
	\subfloat[]{
		\includegraphics[width=0.14\textwidth]{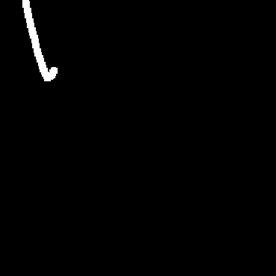}
		\label{method-subfigure-catheter-mask}
	}
	\subfloat[]{
		\includegraphics[width=0.14\textwidth]{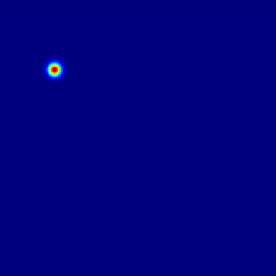}
		\label{method-subfigure-detection-heatmap}
	}
	\caption{Input and ground truth labels for the deep neural network: (a) an input X-ray fluoroscopic image, (b) the binary catheter mask of (a) for catheter segmentation, (c) a 2D Gaussian PDF ($\sigma=4$ px) for likelihood estimation for (a).}
	\label{method-figure-labels}
\end{figure}

To obtain the training labels, we assume that there exists a mapping $h_k$, such that the training label can be defined as a distance-based probability map, i.e. the farther away $\textbf{x}_k$ is from the ground truth tip location in the image $\textbf{z}_k$, the less possible it is to observe $\textbf{z}_k$ given $\textbf{x}_k$ through the process $h_k$. This definition matches the intuition that from a location $\textbf{x}_k$ that is far from the ground truth tip location, the probability of observing a $\textbf{z}_k$ with the catheter tip being located at the ground truth position should be low. For simplicity, a 2D Gaussian probability density function (PDF) $\mathcal{N}(\textbf{x}_k;\textbf{x}_k^{\prime}, \sigma^2\mathbf{\mathit{I}})$ centered at the ground truth tip location $\textbf{x}_k^{\prime}$ with variance $\sigma^2\mathbf{\mathit{I}}$ in the image space is used as the label to train the network (Figure \ref{method-subfigure-detection-heatmap}). Note that this training label makes the estimation of $p(\textbf{z}_k|\textbf{x}_k)$ equivalent to a catheter tip detection problem such that the deep neural network learns features of catheter tip and outputs high probability at locations where the features are present. Due to this reason, we also call $p(\textbf{z}_k|\textbf{x}_k)$ ``detection output'' or ``detection probability'' and call the estimation of $p(\textbf{z}_k|\textbf{x}_k)$ ``catheter tip detection'' in the context of this paper.

\begin{figure*}[t]
	\centering
	\includegraphics[width=0.98\textwidth]{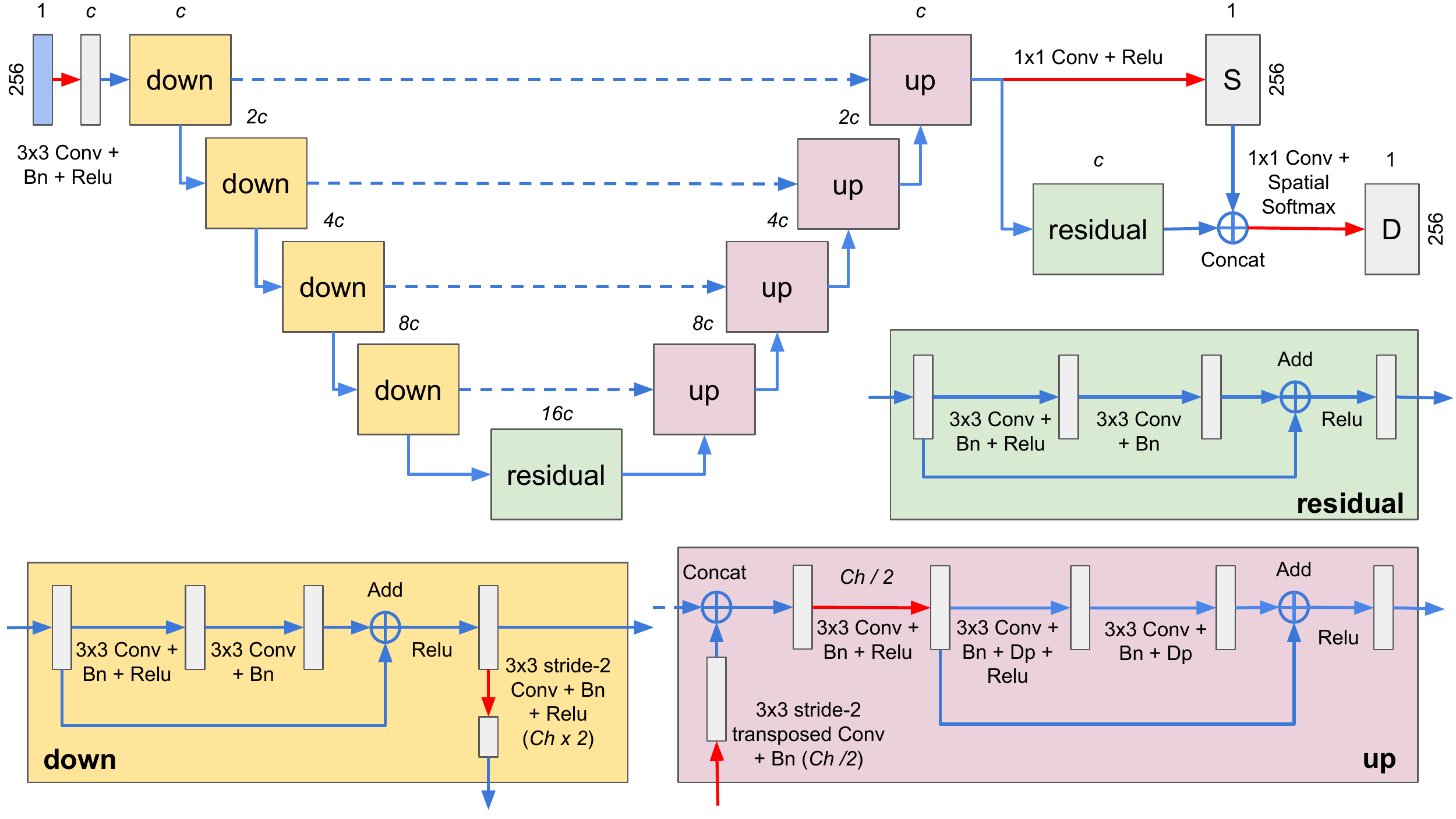}
	\caption{A joint segmentation and detection network for catheter tip detection. This figure shows an example network with 4 levels of depth (the number of down or up blocks). Meaning of abbreviations: \textit{Conv}, 2D convolution; \textit{Bn}, batch normalization; \textit{Relu}, ReLU activation; \textit{Dp}, dropout; \textit{Concat}, concatenation; \textit{Ch}, number of channels; \textit{S}, segmentation output; \textit{D}, detection output. The number above an image or feature maps indicates the number of channels; the number of channels in the residual network in a block is shown above the block; \textit{c} is the basic number of channels, the channel number in the first down block. The number next to a rectangle denotes the size of the image or feature maps. Red arrows indicate a change of number of channels.}
	\label{method-figure-network}
\end{figure*}

\indent The network that we use follows a encoder-decoder architecture with skip connections similar to U-net (\cite{ronneberger2015u}). Additionally, similar to the work by \cite{milletari2016v}, residual blocks (\cite{he2016deep}) are adopted at each resolution level in the encoder and decoder to ease gradient propagation in a deep network. The encoder consists of 4 \textit{down} blocks in which a residual block followed by a stride-2 convolution is used for extraction and down-scaling of feature maps. The number of feature maps is doubled in each downsampling step. The decoder has 4 \textit{up} blocks where a transposed convolution of stride-2 is used for upsampling of the input feature maps. Dropout is used in the residual unit of the \textit{up} block for regularization of the network. Between the encoder and the decoder, another residual block is used to process the feature maps extracted by the encoder. The detailed network architecture is shown in Figure \ref{method-figure-network}. 

\indent Due to similar appearance between a guiding catheter tip and corners of a background structure, such as vertebral bones, lung tissue, stitches or guidewires, ambiguity may exist when the network is expected to output only one blob in the probability map. To alleviate the issue, we adopt a similar strategy as \cite{laina2017concurrent}, using a catheter mask (Figure \ref{method-subfigure-catheter-mask}) as an additional label to jointly train the network to output both the catheter segmentation heatmap and the likelihood probability map. The segmentation heatmap is obtained by applying a $1\times1$ convolution with ReLU activation on the feature maps of the last \textit{up} block. To compute the likelihood probability map, a residual block is firstly applied on the feature maps of the last \textit{up} block. The output feature maps are then concatenated with the segmentation heatmap as one additional channel, followed by a $1\times1$ convolution. Finally, to ensure the network detection output fits the definition of a probability map on image locations, following the $1\times1$ convolution, a spatial softmax layer is computed as Eq.(\ref{method-eq-spatial-softmax}):

\begin{equation}
	\label{method-eq-spatial-softmax}
	D_{k,l} = \frac{e^{A_{k,l}}}{\sum_{i,j}^{}e^{A_{i,j}}} , 
\end{equation}

\noindent where $A$ is the output feature map of the $1\times1$ convolution, $A_{i,j}$ denotes the value of $A$ at location $(i,j)$, $D$ is the final output of the detection network, a 2D probability map representing $p(\textbf{z}_k|\textbf{x}_k)$. The details are shown in Figure \ref{method-figure-network}. 

\indent The training loss is defined as a combination of the segmentation loss and the detection loss. The segmentation loss $L_s$ in this work is a Dice loss defined by Eq.(\ref{method-eq-dice-loss}):

\begin{equation}
	\label{method-eq-dice-loss}
	L_s = 1 - \frac{2\sum_{i,j}M_{i,j}S_{i,j}}{\sum_{i,j}M_{i,j}^2+\sum_{i,j}S_{i,j}^2}
\end{equation}

\noindent where $M$ denotes the ground truth binary catheter masks, $S$ is the segmentation heatmap. The loss function for detection $L_d$ is mean square error (MSE) given by Eq.(\ref{method-eq-mse-loss}):

\begin{equation}
	\label{method-eq-mse-loss}
	L_d = \frac{1}{w \times h}\sum_{i \leq w,j \leq h}|T_{i,j}-D_{i,j}|^2
\end{equation}

\noindent where $T$ denotes the ground truth PDF, $w$ and $h$ are the width and height of an image. The total training loss $L$ is defined as Eq.(\ref{method-eq-total-loss}):

\begin{equation}
	\label{method-eq-total-loss}
	L = L_s + \lambda L_d
\end{equation}

\noindent where $\lambda$ is a weight to balance $L_s$ and $L_d$.

\subsection{Approximation of the Posterior with Particle Filter}
\label{method-subsec-posterior}


Once the deep neural network in Section \ref{method-subsec-likelihood} is trained, its weights are fixed during inference for computing the posterior $p(\textbf{x}_k|\textbf{z}_{0:k})$ for new data. Idealy, the network detection output $p(\textbf{z}_k|\textbf{x}_k)$ should be a Gaussian PDF during inference, as it is trained with labels of Gaussian PDFs. However, due to similar appearance of background structures or contrast residual, the detection output is unlikely to be a perfect Gaussian (possibly non-Gaussian or having multiple modes), which prevents the posterior $p(\textbf{x}_k|\textbf{z}_{0:k})$ in Eq.(\ref{method-eq-bayes-filter}) being solved with an analytical method. In practice, the posterior can be approximated using a particle filter method (\cite{arulampalam2002tutorial}). 

Particle filter methods approximate the posterior PDF by a set of $N_s$ random samples with associated weights $\{\textbf{x}_k^i, w_k^i\}_{i=1}^{N_s}$ (\cite{arulampalam2002tutorial}). As $N_s$ becomes very large, this discrete representation approaches the true posterior. According to \cite{arulampalam2002tutorial}, the approximation of the posterior $p(\textbf{x}_k|\textbf{z}_{0:k})$ is given by Eq.(\ref{method-eq-approximation}):

\begin{equation}
	\label{method-eq-approximation}
	p(\textbf{x}_k|\textbf{z}_{0:k}) \approx \sum_{i=1}^{N_s}w_k^i\delta(\textbf{x}_k-\textbf{x}_k^i)
\end{equation}

\noindent where $\delta(\cdot)$ is the Dirac delta function. The weight $w_k^i$ can be computed in a recursive manner as Eq.(\ref{method-eq-weight-update}) once $w_{k-1}^i$ is known (\cite{arulampalam2002tutorial}):

\begin{equation}
	\label{method-eq-weight-update}
	w_k^i \propto w_{k-1}^i\frac{p(\textbf{z}_k|\textbf{x}_k^i)p(\textbf{x}_k^i|\textbf{x}_{k-1}^i)}{q(\textbf{x}_k^i|\textbf{x}_{k-1}^i, \textbf{z}_k)}
\end{equation}

\noindent where $q(\textbf{x}_k|\textbf{x}_{k-1}^i, \textbf{z}_k)$ is an importance density from which it should be possible to sample $\textbf{x}_k^i$ easily. For simplicity, a good and convenient choice of the importance density is the prior $p(\textbf{x}_k|\textbf{x}_{k-1}^i)$ (\cite{arulampalam2002tutorial}), so that the weight update rule (\ref{method-eq-weight-update}) becomes $w_k^i \propto w_{k-1}^i p(\textbf{z}_k|\textbf{x}_k^i)$.

A sample can be drawn from $p(\textbf{x}_k|\textbf{x}_{k-1}^i)$ in the following way. First, a process noise sample $\textbf{v}_{k-1}^i$ is sampled from $p_v(\textbf{v}_{k-1})$, the PDF of $\textbf{v}_{k-1}$; then $\textbf{x}_k^i$ is generated from $\textbf{x}_{k-1}^i$ via the state transition model $\textbf{x}_k^i=f_k(\textbf{x}_{k-1}^i, \textbf{v}_{k-1}^i)$. In this work, $p_v(\textbf{v}_{k-1})$ is set to be a Gaussian $\mathcal{N}(\textbf{0},\, \sigma_v^2\mathbf{\mathit{I}})$. The choice of motion model for $f_k$ is important for an accurate representation of the true state transition prior $p(\textbf{x}_k|\textbf{x}_{k-1})$. A random motion cannot characterize well the motion of catheter tip in XA frames. In this work, we estimated the motion from adjacent frames using an optical flow method, as this approach 1) takes into account of the observation $\textbf{z}_k$, which results in a better guess of the catheter tip motion, and 2) enables estimation of a dense motion field where the motion of a sample $\textbf{x}_k^i$ can be efficiently obtained. Therefore, $f_k$ is defined as Eq.(\ref{method-eq-state-transition-flow}):

\begin{equation}
	\label{method-eq-state-transition-flow}
	\textbf{x}_k=\textbf{x}_{k-1}+\textbf{u}_{k-1}(\textbf{x}_{k-1})+\textbf{v}_{k-1}
\end{equation}

\noindent where $\textbf{u}_{k-1}(\cdot)$ is the motion from frame $k\!-\!1$ to frame $k$ estimated with optical flow using the method by \cite{farneback2003two}, $\textbf{u}_{k-1}(\textbf{x}_{k-1})$ is the motion from state $\textbf{x}_{k-1}$.

Once samples are drawn and their weights are updated, the so-called ``resampling'' of the samples should be performed to prevent the degenaracy problem, where all but one sample will have negligible weight after a few iterations (\cite{arulampalam2002tutorial}). The resampling step resamples the existing samples according to their updated weights and then resets all sample weights to be $1/N_s$, so the number of effective samples which have actual contribution to approximate $p(\textbf{x}_k|\textbf{z}_{0:k})$ is maximized (\cite{arulampalam2002tutorial}). If the resampling is applied at every time step, the particle filter becomes a sampling importance resampling (SIR) filter, and the weight update rule follows Eq.(\ref{method-eq-weight-update-simplest}).

\begin{equation}
	\label{method-eq-weight-update-simplest}
	w_k^i \propto p(\textbf{z}_k|\textbf{x}_k^i)
\end{equation}

\noindent The final decision on catheter tip location in frame $k$  can then be computed as the expectation of $\textbf{x}_k$, $\hat{\textbf{x}}_k =\int\textbf{x}_kp(\textbf{x}_k|\textbf{z}_{0:k})d\textbf{x}_k$, which is in this case, the weighted sum of all samples:

\begin{equation}
	\label{method-eq-final-tip-location}
	\hat{\textbf{x}}_k = \sum_{i=1}^{N_s}w_k^i\textbf{x}_k^i.
\end{equation}

\subsection{Summary}
\label{method-subsec-summary}


The overall catheter tip tracking using a deep learning based Bayesian filtering method is summarized in Algorithm \ref{method-alg-dl-bayes}.

%
\begin{algorithm*}
	\caption{Deep learning based Bayesian filtering for online tracking of catheter tip in X-ray fluoroscopy}
	\label{method-alg-dl-bayes}
	\begin{algorithmic}[1]
		\REQUIRE $\{\textbf{z}_0, \dots, \textbf{z}_T\}$ (sequentially observed frames), $\mathcal{D}$ (A trained network from Section \ref{method-subsec-likelihood}), $p(\textbf{x}_0)$ (the initial PDF), $\sigma_v^2$ (the variance of $\textbf{v}_{k-1},\, k=1, \dots, T$), $T$ (number of frames for tracking), $N_s$ (number of samples)
		\STATE{Draw $\textbf{x}_0^i \sim p(\textbf{x}_0)$, set $w_0^i=1/N_s$, $\forall\, i=1, \dots, N_s$}
		\FOR{$k$ = 1 \TO $T$}
			\STATE{Compute $\textbf{u}_{k-1}$ from $\textbf{z}_{k-1}$ to $\textbf{z}_k$ using the optical flow method of \cite{farneback2003two}}
			\FOR{$i$ = 1 \TO $N_s$}
				\STATE{Draw $\textbf{v}_{k-1}^i \sim \mathcal{N}(\textbf{0},\, \sigma_v^2\mathbf{\mathit{I}})$}
				\STATE{Compute the motion of $\textbf{x}_{k-1}^i$: $\textbf{u}_{k-1}^i = \textbf{u}_{k-1}(\textbf{x}_{k-1}^i)$}
				\STATE{Draw $\textbf{x}_k^i \sim p(\textbf{x}_k|\textbf{x}_{k-1}^i)$: $\textbf{x}_k^i = \textbf{x}_{k-1}^i+\textbf{u}_{k-1}^i+\textbf{v}_{k-1}^i$}
				\STATE{Update weight $w_k^i = p(\textbf{z}_k|\textbf{x}_k^i) = \mathcal{D}_{\textbf{z}_k}(\textbf{x}_k^i)$}
			\ENDFOR
			\STATE{Normalize $w_k^i \leftarrow w_k^i/\sum_{i=1}^{N_s}w_k^i$, $\forall\, i=1, \dots, N_s$}
			\STATE{Prediciton in frame $k$:  $\hat{\textbf{x}}_k = \sum_{i=1}^{N_s}w_k^i\textbf{x}_k^i$ }
			\STATE{Resample $\{\textbf{x}_k^i, w_k^i\}_{i=1}^{N_s}$ using the method of \cite{arulampalam2002tutorial}} (so all $w_k^i$ are set to $1/N_s$ again)
		\ENDFOR
		
	\end{algorithmic}
\end{algorithm*}

%% file: sections/setup/setup.tex

\section{Experimental Setup}


\subsection{Data}
\label{setup-subsec-data}

Anonymized clinical imaging data were used for our experiments. The data were acquired with standard clinical protocol using Siemens AXIOM-Artis system, and are from 55 patients who underwent a PCI procedure at the Department of Cardiology at Erasmus MC in Rotterdam, Netherlands. Out of these data, we selected data from 37 patients which were acquired since the year 2014 to develop our method, and used the data from the other 18 patients acquired before the year 2013 for evaluation. The detailed information about the data is listed in Table \ref{setup-tab-data}.

\begin{table}
	\centering
	\caption{Basic information of the acquired X-ray image data for our experiments. The number in the parenthesis next to the pixel size indicates the possible image size.}
	\begin{tabular}{l l l}
		\toprule
	 	Data & Development & Evaluation \\
	 	\midrule
	 	No. patients & 37 & 18 \\
	 	No. sequences & 354 & 34 \\
	 	Frame rate (fps) & 15 & 15 \\
	 	Image size (px) & 512 $\times$ 512 & 512 $\times$ 512 \\
	 	                & 600 $\times$ 600 & 600 $\times$ 600 \\
	 	                & 776 $\times$ 776 & 776 $\times$ 776 \\
	 	                & 960 $\times$ 960 & 1024 $\times$ 1024 \\
	 	                & 1024 $\times$ 1024 &                    \\
	 	Pixel size (mm) & 0.108 (1024) & 0.139 (1024) \\
	 	                & 0.139 (1024) & 0.184 (600) \\
	 	                & 0.184 (600) & 0.184 (776) \\
	 	                & 0.184 (776) & 0.184 (1024) \\
	 	                & 0.184 (960) & 0.216 (512) \\
	 	                & 0.184 (1024) & 0.279 (512) \\
	 	                & 0.216 (512) &              \\
	 	\bottomrule	
	\end{tabular}
	\label{setup-tab-data}
\end{table}

In order to evaluate the proposed roadmapping method, for which an off-line angiographic sequence is required for roadmap preparation and an online fluoroscopic sequence taken from the same C-arm position is needed for performing the actual roadmapping (see Section \ref{method-sec-overview}), we selected the contrast frames from a real clinical sequence to simulate the off-line sequence, and chose the non-contrast frames from the same clinical sequence to simulate the online sequence. The selected contrast sequence were ensured sufficiently long to cover at least one complete cardiac cycle.

\subsection{Data Split for Catheter Tip Detection and Tracking}
\label{setup-subsec-tracking}

To develop the catheter tip tracking method, 1086 X-ray fluoroscopic images selected from 260 non-contrast sequences of 25 patients from the development set were used for training the network from Figure \ref{method-figure-network}; 404 images from 94 non-contrast sequences of another 12 patients from the development set were used as validation set for the network model and hyperparameter selection. In the training and validation sets, 4-5 frames were randomly selected from each sequence, which are not necessarily continuous. To tune the parameters for tracking, 1583 images from 88 sequences out of the 94 from the same 12 patients of the validation set were used (6 sequences were not selected for this task due to very short sequence length not more than 5 frames). Finally, to evaluate catheter tip tracking accuracy, 1355 images from 34 non-contrast sequences of 18 patients from the evaluation set were used for testing. The frames selected for tracking from each sequence must be continuous; the number of selected frames for tracking might vary, depending on the number of the non-contrast frames in the sequences. Details of the datasets for training, validation and test are listed in Table \ref{setup-tab-tip-experiment}. 

\begin{table*}
	\centering
	\caption{Dataset of training, validation and test for detection and tracking of catheter tip in X-ray fluoroscopic frames.}
	\begin{tabular}{l l l l l}
		\toprule
	 	 & Training & Validation & Validation & Test \\
	 	 & (detection) & (detection) & (tracking) & (tracking) \\
		\midrule
		No. patients & 25 & 12 & 12 & 18 \\
		No. sequences & 260 & 94 & 88 & 34 \\
		No. frames   & 1086 & 404 & 1583 & 1355 \\
		Continous frames? & No & No & Yes & Yes \\
		\bottomrule	
	\end{tabular}
	\label{setup-tab-tip-experiment}
\end{table*}

\subsection{Experimental Settings for Training the Deep Network}
\label{setup-subsec-trainingnetwork}

This section describes the basic experimental settings for training the deep neural network. Details of the training setup can be found in Appendix \ref{app-subsec-training-setup}.

\subsubsection{Preprocessing}
As the image data have different size ranging from $512\times512$ to $1024\times1024$, all images were resampled to a grid of $256\times256$ before being processed by the neural network. In addition, the image intensities were rescaled to the range from 0 to 1.

\subsubsection{Training label}
The standard deviation $\sigma$ of the Gaussian PDF for the training label of the detection network was set to 4 pixels in the resampled image space ($256\times256$). This choice corresponds to the estimation of the maximal possible catheter tip radius. An example of the Gaussian PDF is shown in Figure \ref{method-subfigure-detection-heatmap}.

\subsubsection{Evaluation Metric}
To select hyperparameters and model weights in training, an evaluation metric is required. As the deep network is essentially a catheter tip detector, accurate detection of the tip location is desired. Therefore, we chose the location with the highest value in the detection output, and computed the Euclidean distance between the chosen location and the ground truth tip coordinate as the evaluation metric to tune the deep network.

\subsection{Setup for Evaluating Dynamic Coronary Roadmapping}
\label{setup-subsec-roadmapping}

It is in general a challenge to evaluate the roadmapping accuracy, as the structure of interest, e.g. coronary arteries in our case, is not directly visible in the target image. One possible choice introduced by \cite{zhu2010image} is to use the guidewire as a surrogate of the target vessel centerline in non-contrast images, as guidewire is always inside vessels and commonly present in image sequences during interventions. In this work, we follow a similar strategy to evaluate the accuracy of dynamic coronary roadmapping.

The first step is to select frames for roadmapping evaluation. From each non-contrast sequence in the test set for tracking in Section \ref{setup-subsec-tracking}, we uniformly select 8-20 frames to annotate guidewire. The number of the selected frames from each sequence depends on the sequence length, the frame interval size and guidewire visibility. For some rare cases in our data where no guidewire is present in the image, we discarded that non-contrast frame, and chose those frames with little vessel contrast from the same sequence and annotated the vessel centerline. The selection results in 409 frames from 34 sequences in total. Once the target non-contrast frames for evaluating roadmapping are chosen, their corresponding angiographic frames were found using the ECG matching method in Section \ref{method-sec-roadmap}. We then annotated the centerline of the vessel corresponding to the guidewire in the non-contrast frames.

The next step is performing the transformation of the labelled vessel centerline from the angiographic frame to its corresponding target non-contrast frame via displacement of catheter tip in the two frames. This step simulates the roadmapping transformation in the last step in Figure \ref{method-figure-overview}. 

\begin{figure}[h]
	\centering
	\subfloat[]{
		\includegraphics[width=0.15\textwidth]{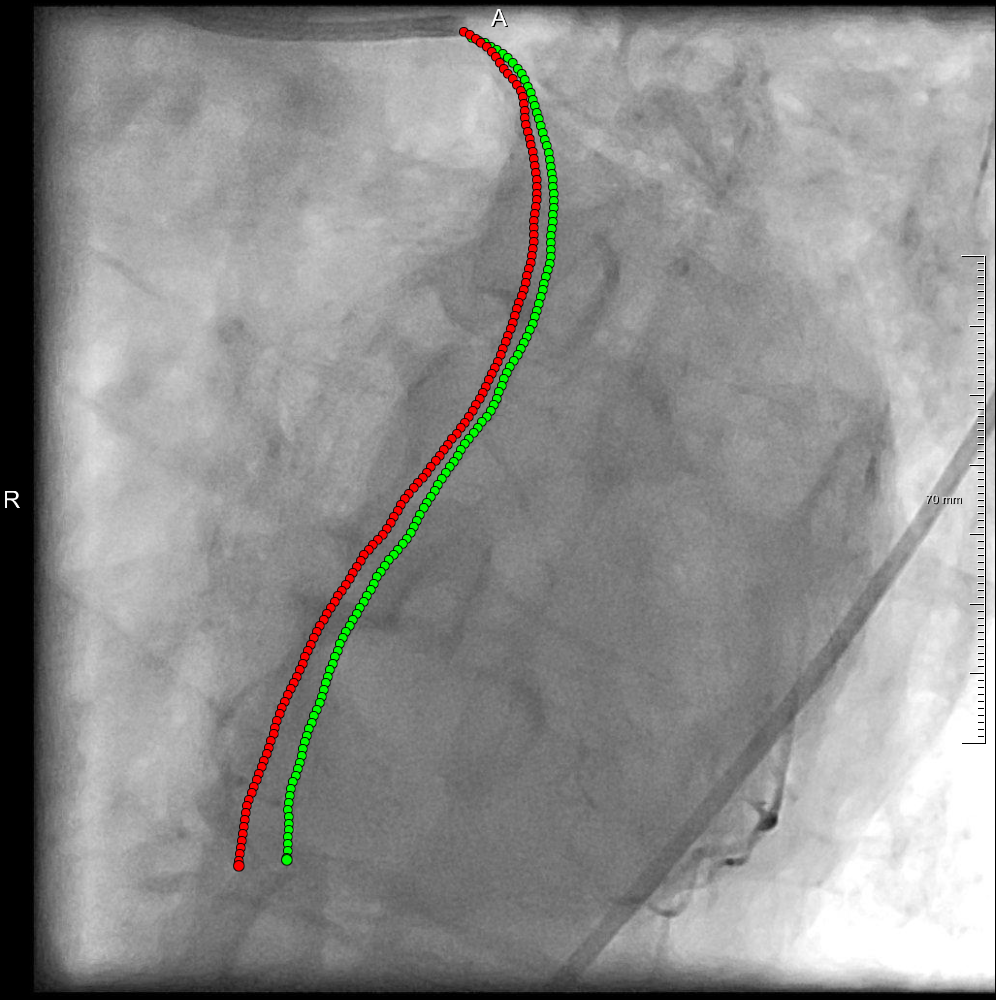}
		\label{setup-subfigure-without-vector}
	}
	\subfloat[]{
		\includegraphics[width=0.15\textwidth]{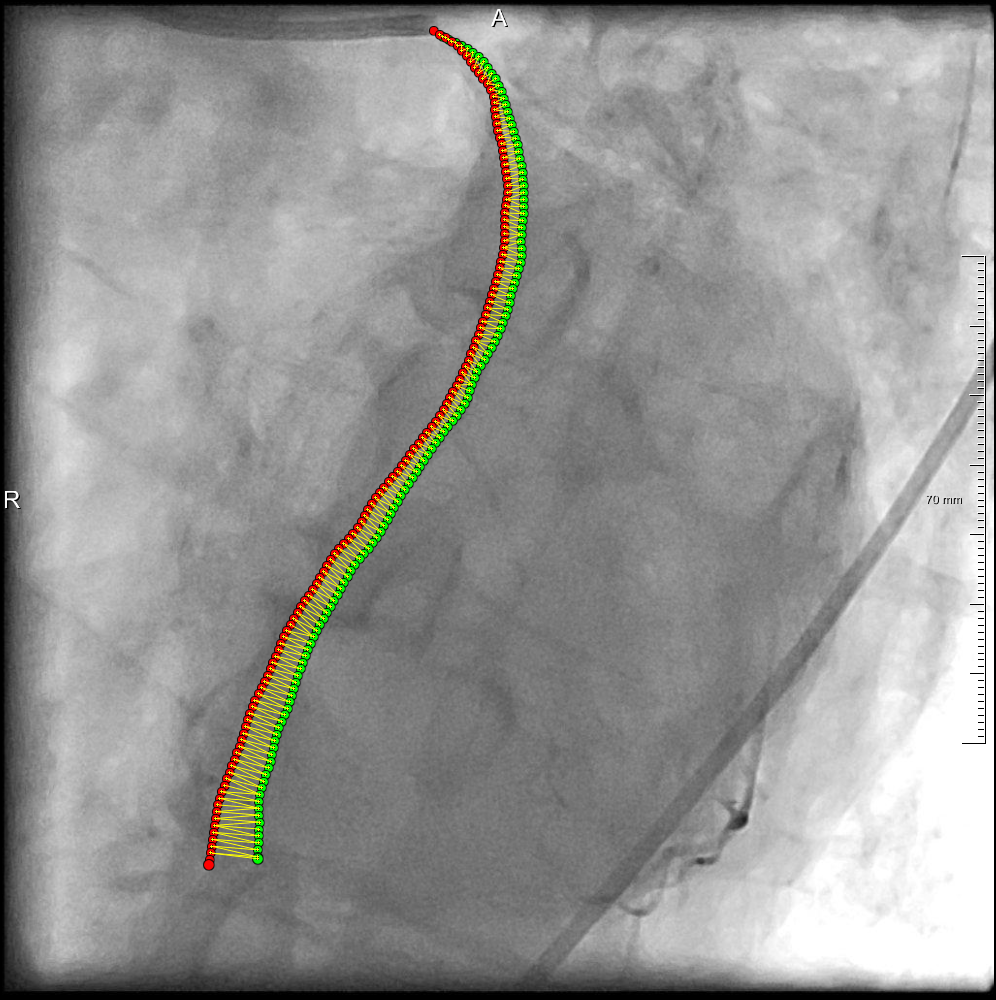}
		\label{setup-subfigure-with-vector}
	}
	\subfloat[]{
		\includegraphics[width=0.15\textwidth]{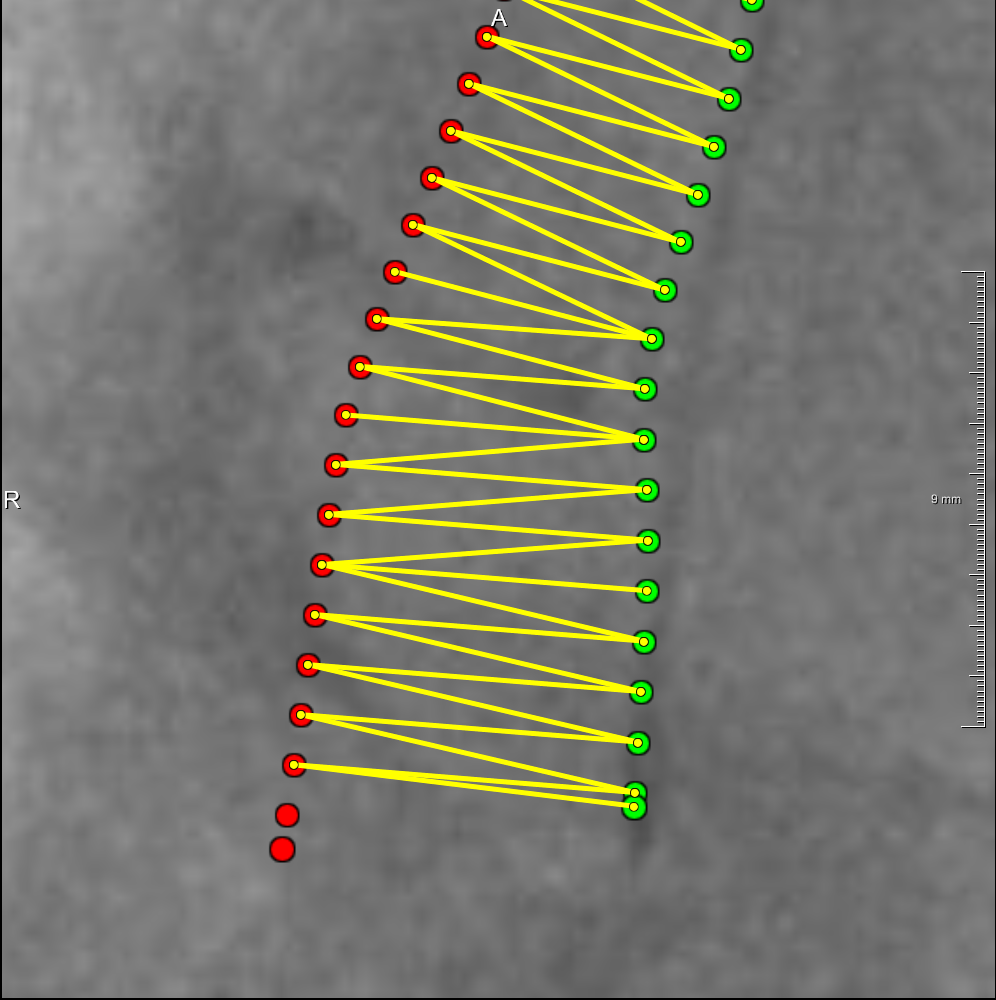}
		\label{setup-subfigure-with-vector-zoomin}
	}
	\caption{Correspondence between the labelled guidewire (green) and the transformed vessel centerline (red). The yellow lines connecting the two point sets illustrate the correspondence between red and green points.}
	\label{setup-figure-distance-vector}
\end{figure}

Finally, the distance between the guidewire annotation in the target frame and the transformed vessel centerline is reported as the roadmapping accuracy. In order to compute the distance between two point sets of annotations (e.g. Figure \ref{setup-subfigure-without-vector}), point-point correspondence between the two sets is required (Figure \ref{setup-subfigure-with-vector}). The point sets were firstly resampled with the point interval being 1 mm. We then followed the approach of \cite{van2008averaging} to find such correspondences which minimizes the sum of the Euclidean distance of all valid point-point correspondence paths. This way guarantees no cross-over connection and each point in one set is connected to at least one point in the other set. As the annotated point sets may have different size, the point correspondences to endpoints are excluded such that we only focused on the distance between corresponding sections, not the entire centerlines (Figure \ref{setup-subfigure-with-vector-zoomin}). Once the point-point correspondence is available, the distance between the two points in a pair can be used for evaluating the accuracy of DCR.

\subsection{Implementation}
\label{setup-setup-implementation}

The proposed method was developed in Python. The framework used for developing the deep learning approach for likelihood approximation is PyTorch. The major experiments of dynamic coronary roadmapping were performed on a computer with an Intel Xeon E5-2620 v3 2.40 GHz CPU  and 16 GB RAM running Ubuntu 16.04. The deep neural network and the optical flow method were running on an NVIDIA GeForce GTX 1080 GPU. The approach for evaluating dynamic coronary roadmapping was developed and running in MeVisLab on a computer with an Intel Core i7-4800MQ 2.70 GHz CPU and 16 GB RAM running Windows 7.

%% file: sections/experiments/experiments.tex

\section{Experiments and Results}
\label{exp-sec}


The following experiments are performed to assess the methods. First, In Section \ref{exp-subsec-detection}, the training of the deep neural network is described. Then in Section \ref{exp-subsec-tracking}, the accuracy of catheter tip tracking using the optimized trained network and the tuned particle filter is presented. Section \ref{exp-subsec-roadmapping} describes the accuracy evaluation of dynamic coronary roadmapping via the proposed catheter tip tracking method. Finally, in Section \ref{exp-subsec-time}, we measure the processing time of the proposed DCR approach.

\subsection{Training the Deep Neural Network}
\label{exp-subsec-detection}

The purpose of this experiment is to train the deep neural network to output reasonable likelihood probability map. The network hyperparameters were tuned to optimize the detection performance. 

The training and validation data for detection mentioned in Section \ref{setup-subsec-tracking} were used for training the deep neural network. The evaluation metric mentioned in Section \ref{setup-subsec-trainingnetwork}, the mean Euclidean distance between the ground truth and the predicted tip location averaged over all validation frames, was used as the validation criteria for selecting the optimal training epoch and the network hyperparameters. When we evaluated hyperparameter settings, we firstly selected the training epoch with the lowest mean validation error for each setting, then the settings were compared using the model weights (trainable network parameters) of their chosen epochs.

The network hyperparameters we investigated in the experiments include (1) the basic channel number, i.e. the number of channels or feature maps in the first down block, (2) the network depth level, the number of down or up blocks, and (3) the dropout probability.

The validation errors for different hyperparameter settings using the experimental settings in Section \ref{setup-subsec-trainingnetwork} are shown in Table \ref{exp-tab-training-network}. The table shows that the hyperparameter setting with the lowest mean error, which has 4 level in depth and 64 channels in the first down block, achieves a validation error of about 2 mm. The table also shows other good choices of network architecture that have a small validation error (shown in red in Table \ref{exp-tab-training-network}): 32 channels in the first down block with 4 or 5 levels in depth, or 64 channels with 3 or 4 depth levels. The dropout regularization improves the accuracy of the model, compared to the ones without dropout.

\begin{figure}[t]
	\centering
	\subfloat[Total loss]{
		\includegraphics[width=0.23\textwidth]{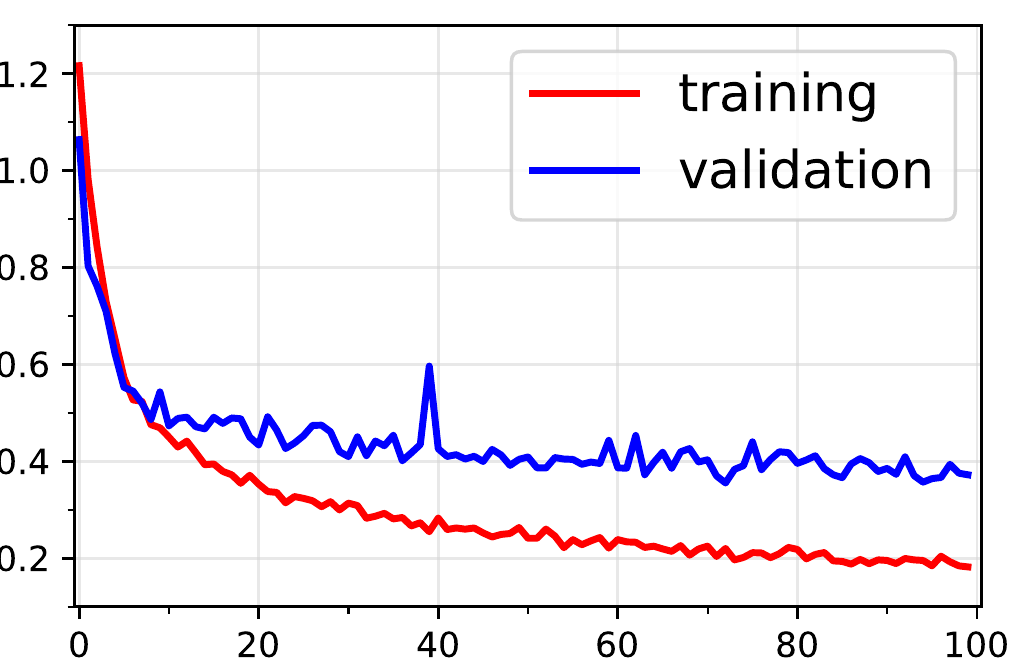}
		\label{exp-subfigure-lineplot-total-loss}
	}
	\subfloat[Detection error (mm)]{
		\includegraphics[width=0.23\textwidth]{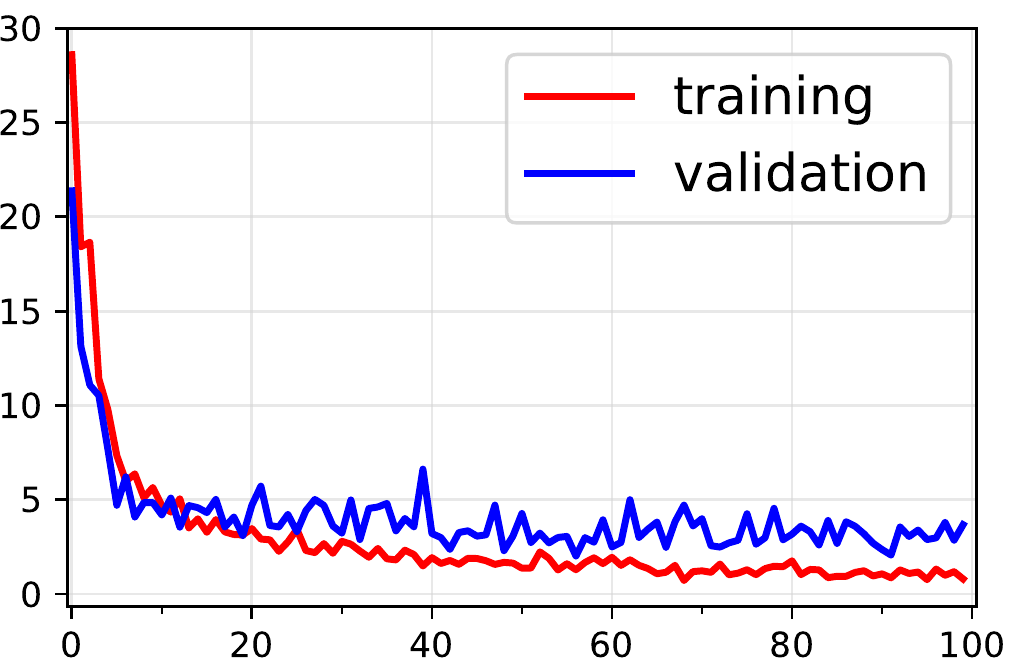}
		\label{exp-subfigure-lineplot-detection-error}
	}
	\\[-1ex]
	\subfloat[Segmentation loss]{
		\includegraphics[width=0.23\textwidth]{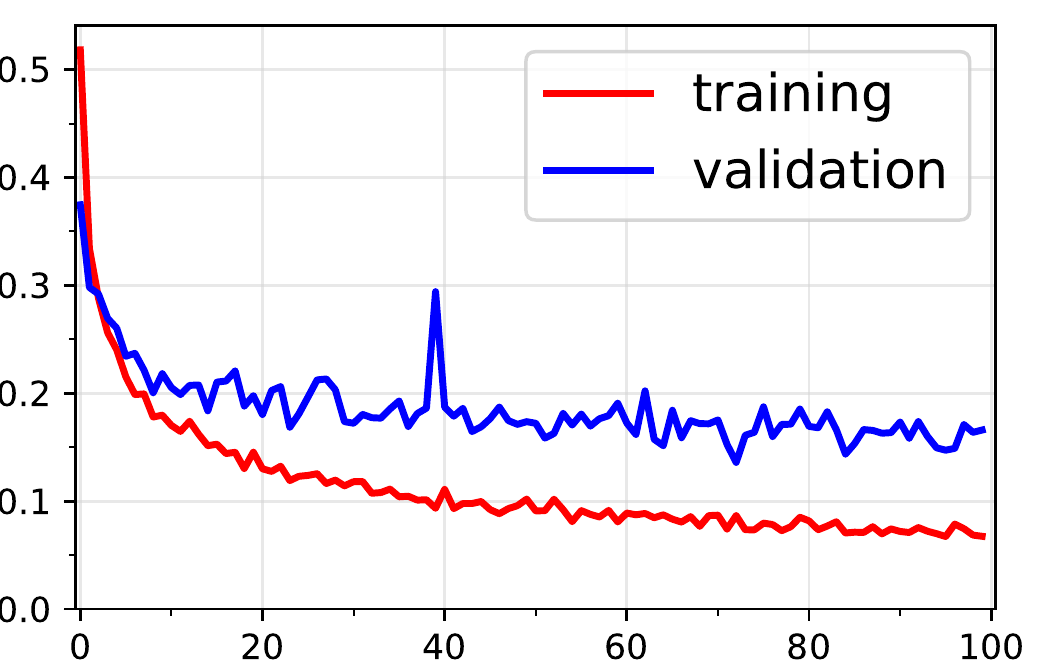}
		\label{exp-subfigure-lineplot-seg-loss}
	}
	\subfloat[Detection loss]{
		\includegraphics[width=0.23\textwidth]{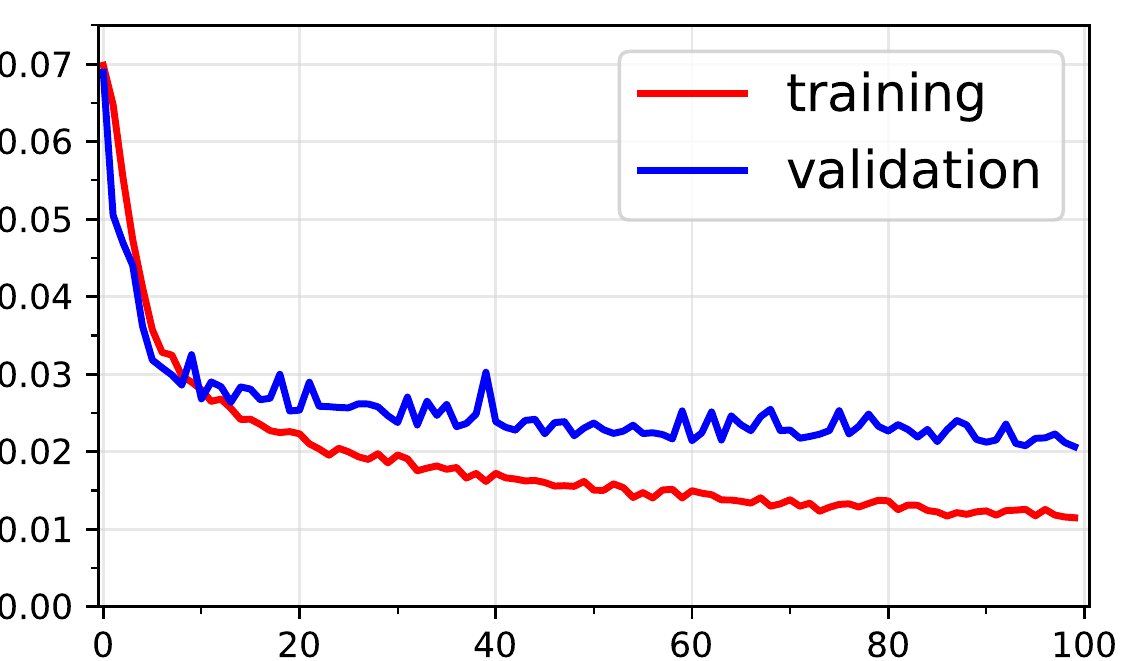}
		\label{exp-subfigure-lineplot-det-loss}
	}
	\caption{Learning curves for the chosen hyperparameter setting.}
	\label{exp-figure-learning-curves}
\end{figure}

The learning curves of the training process with the chosen hyperparameter setting are illustrated in Figure \ref{exp-figure-learning-curves}. The curves indicate that both segmentation and detection reach convergence after training 100 epochs.

We did not investigate a model with more than 64 channels or 5 depth levels, because (1) it will further increase the processing time which makes online applications less feasible; (2) the results in Table \ref{exp-tab-training-network} show that such a setting (64 channels, 5 depth levels) starts increasing the validation error compared to those less complex models.

The subsequent experiments will be based on the network trained with the chosen hyperparameter setting indicated in Table \ref{exp-tab-training-network} (64 channels, 4 depth levels, dropout 0.2, also see Table \ref{exp-tab-all-hyperparameters}).

\begin{table*}[h]
	\centering
	
	\caption{Validation errors (mm) for different hyperparameter settings. Red cells show the settings with the 10 smallest validation errors. \textit{bold} number indicates the setting with the lowest error.}
	
	\begin{tabular}{c c c c c c c c}
		\toprule
		Basic Number & Depth & \multicolumn{6}{c}{Dropout} \\
		\cmidrule(lr){3-8}
		of Channels & Level & none & 0.1 & 0.2 & 0.3 & 0.4 & 0.5 \\
		\midrule 
		8  & 3 & 5.43 & 4.99 & 5.02 & 5.37 & 4.38 & 4.24 \\
		   & 4 & 4.17 & 4.45 & 4.25 & 5.04 & 4.75 & 4.36 \\
		   & 5 & 3 & 4.14 & 3.53 & 4.28 & 3.95 & 4.11 \\
		16 & 3 & 3.74 & 4.29 & 3.57 & 4.11 & 3.74 & 3.4 \\
		   & 4 & 3.36 & 3.11 & 3.63 & 3.33 & 3.36 & 3.78 \\
		   & 5 & 3.38 & 2.89 & 3.16 & 2.52 & 2.71 & 2.74\\
		32 & 3 & 2.99 & 3.02 & 3.26 & 2.82 & 3.26 & 2.56 \\
		   & 4 & 2.87 & 2.34 & 2.46 & 2.6 & 2.65 & \cellcolor{red!30}2.27 \\
		   & 5 & 3.04 & 2.51 & \cellcolor{red!30}2.21 & \cellcolor{red!30}2.29 & 2.3 & \cellcolor{red!30}2.25 \\
		64 & 3 & \cellcolor{red!30}2.19 & 2.54 & 2.34 & \cellcolor{red!30}2.27 & \cellcolor{red!30}2.26 & 2.49 \\
		   & 4 & 2.55 & 2.31 & \cellcolor{red!30}\textbf{2.04} & 2.44 & \cellcolor{red!30}2.22 & \cellcolor{red!30}2.27 \\
		   & 5 & 2.42 & \cellcolor{red!30}2.29 & 2.73 & 2.77 & 2.61 & 2.85 \\
		\bottomrule	
	\end{tabular}
	
	\label{exp-tab-training-network}
\end{table*}

\subsection{Catheter Tip Tracking}
\label{exp-subsec-tracking}

The purpose of this experiment is to assess the accuracy of catheter tip tracking with the proposed method in Section \ref{method-sec-bayesian-filter}. Guiding catheter tip is tracked in X-ray fluoroscopy using Algorithm \ref{method-alg-dl-bayes} based on a trained network with the optimal hyperparameter setting from Section \ref{exp-subsec-detection}. First, the parameters of the optical flow method used in Algorithm \ref{method-alg-dl-bayes} and particle filter were tuned on the validation data for tracking in Section \ref{setup-subsec-tracking} (see Appendix \ref{app-subsec-parameter-tuning} for details). Then in Section \ref{exp-subsubsec-tracking}, we evaluated the tracking accuracy with the tuned optimal parameter setting (see Table \ref{exp-tab-all-hyperparameters}) on the test dataset, and compared the proposed tracking method with alternative approaches using only the detection network in Section \ref{method-subsec-likelihood} or using only optical flow. Finally, in Section \ref{exp-subsubsec-initialization}, we investigated tracking accuracy with different ways of tip initialization in the first frame.


\begin{table}[h]
	\centering
	
	\caption{The chosen (hyper-)parameters for different building blocks of the catheter tip tracking algorithm. The parameters of the optical flow method can be found in Appendix \ref{exp-subsubsec-of}.}
	
	\begin{tabular}{l l l}
		\toprule
		Building block & (hyper-)parameters & value \\ 
		\midrule
		Deep learning & Basic channel number & 64 \\
		                         & Depth  & 4\\
		                         & Dropout & 0.2 \\ 
		\midrule
		Particle filter & $\sigma_v$ (px) & 5  \\
		                       & $N_s$ & 1000 \\ 
		\bottomrule
	\end{tabular}
	
	\label{exp-tab-all-hyperparameters}
\end{table}

\subsubsection{Tracking Methods Evaluation}
\label{exp-subsubsec-tracking}

In this experiment, the proposed tracking method in Algorithm \ref{method-alg-dl-bayes} uses the ground truth tip probability map of the first frame as the initial PDF $p(\textbf{x}_0)$ to draw samples. This method is referred to as ``Tracking''. In addition, we compared the proposed method with three alternatives. The first one tracks catheter tip using only the detection network in Section \ref{method-subsec-likelihood} with the chosen network architecture and trained parameters in Section \ref{exp-subsec-detection}, therefore, no temporal information is used. This method is referred to as ``Detection (Net)''. The other two methods in this experiment use only optical flow to track catheter tip starting from the ground truth tip position in the first frame. The motion field towards the current frame, estimated by the two methods, was based on the deformation from the previous frame or the first frame in the sequence, respectively. The same implementation setting as in Appendix \ref{exp-subsubsec-of} was used for these two methods. They are called ``Optical Flow (previous)'' and ``Optical Flow (first)'', or in short form, ``OF (pre)'' and ``OF (1st)''. Additionally, we refer the interested readers to Appendix \ref{app-subsubsec-detection-tracking} where the influence of catheter segmentation on the detection and tracking approaches is reported.

The tracking accuracies of all methods reported in this section were obtained on the test set from Table \ref{setup-tab-tip-experiment}. The mean, the median and the maximal tracking error between the predicted and the ground truth tip position of all test images are reported in Table \ref{exp-tab-tracking-test}. In addition, as the sequences in the test set have different lengths, we also computed the mean and the median error per sequence, and report the the average of the sequence mean and median errors, so that each sequence contributes equally in these metrics. Table \ref{exp-tab-tracking-test} shows that the results from the detection network have large average errors which are caused by some completely failed cases. The proposed tracking method has median errors of about 1 mm and mean errors of about 1.3 mm. It achieves the lowest errors compared to the other 3 methods on all listed evaluation criteria.

\begin{table*}[h]
	\centering
	
	\caption{Catheter tip tracking errors (mm) of the 4 methods on the test (tracking) dataset. $\dagger$ indicates that the difference between that method and the ``Tracking'' method are statistically highly significant with the two-sided Wilcoxon signed-rank test ($p < 0.001$).}
	
	\begin{tabular}{l c c c c}
		\toprule
		\multirow{2}{*}{Evaluation Metrics} & Optical Flow$\dagger$ & Optical Flow$\dagger$ & Detection Net$\dagger$ & Tracking \\
		& (previous) & (first) & (Section \ref{method-subsec-likelihood}) &  \\
		\midrule 
		Maximal error of all images & 29.16 & 20.83 & 108.20 & \textbf{17.72} \\
		Median error of all images & 1.78 & 1.22 & \textbf{0.96} & \textbf{0.96} \\
		Mean error of all images & 3.74 $\pm$ 4.93 & 3.05 $\pm$ 4.05 & 5.62 $\pm$ 15.91 & \textbf{1.29 $\pm$ 1.76} \\
		\midrule
		Average of sequence median error & 2.35 $\pm$ 2.52 & 2.64 $\pm$ 3.52 & 6.26 $\pm$ 17.11 & \textbf{1.03 $\pm$ 0.49} \\
		Average of sequence mean error & 2.59 $\pm$ 2.69 & 3.31 $\pm$ 2.81 & 6.83 $\pm$ 13.88 & \textbf{1.29 $\pm$ 0.94} \\
		\bottomrule	
	\end{tabular}
	
	\label{exp-tab-tracking-test}
\end{table*}

\begin{figure}[h]
	\centering
	\subfloat[Overall view of tracking errors]{
		\includegraphics[width=0.23\textwidth]{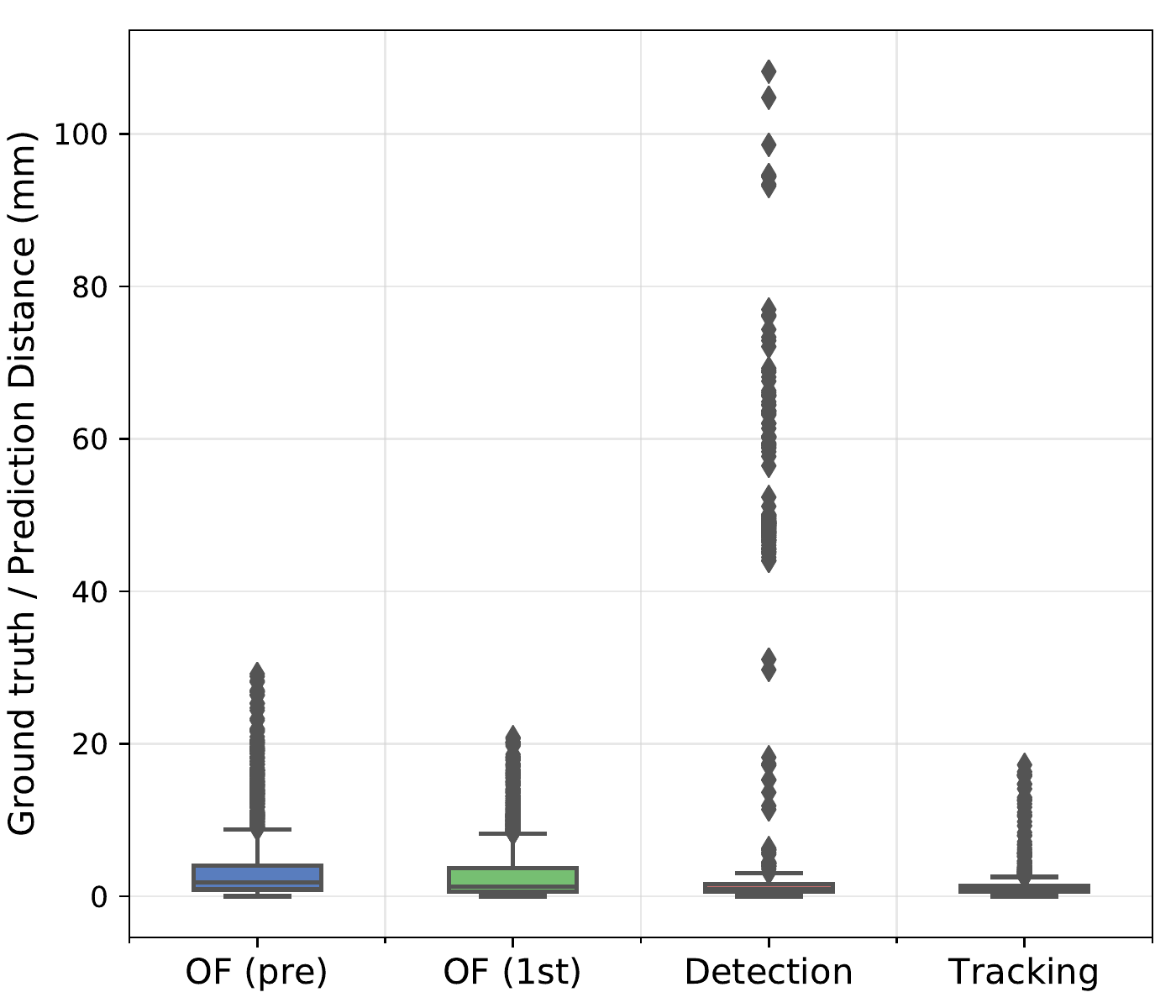}
		\label{exp-subfigure-boxplot-all-dist}
	}
	\subfloat[A zoom-in view of (a)]{
		\includegraphics[width=0.23\textwidth]{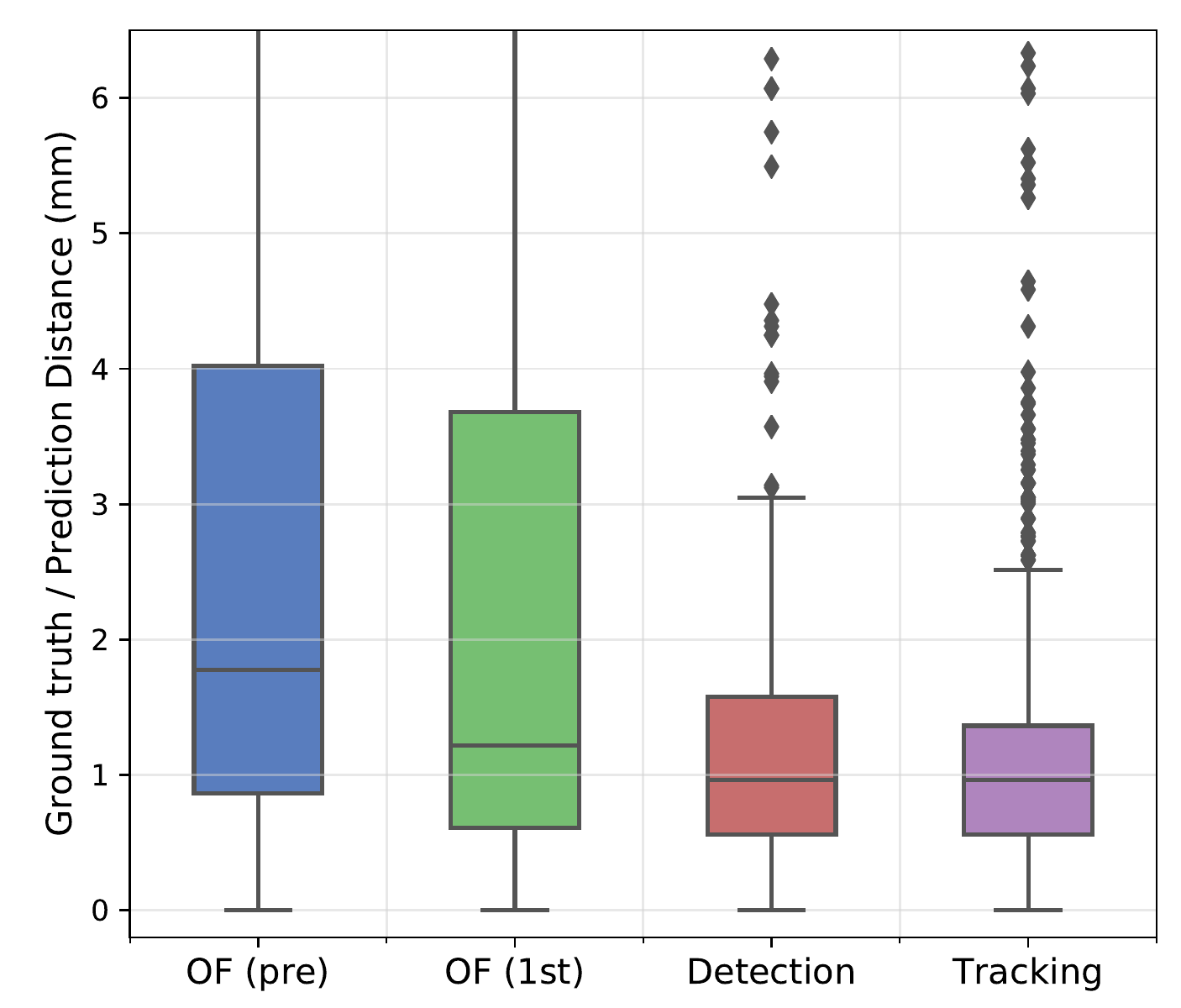}
		\label{exp-subfigure-boxplot-all-dist-zoom-in}
	}
	\caption{Tracking errors for the 4 methods on all test images.}
	\label{exp-figure-boxplot-all_dist}
\end{figure}

Figure \ref{exp-figure-boxplot-all_dist} illustrates the boxplots of tracking errors made by the 4 methods on all test images. It shows that the proposed tracking approach outperforms the detection method by avoiding making extremely large errors (Figure \ref{exp-subfigure-boxplot-all-dist}); meanwhile, it maintains as accurate as the detection method for cases with small errors, and is more accurate than the methods based solely on optical flow (Figure \ref{exp-subfigure-boxplot-all-dist-zoom-in}).

\begin{figure*}[t]
	\centering
	\subfloat{
		\parbox[t]{0.23\textwidth}{
		\centering
		{\normalsize Optical Flow (previous)} \\[1ex]
		\includegraphics[width=0.23\textwidth]{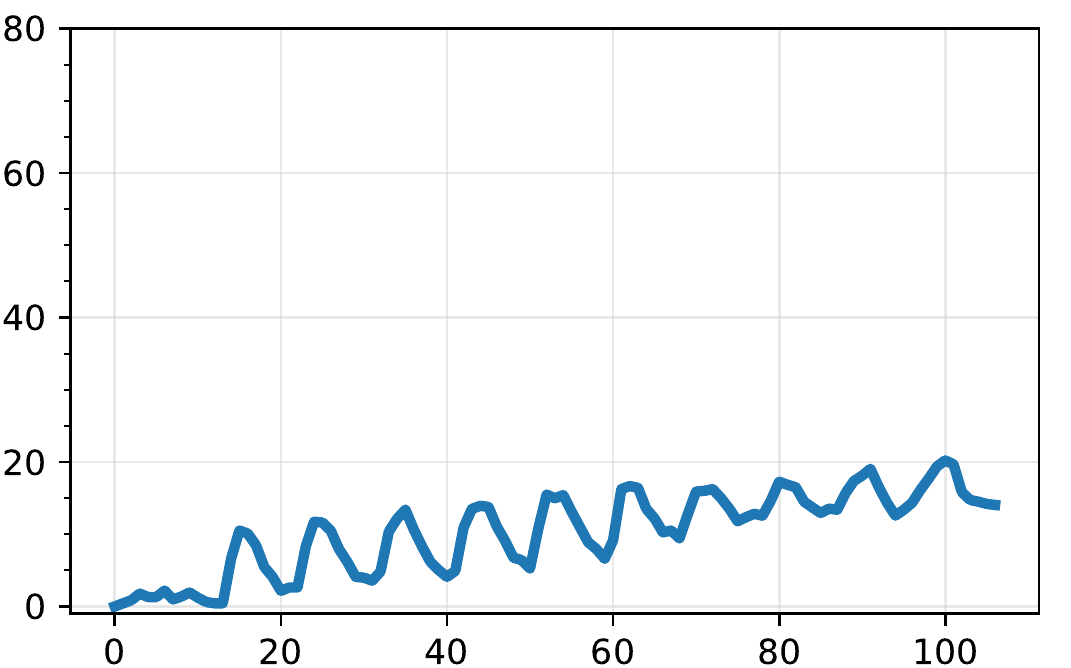}
		}
	}
	\subfloat{
		\parbox[t]{0.23\textwidth}{
		\centering
		{\normalsize Optical Flow (first)} \\[1ex]
		\includegraphics[width=0.23\textwidth]{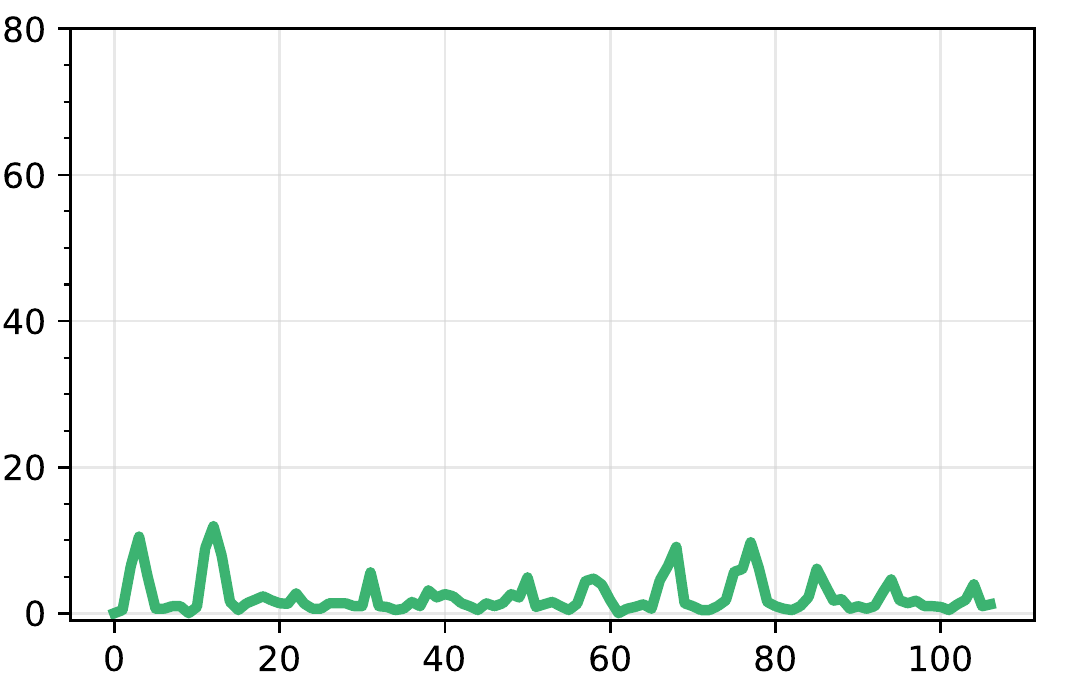}
		}
	}
	\subfloat{
		\parbox[t]{0.23\textwidth}{
		\centering
		{\normalsize Detection} \\[1ex]
		\includegraphics[width=0.23\textwidth]{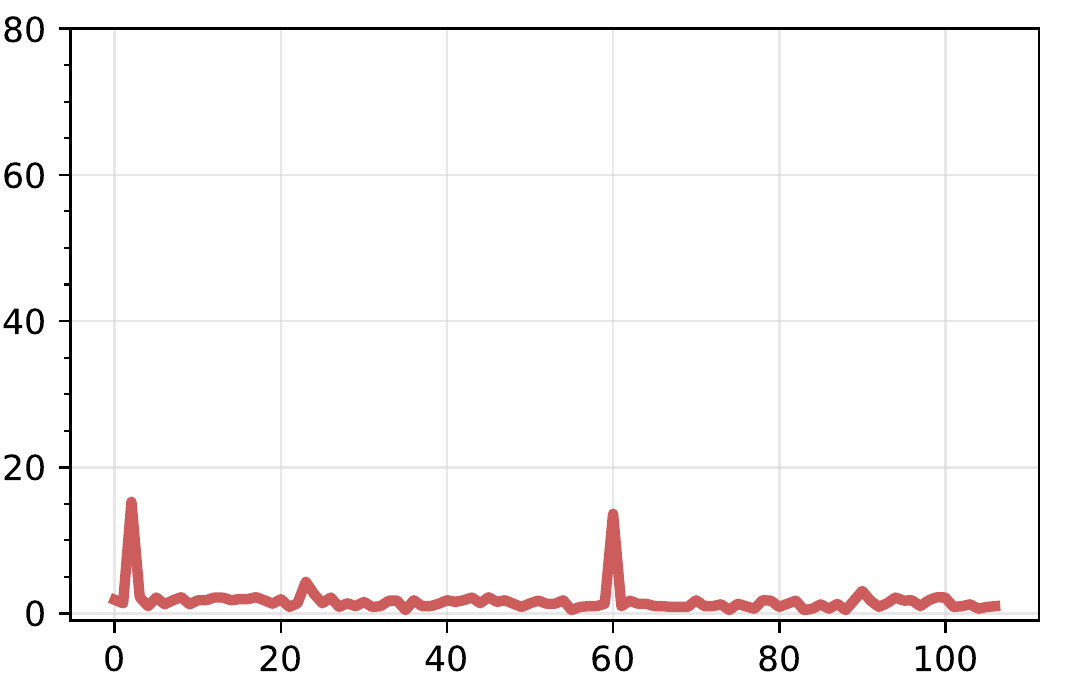}
		}
	}
	\subfloat{
		\parbox[t]{0.23\textwidth}{
		\centering
		{\normalsize Tracking} \\[1ex]
		\includegraphics[width=0.23\textwidth]{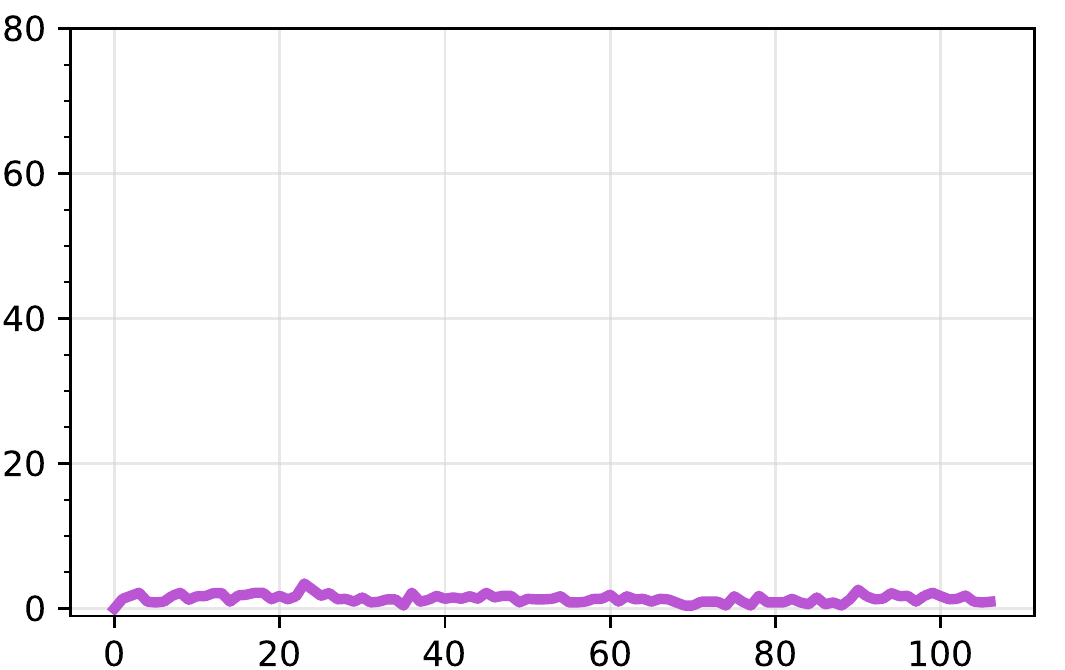}
		}
	}
	\\[-2ex]
	\subfloat{
		\includegraphics[width=0.23\textwidth]{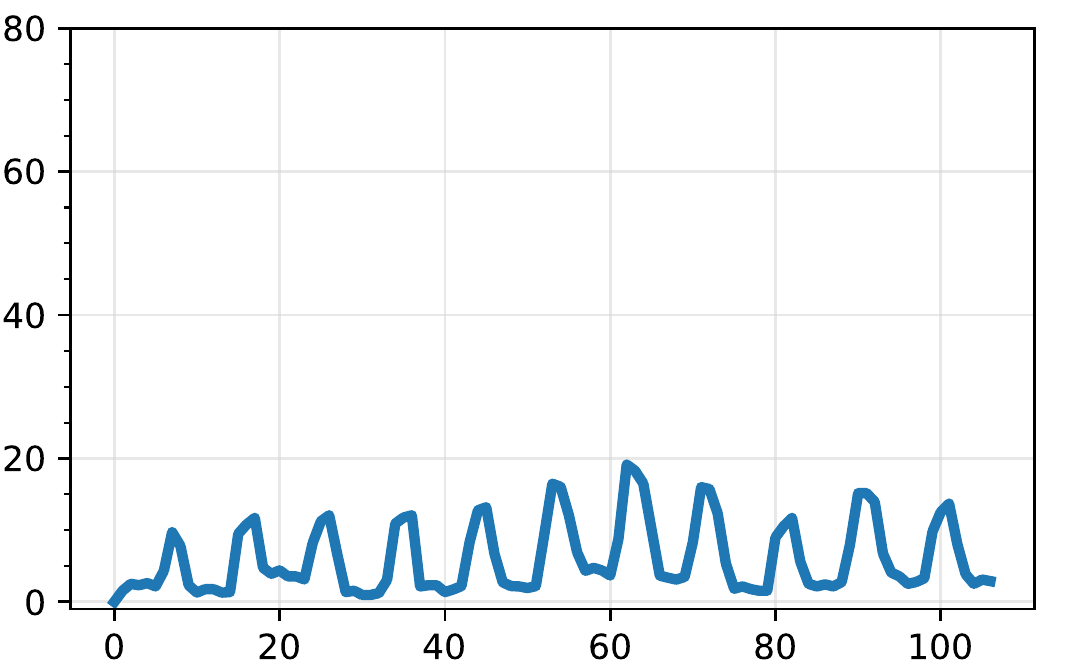}
	}
	\subfloat{
		\includegraphics[width=0.23\textwidth]{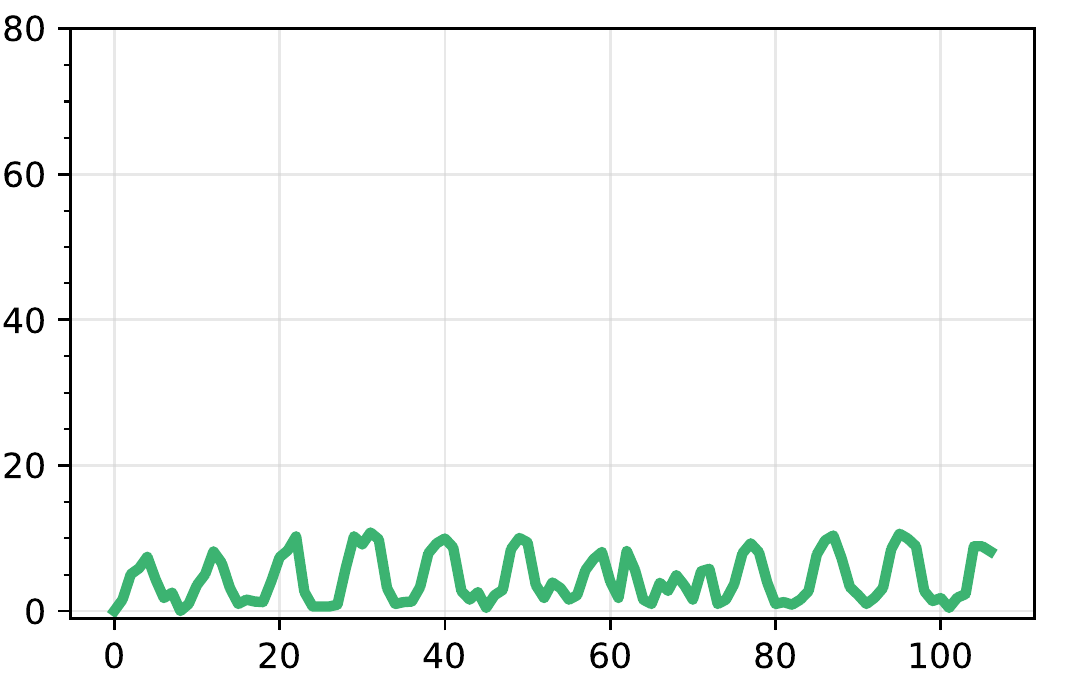}
	}
	\subfloat{
		\includegraphics[width=0.23\textwidth]{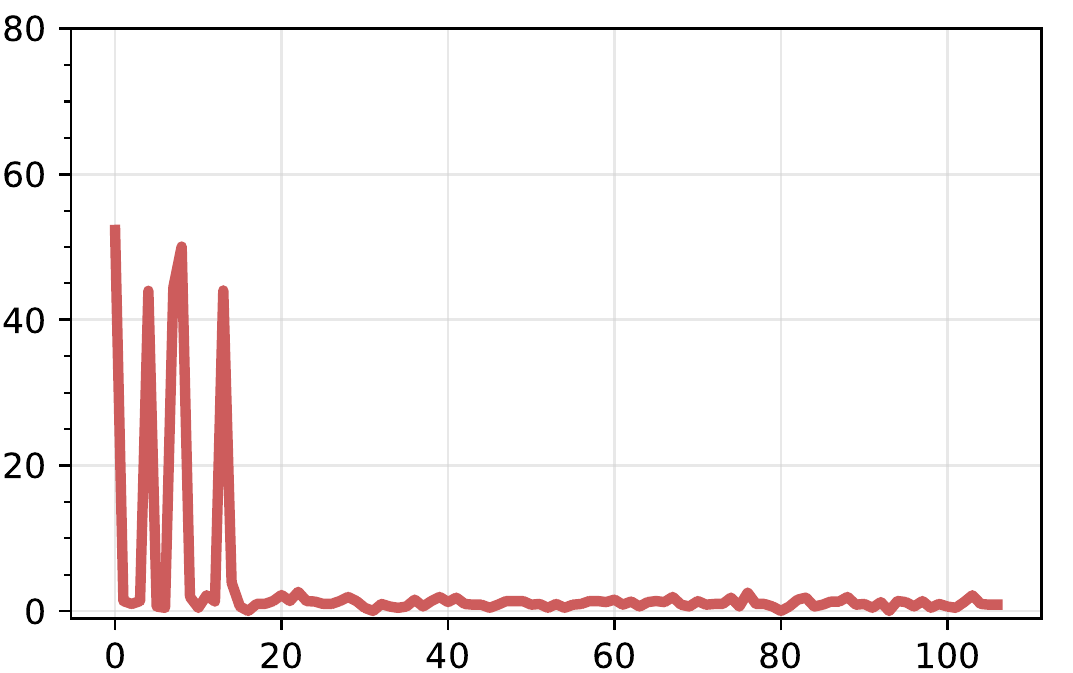}
	}
	\subfloat{
		\includegraphics[width=0.23\textwidth]{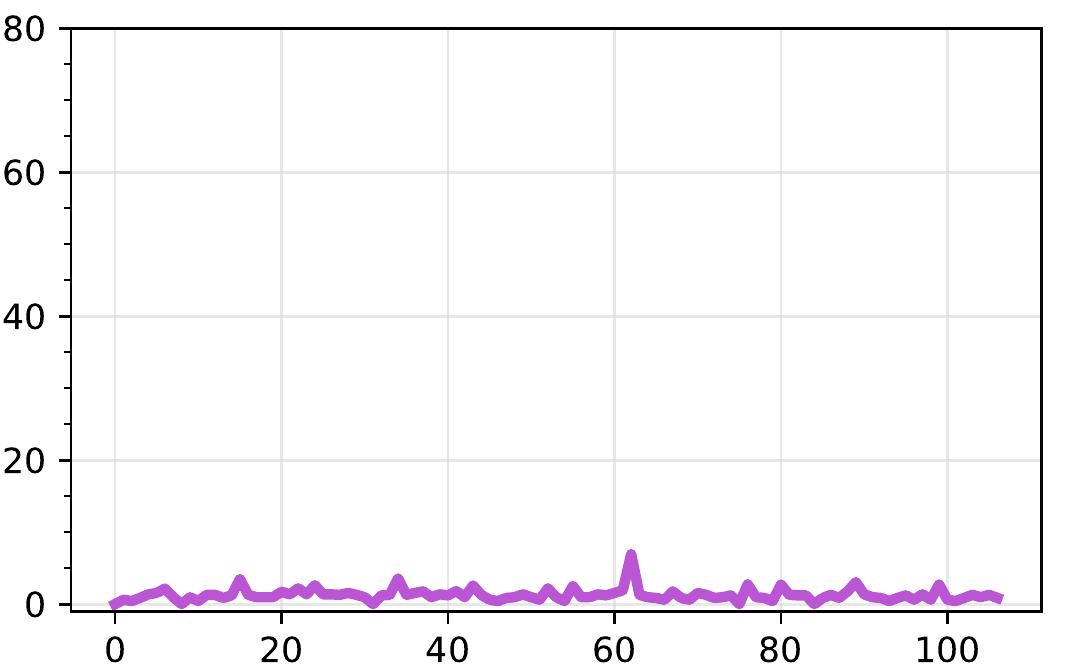}
	}
	\\[-2ex]
	\subfloat{
		\includegraphics[width=0.23\textwidth]{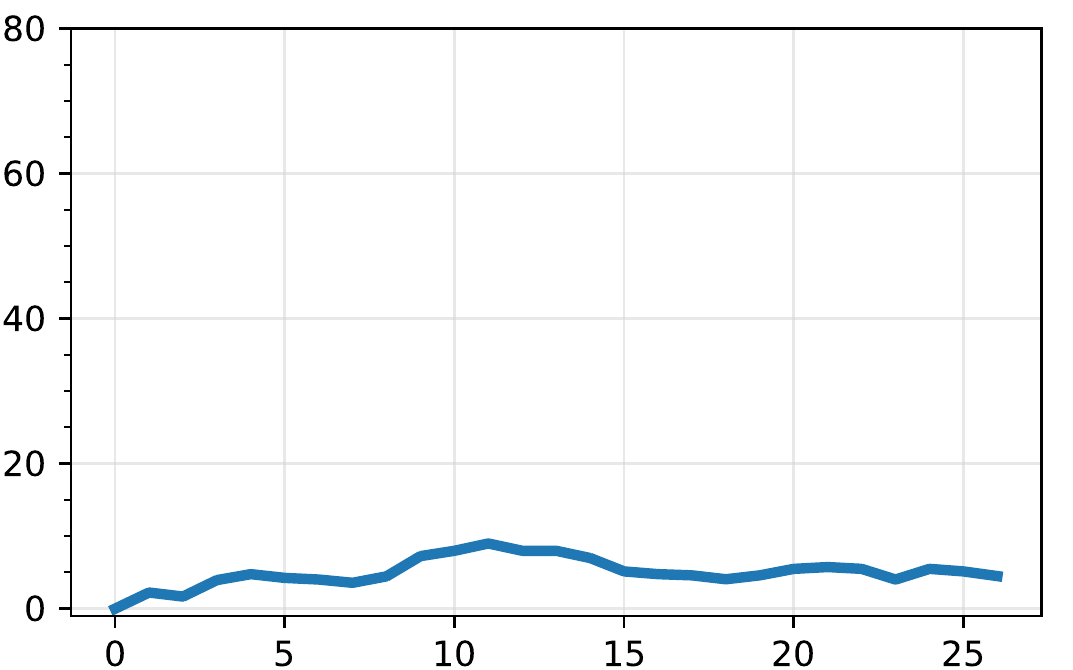}
	}
	\subfloat{
		\includegraphics[width=0.23\textwidth]{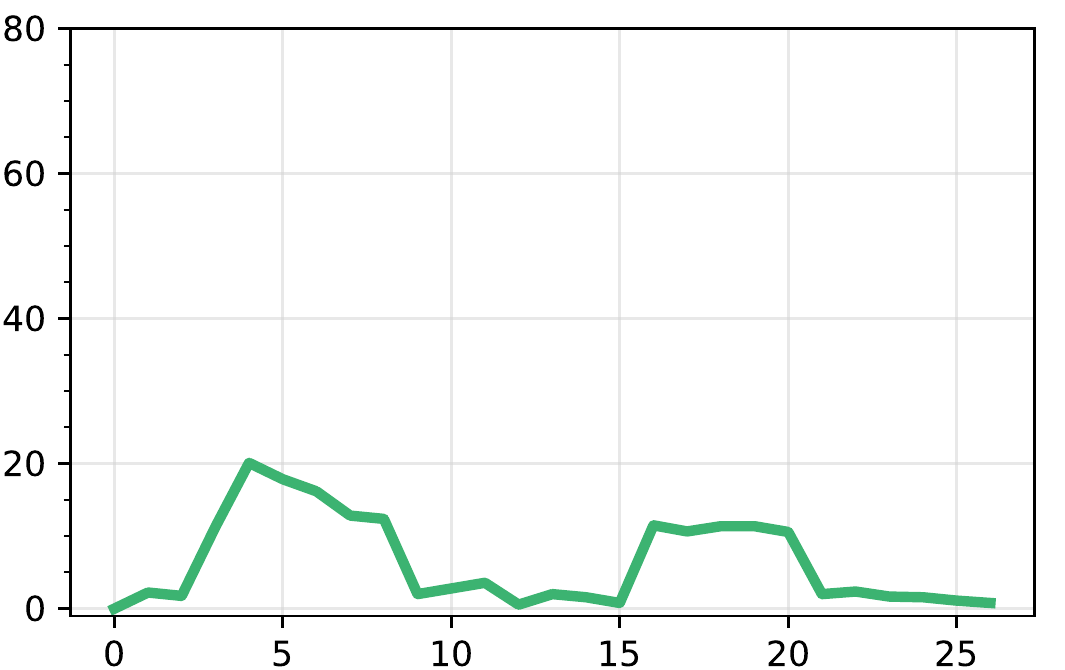}
	}
	\subfloat{
		\includegraphics[width=0.23\textwidth]{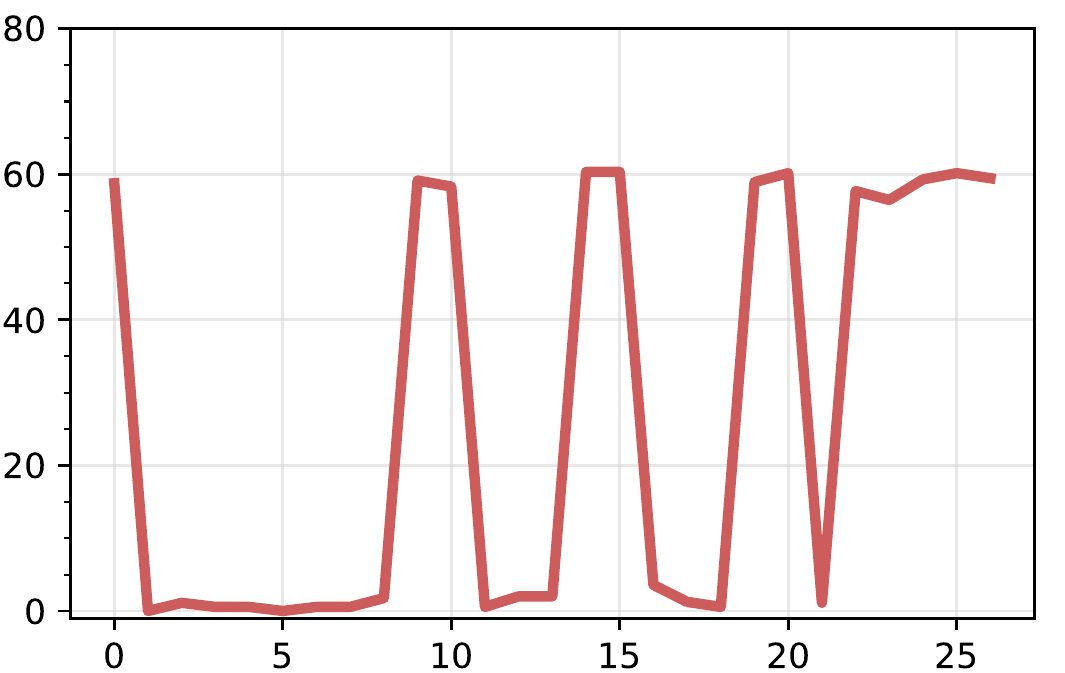}
	}
	\subfloat{
		\includegraphics[width=0.23\textwidth]{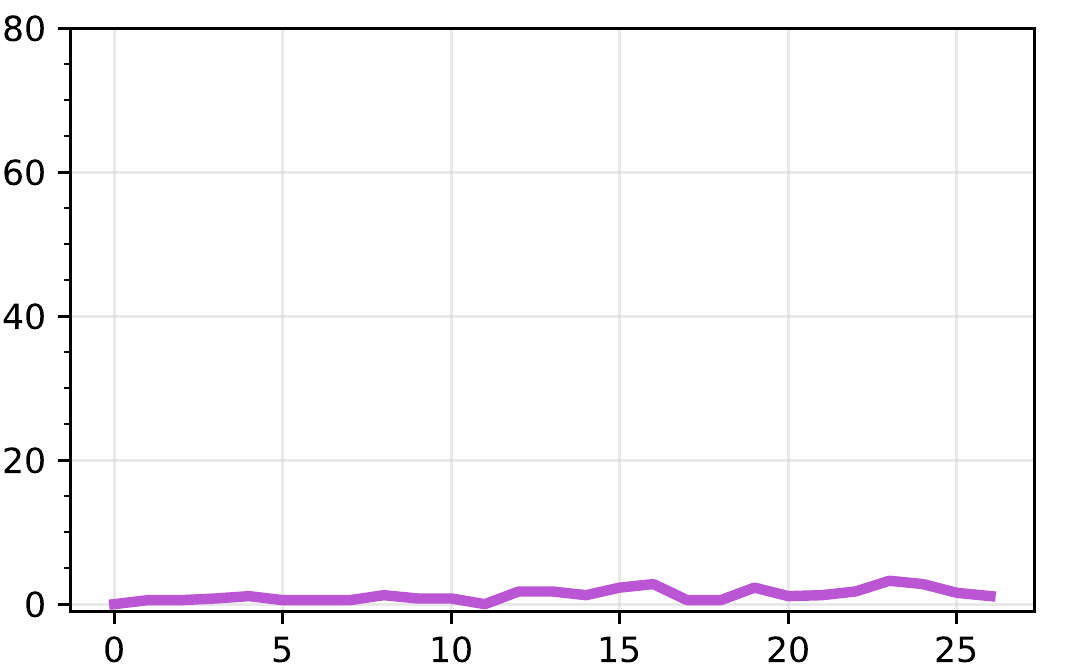}
	}
	\\[-2ex]
	\subfloat{
		\includegraphics[width=0.23\textwidth]{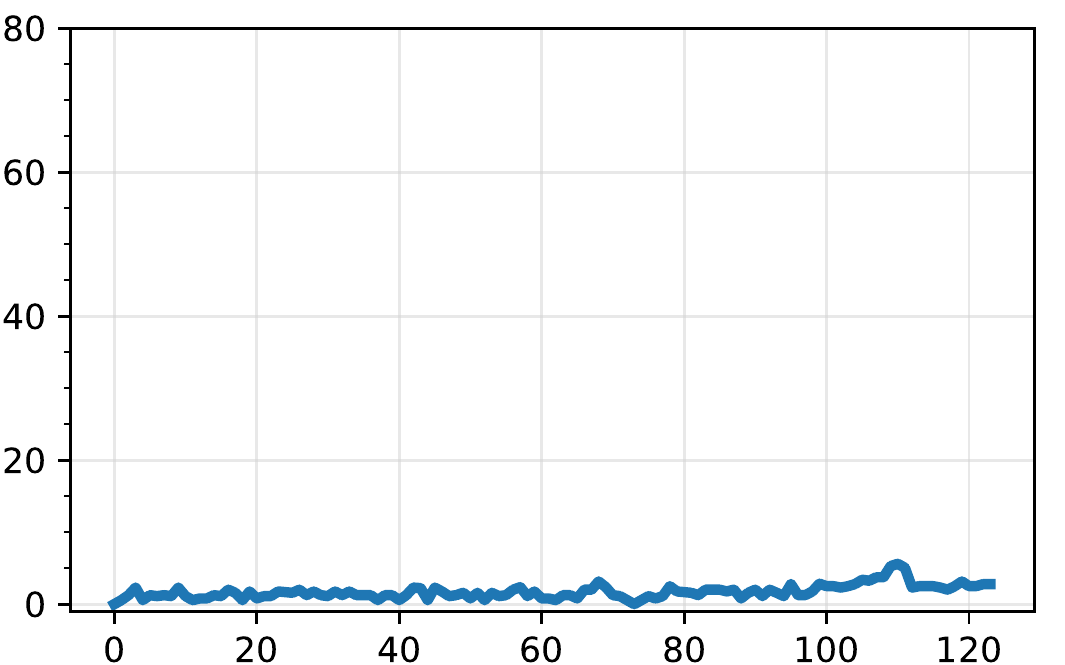}
	}
	\subfloat{
		\includegraphics[width=0.23\textwidth]{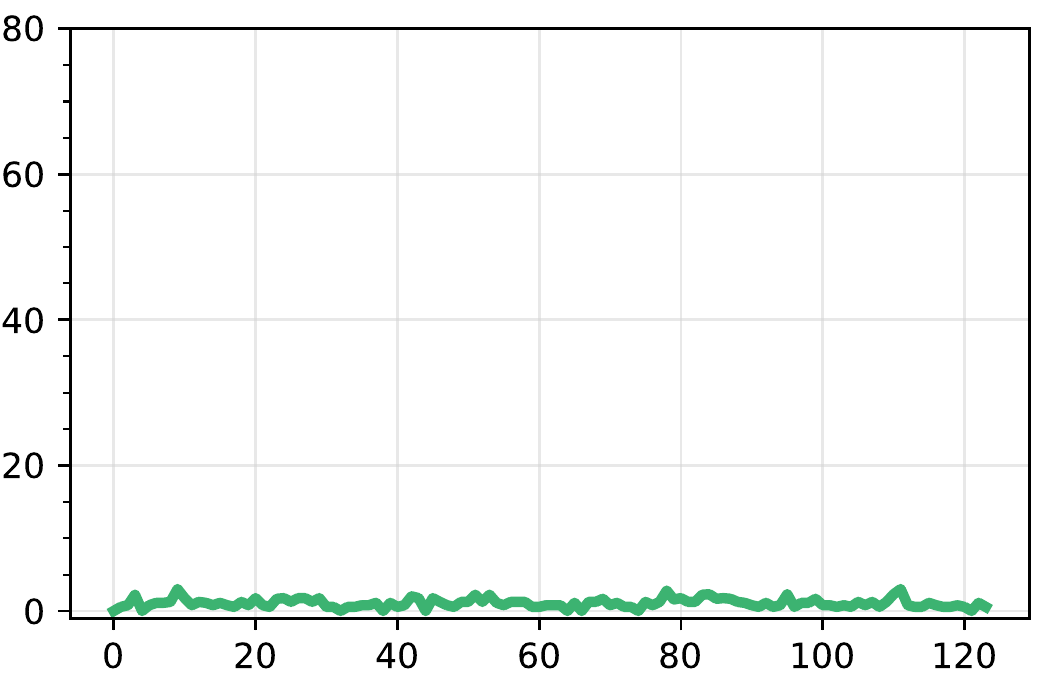}
	}
	\subfloat{
		\includegraphics[width=0.23\textwidth]{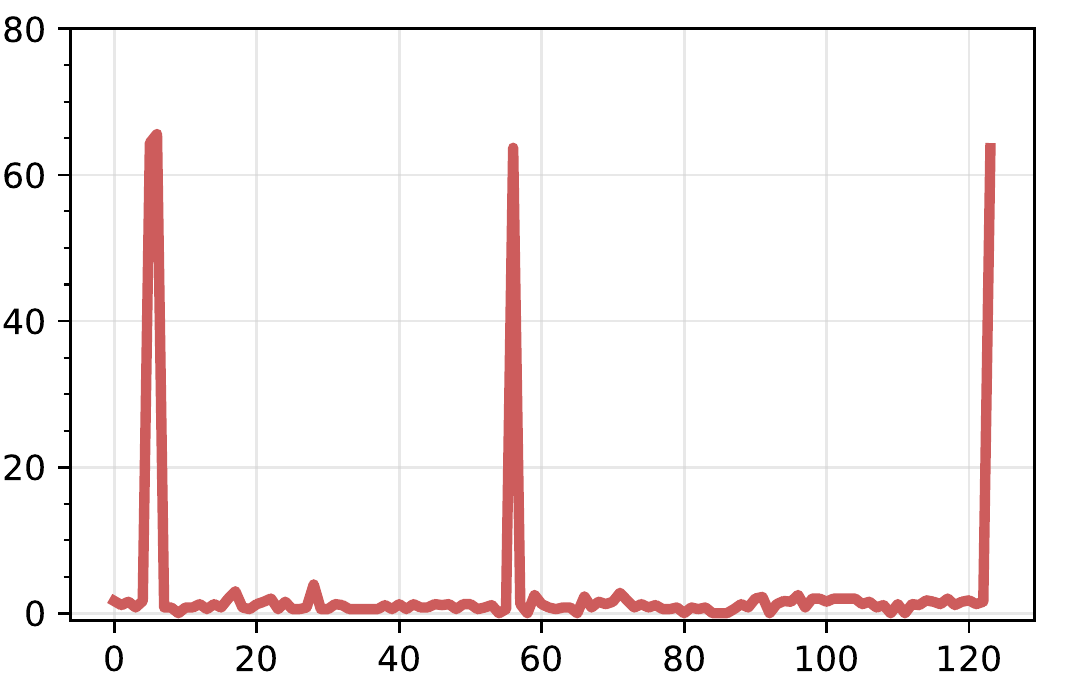}
	}
	\subfloat{
		\includegraphics[width=0.23\textwidth]{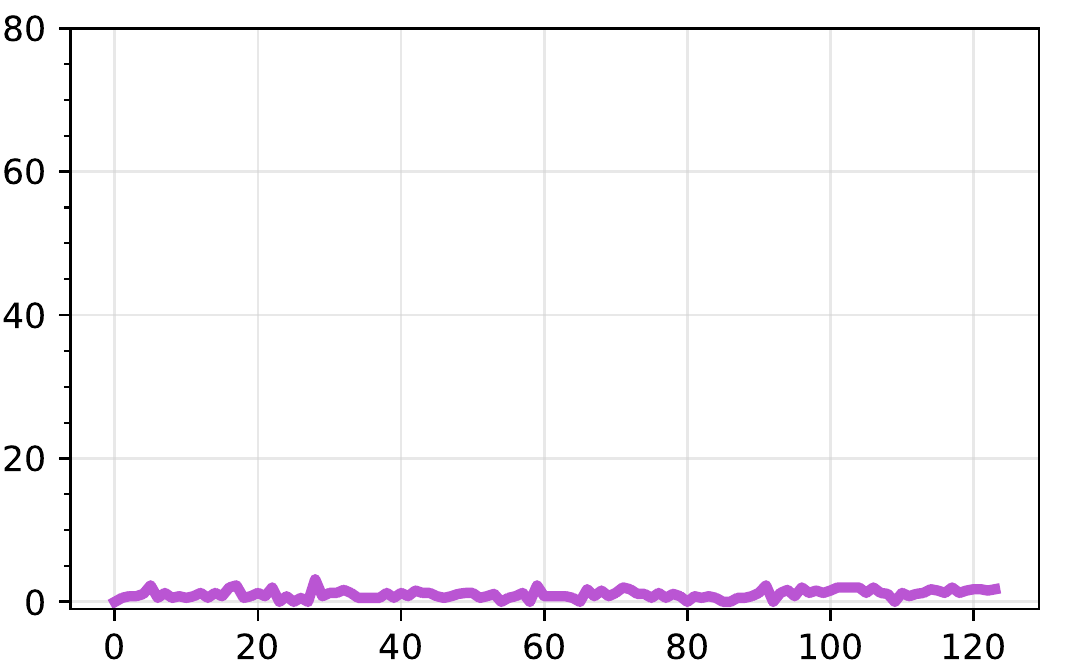}
	}
	
	\caption{Longitudinal view of tracking errors made by the 4 methods on 4 test sequences (one sequence per row). The x-axis denotes the time steps of a sequence, the y-axis is the tracking error (mm).}
	\label{exp-figure-line-plot}
\end{figure*}

\begin{figure*}[h]
	\centering
	\includegraphics[width=0.9\textwidth]{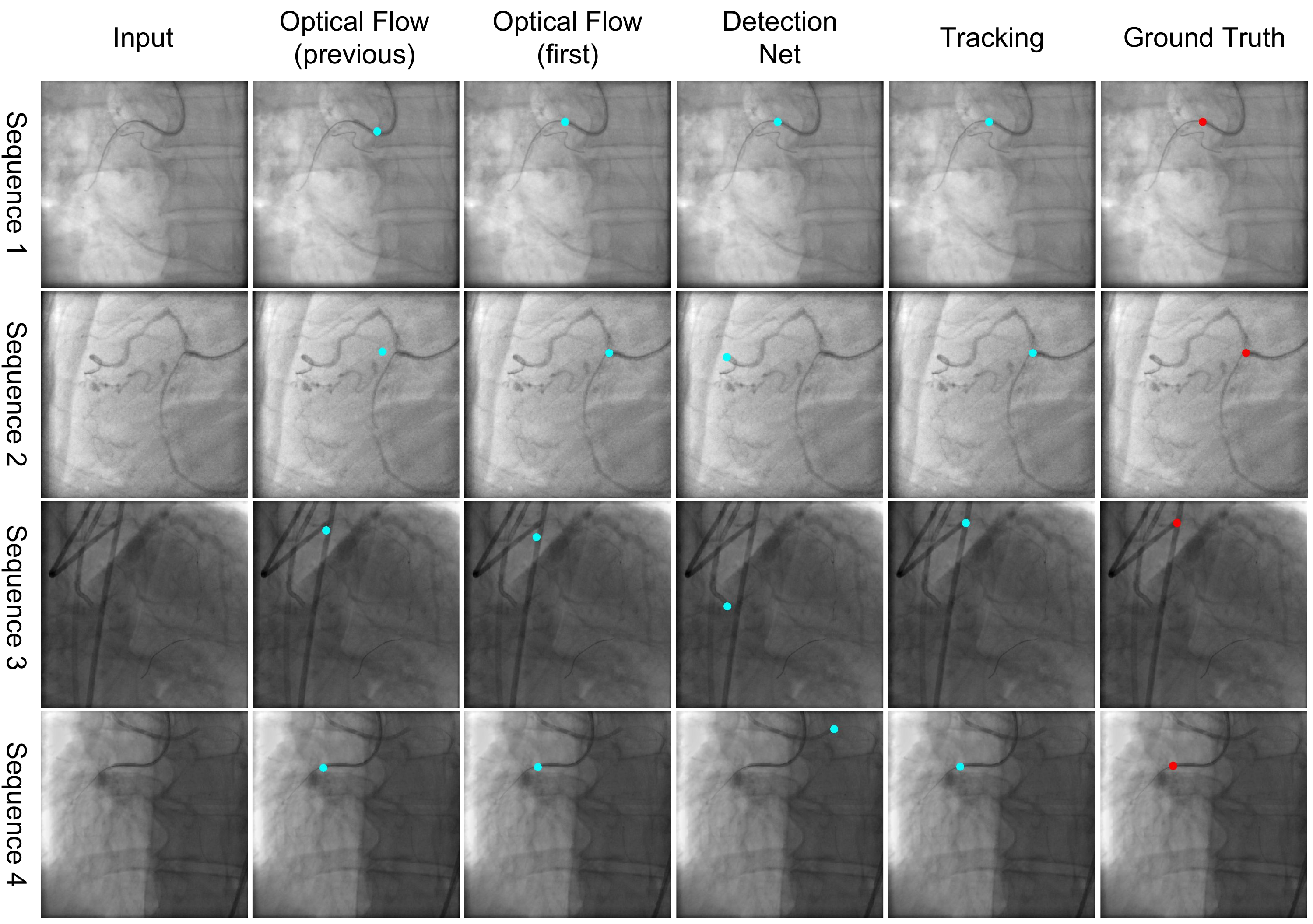}
	\caption{Tracking results on example frames from the same 4 sequences in Figure \ref{exp-figure-line-plot}. The blue point indicates the predicted catheter tip location; the red point shows the ground truth location. (Best viewed in color)}
	\label{exp-figure-tracking-compare-examples-small-2}
\end{figure*}

Figure \ref{exp-figure-line-plot} shows longitudinal views of tracking errors of the 4 methods on 4 example sequences. Although the optical flow methods show high accuracy when the target is on the track (row 4), they present periodic error patterns in two sequences due to large cardiac motion. The detection method shows peaks of large errors, this is because temporal relation between frames is not modeled by the approach, thus the detection on different frames is independent of each other. The proposed tracking method overcomes the problems that other methods have and presents accurate detection on these 4 sequences. The tracking results of the 4 methods on example frames from the 4 sequences are illustrated in Figure \ref{exp-figure-tracking-compare-examples-small-2}.

\begin{figure*}[h]
	\centering
	\includegraphics[width=0.99\textwidth]{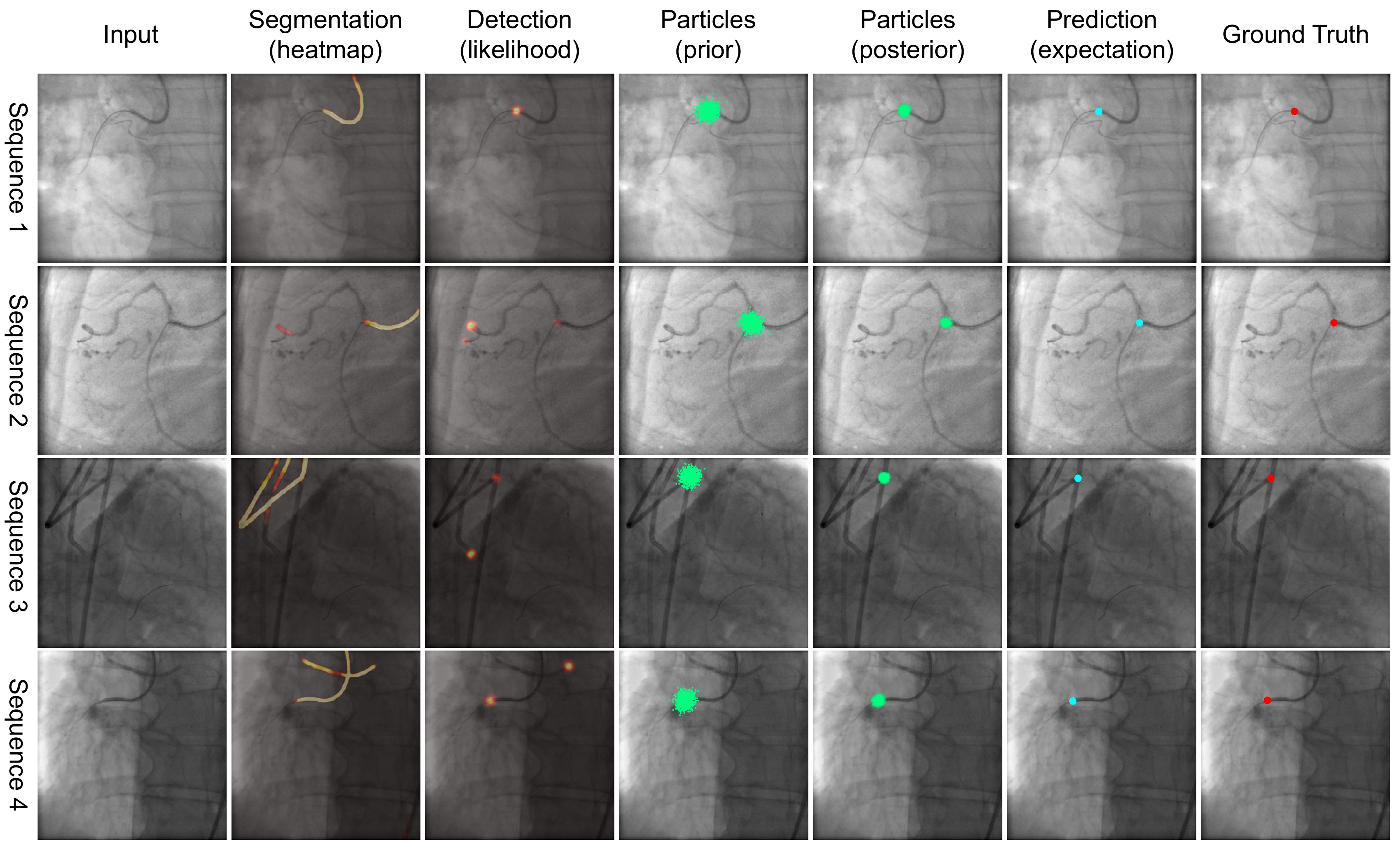}
	\caption{Workflow of the proposed tracking method on the same 4 frames in Figure \ref{exp-figure-tracking-compare-examples-small-2}. The high probability is shown with bright color in the detection map. Samples or particles are presented as green dots. The blue point indicates the predicted catheter tip location; the red point shows the ground truth location. (Best viewed in color)} 
	\label{exp-figure-tracking-examples-small-2}
\end{figure*}

Figure \ref{exp-figure-tracking-examples-small-2} illustrates how the proposed tracking method works on the same 4 frames in Figure \ref{exp-figure-tracking-compare-examples-small-2}. It shows that the prior hypotheses (samples) assists to focus on the correct target location and results in reliable posterior estimation, especially when the detection produces ambiguity in cases of multiple catheters or contrast residual presented in images.

\subsubsection{Catheter Tip Initialization}
\label{exp-subsubsec-initialization}

In this experiment, the initial PDF $p(\textbf{x}_0)$ from which samples are drawn in the proposed tracking is investigated (Algorithm \ref{method-alg-dl-bayes}). In particular, we explored and evaluated the tracking accuracy with an automatic initialization using the probability map obtained from the trained detection network in Section \ref{method-subsec-likelihood} with the chosen setting in Section \ref{exp-subsec-detection}.

\begin{figure}[t]
	\centering
	\includegraphics[width=0.48\textwidth]{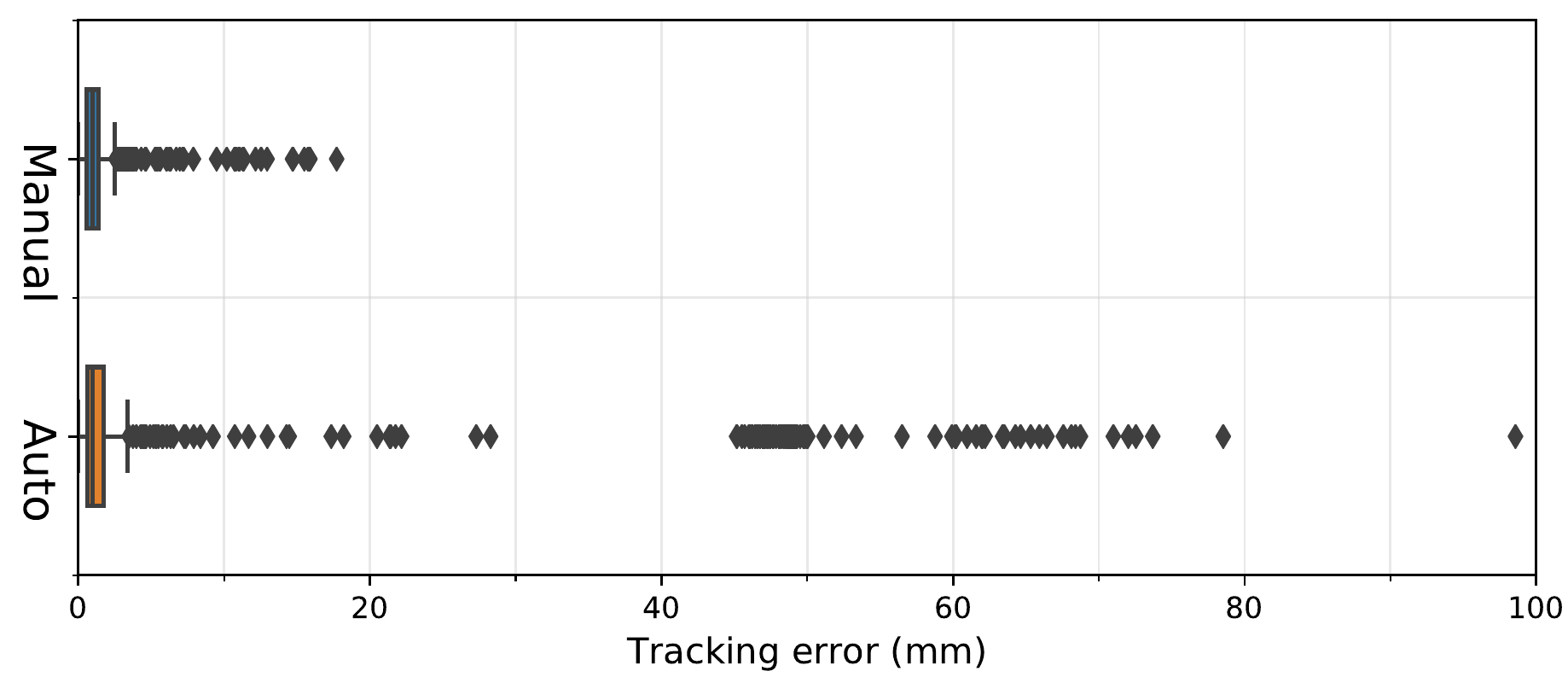}
	\caption{Catheter tip tracking errors (mm) with manual and automatic initialization.}
	\label{exp-figure-tip-init}
\end{figure}

Figure \ref{exp-figure-tip-init} shows the boxplot of tracking errors on all test images with automatic initialization (Auto) and manual initialization (Manual) for which the ground truth tip probability map of the first frame was used. The tracking with automatic initialization presents an accuracy similar to the one with manual initialization for small tracking errors, but has more large tracking errors which influence the mean error over all test images (Table \ref{exp-tab-auto-init}). We, therefore, defined the tracking errors on the right side of the gap in the boxplot ($>$ 40 mm) as outliers, and explored the statistics without those outliers. 

Table \ref{exp-tab-auto-init} indicates that, the mean and median error of the tracking with automatic initialization excluding the outliers are only slightly higher than the tracking with manual initialization and the detection method. While the tracking with automatic initialization has 100 outliers in total from 6 sequences, the detection method that has 10 sequences containing 106 outliers. 

\begin{figure}[t]
	\centering
	\subfloat{
		\includegraphics[width=0.15\textwidth]{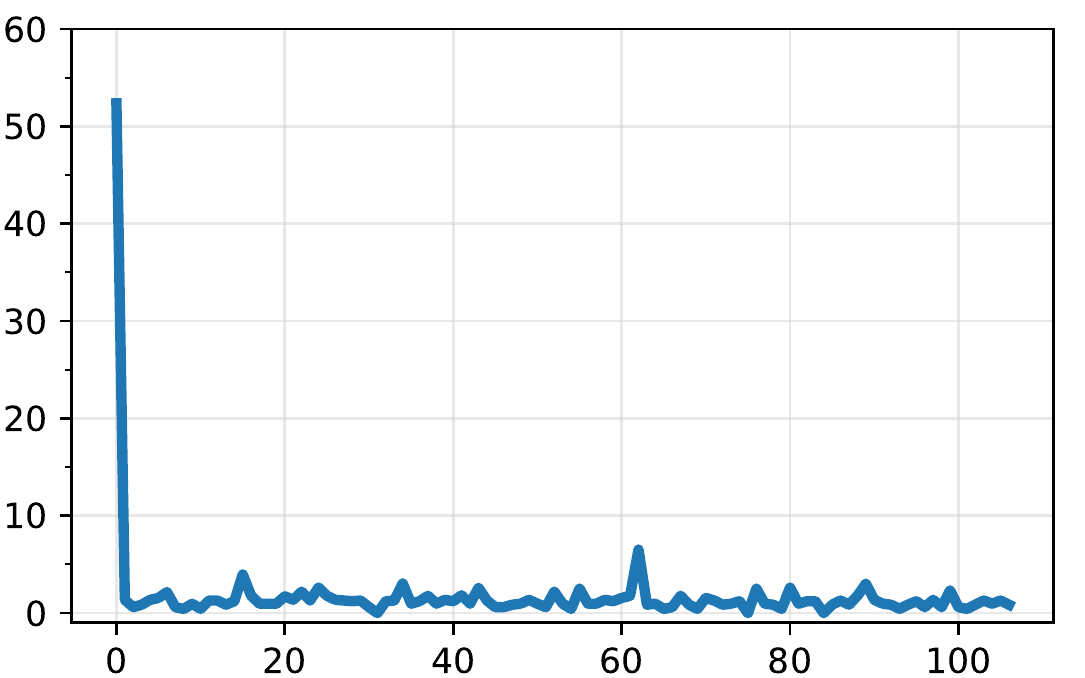}
	}
	\subfloat{
		\includegraphics[width=0.15\textwidth]{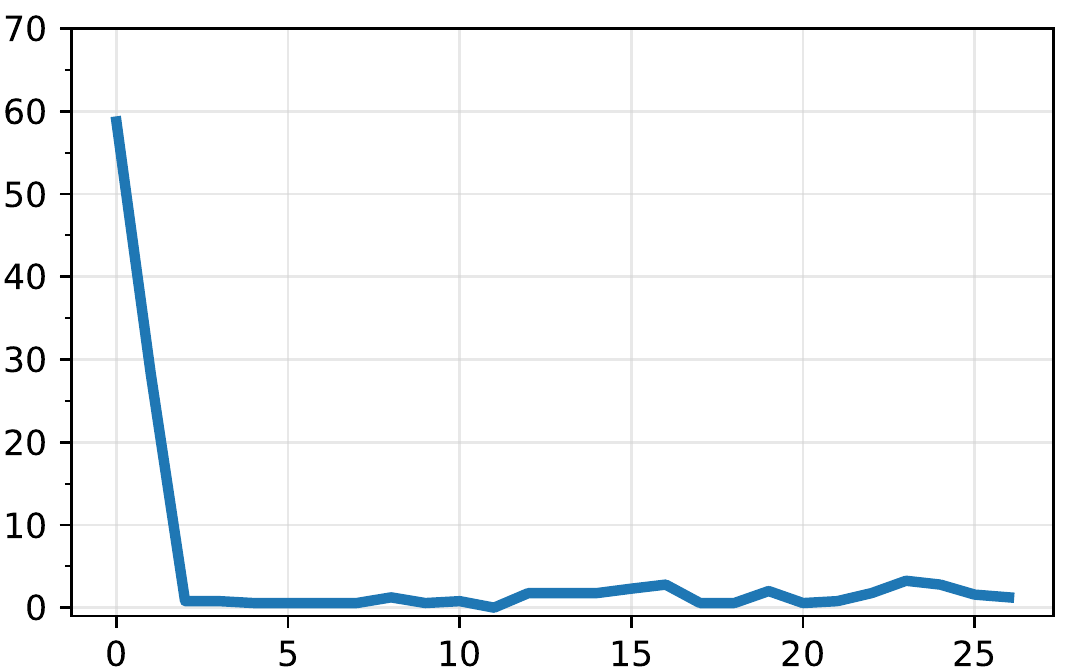}
	}
	\subfloat{
		\includegraphics[width=0.15\textwidth]{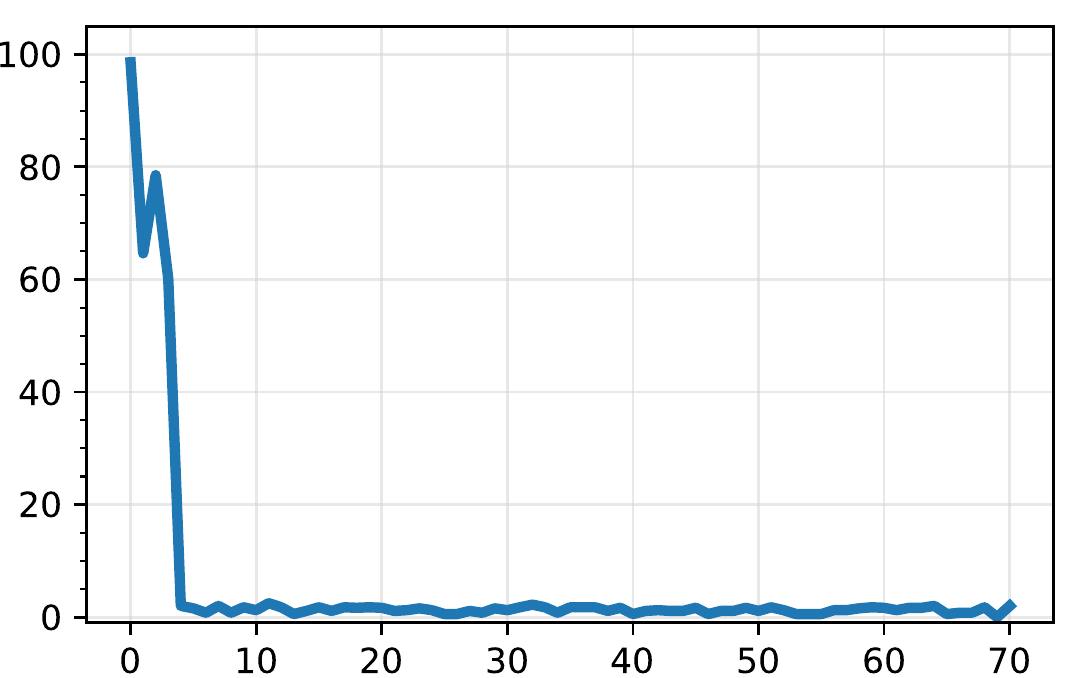}
	}
	\\[-2ex]
	\subfloat{
		\includegraphics[width=0.15\textwidth]{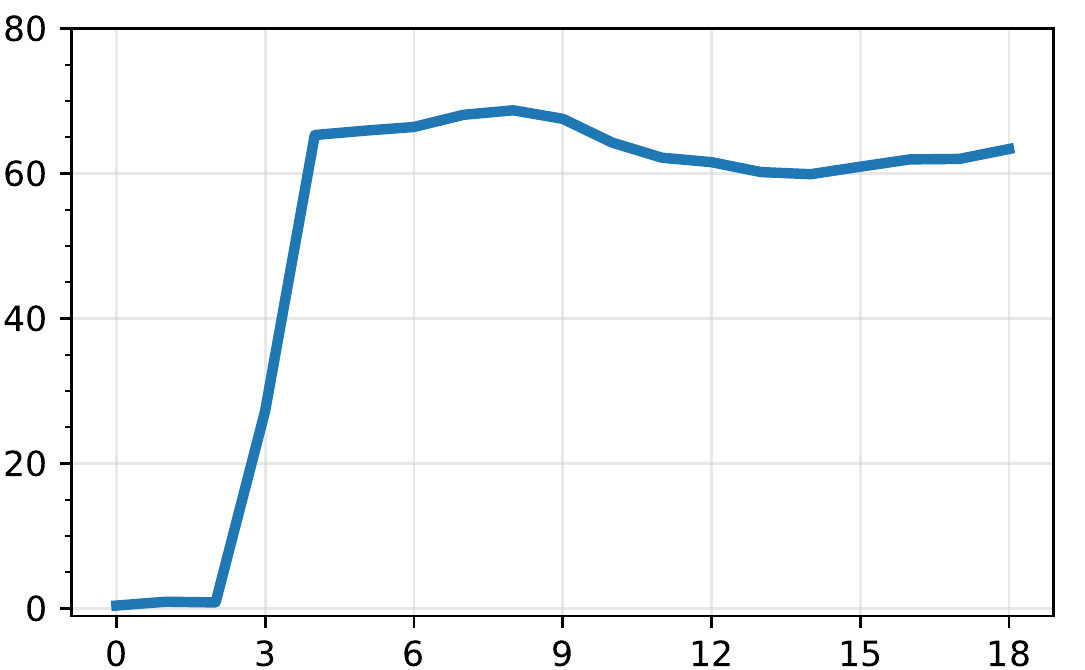}
	}
	\subfloat{
		\includegraphics[width=0.15\textwidth]{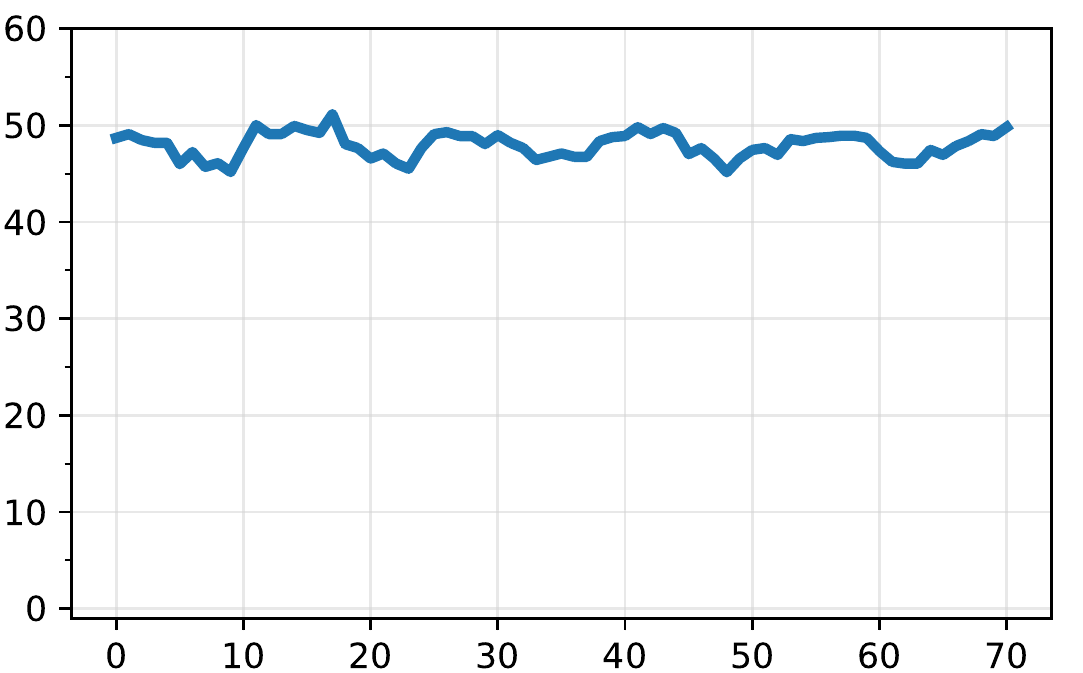}
	}
	\subfloat{
		\includegraphics[width=0.15\textwidth]{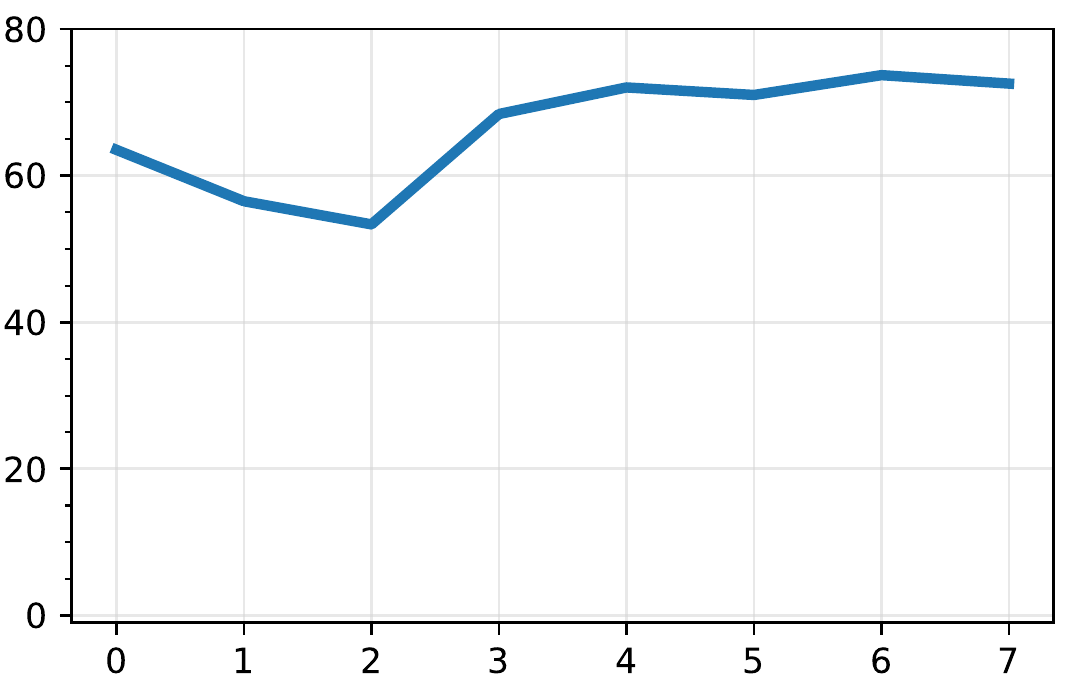}
	}
	
	\caption{Longitudinal views of tracking errors (mm) for the 6 sequences with outliers using automatic initialization.}
	\label{exp-figure-auto-init}
\end{figure}

Unlike the detection method for which the outliers are mainly presented as the peaks in the longitudinal views (Figure \ref{exp-figure-line-plot}), the outliers for the tracking with automatic initialization are more consistent over time. Figure \ref{exp-figure-auto-init} shows the temporal change of tracking errors for the 6 sequences with outliers using the tracking with automatic initialization. For the 3 sequences on the top row, the tracking with automatic initialization makes large errors at the beginning, but becomes accurate very fast in a few frames; for the 3 sequences on the bottom row, however, the tracking errors remain large till the end of the sequences.

\begin{figure}
	\centering
	\subfloat[Sequence 1-3 on the top row in Figure \ref{exp-figure-auto-init}]{
		\includegraphics[width=0.48\textwidth]{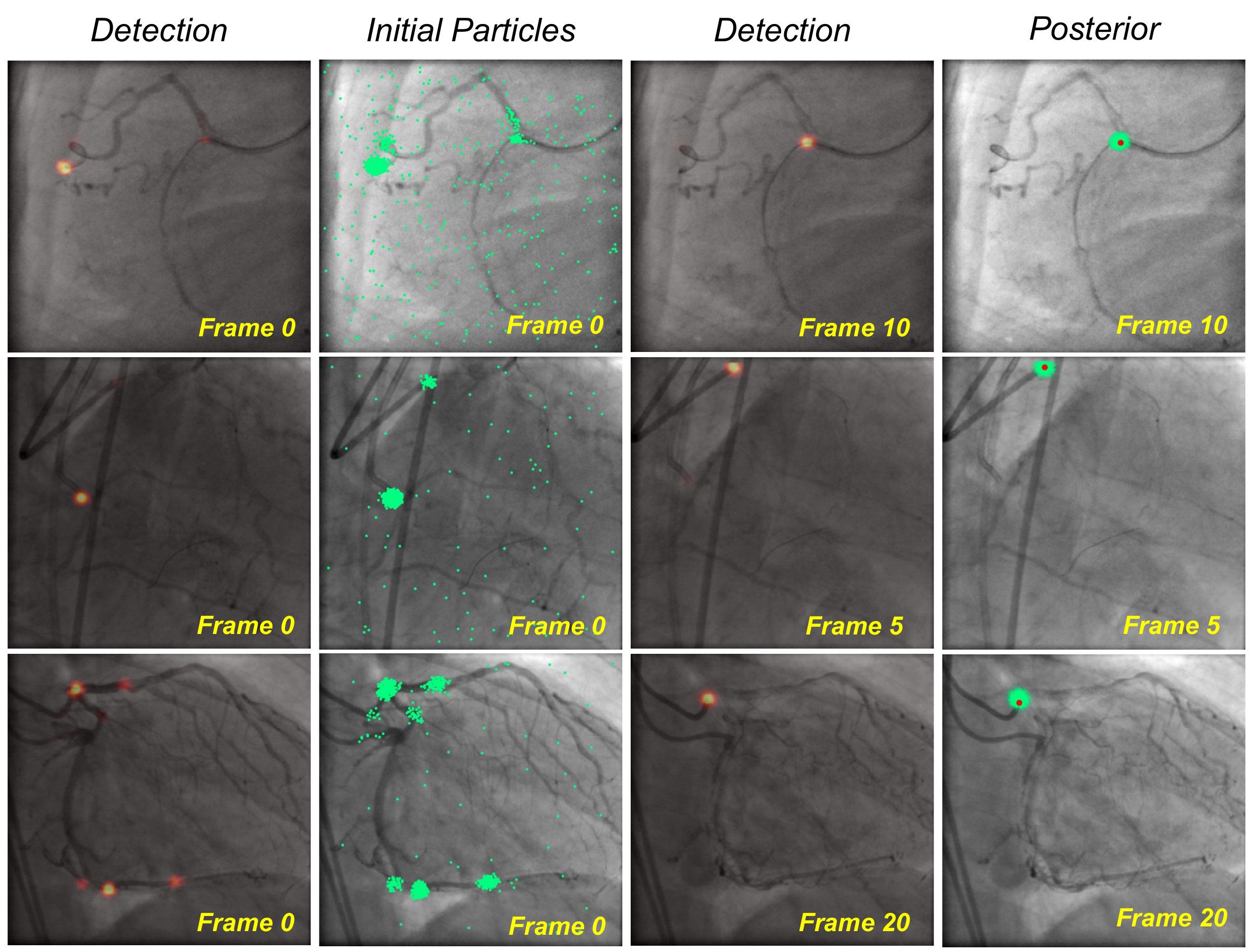}
		\label{exp-subfigure-auto-init-examples-1}
	}
	\\
	\subfloat[Sequence 4-6 on the bottom row in Figure \ref{exp-figure-auto-init}]{
		\includegraphics[width=0.48\textwidth]{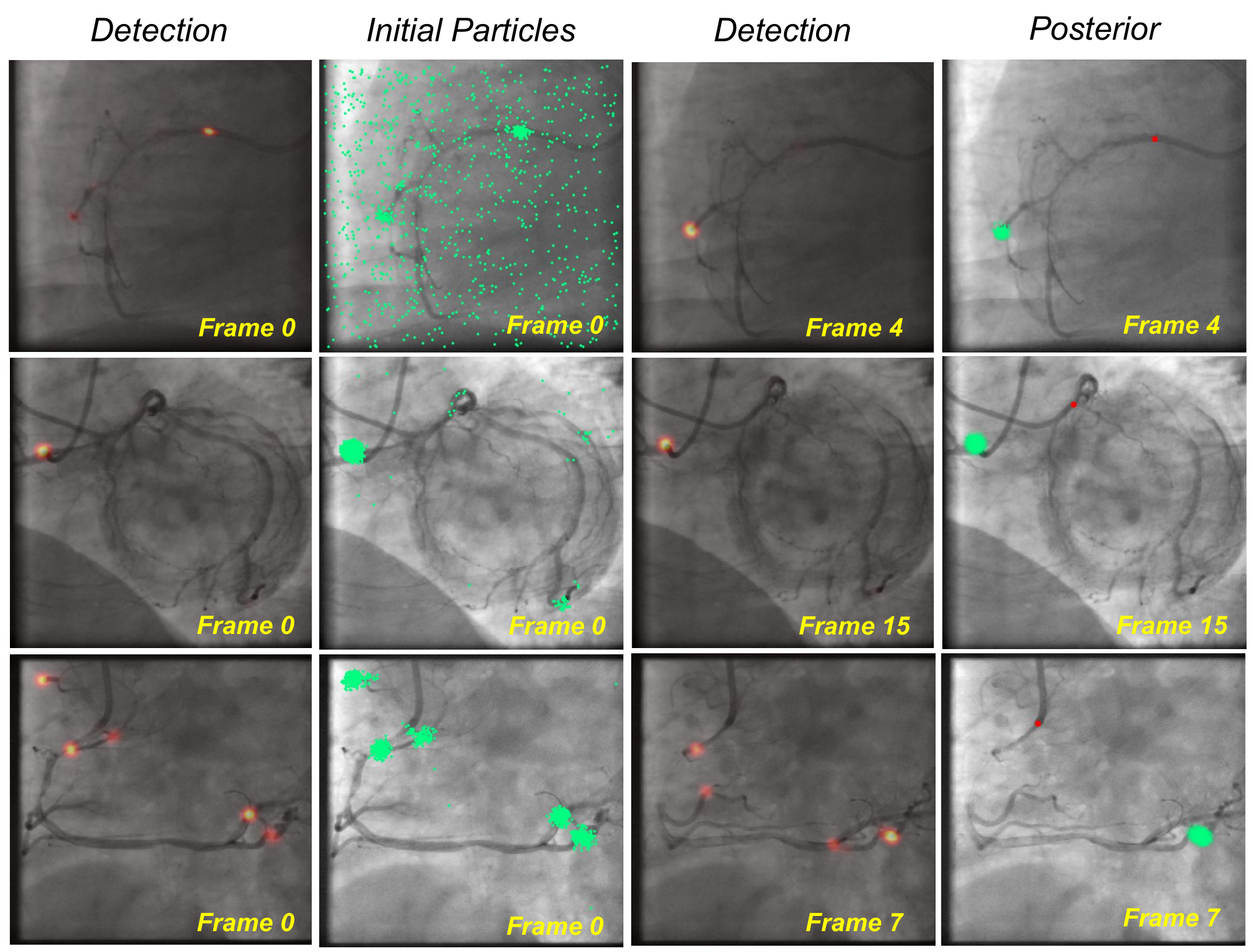}
		\label{exp-subfigure-auto-init-examples-2}
	}
	\caption{Examples frames from the 6 sequences in Figure \ref{exp-figure-auto-init}. The high probability in the detection heatmap is highlighted as bright color. Particles are presented as green dots. The red dots in the last column indicate the ground truth tip location. (Best viewed in color)}
	\label{exp-figure-auto-init-examples}
\end{figure}

Figure \ref{exp-figure-auto-init-examples} shows example frames to give an insight of the tracking with automatic initialization on the 6 sequences in Figure \ref{exp-figure-auto-init}. For the 3 sequences on the top row (Figure \ref{exp-subfigure-auto-init-examples-1}), although the initialization on the first frame (frame 0) is overall not correct, the true tip positions are still covered by some samples; once the detection in subsequent frames is correct, the tracker can still converge to the right target. For the 3 sequence on the bottom row (Figure \ref{exp-subfigure-auto-init-examples-2}), the initializations of samples are ambiguous in frame 0; the detection in subsequent frames focuses on a wrong area also given by the initial samples due to residual of contrast agent or multiple catheters, the tracker then tends to find the wrong target.

\begin{table*}[h]
	\centering
	
	\caption{Catheter tip tracking errors (mm) of detection and tracking with manual and automatic initialization}
	
	\begin{tabular}{l c c c}
		\toprule
		& \multirow{2}{*}{Detection} & \multicolumn{2}{c}{Tracking} \\
		\cmidrule(lr){3-4}
		 &  & Manual init. & Automatic init. \\
		\midrule 
		Maximal error & 108.20 & 17.23 & 98.58 \\
		Median error & 0.96 & 0.96 & 0.96 \\
		Mean error & 5.62 $\pm$ 15.91 & 1.29 $\pm$ 1.76 & 5.16 $\pm$ 13.91 \\ 
		\midrule
		No. of outliers ($>$ 40 mm) & 106 & 0 & 100 \\
		No. of sequences with outliers & 10 & 0 & 6 \\
		\midrule
		Maximal error of inliers & 31.06 & 17.23 & 28.28 \\
		Median error of inliers & 0.96 & 0.96 & 0.96 \\
		Mean error of inliers & 1.17 $\pm$ 1.78 & 1.29 $\pm$ 1.76 & 1.34 $\pm$ 2.15 \\
		\bottomrule	
	\end{tabular}
	
	\label{exp-tab-auto-init}
\end{table*}


\subsection{Dynamic Coronary Roadmapping}
\label{exp-subsec-roadmapping}


In this experiment, the accuracy of dynamic coronary roadmapping using the proposed method with manual tip initialization was evaluated. For roadmap selection with ECG matching (Section \ref{method-sec-roadmap}), the number of online ECG signal points $N_{ECG}$ was manually determined so that the ECG signal stored in the buffer corresponding to 12 X-ray frames (0.8 second in acquisition time). Following the setup in Section \ref{setup-subsec-roadmapping}, we used the distance between the two points in each point pair as the evaluation metric for DCR (the length of a yellow line segment in Figure \ref{setup-figure-distance-vector}). As each frame may have different numbers of point pairs, depending on the length of the target guidewire, the average point pair distance per frame was also computed for evaluation. These distances were evaluated on 409 selected frames with manual annotation of guidewires and vessel centerlines (Section \ref{setup-subsec-roadmapping}).

\begin{figure}
	\centering
	\includegraphics[width=0.48\textwidth]{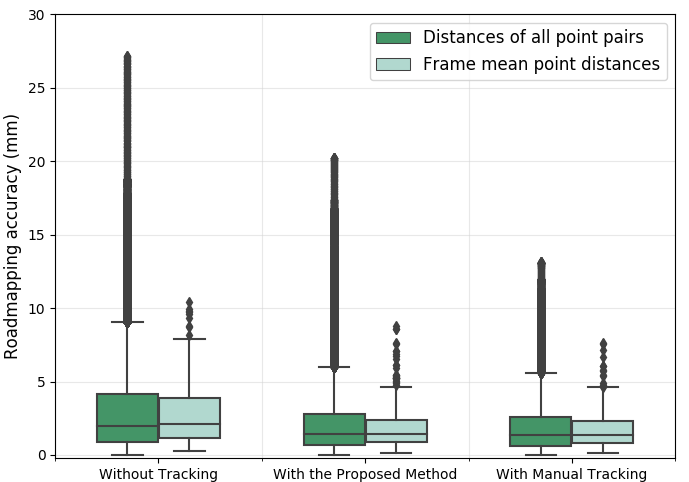}
	\caption{Accuracy (mm) of DCR with three different tracking scenarios.}
	\label{exp-figure-roadmapping-boxplot}
\end{figure}

In the experiment, we compared the DCR with the proposed tracking method to those with manual tip tracking and without tracking. All three approaches were based on the same ECG matching method (Section \ref{method-sec-roadmap}) for selecting roadmaps. The accuracy of the DCR without tracking in Table \ref{exp-tab-roadmapping} shows that the mean distances are reduced to less than 3 mm by compensating only cardiac motion via roadmap selection with ECG matching. Table \ref{exp-tab-roadmapping} also shows that the DCR with the proposed method achieves median distances of about 1.4 mm and mean distances of about 2 mm. The boxplots of the distances of all point pairs and the frame mean point distances of all 409 evaluation frames are illustrated in Figure \ref{exp-figure-roadmapping-boxplot}. The comparison of the three DCR approaches from Table \ref{exp-tab-roadmapping} and Figure \ref{exp-figure-roadmapping-boxplot} indicates that the accuracy of the proposed DCR method has shown improvement over the DCR without tracking, and is only slightly less than the DCR with manual tip tracking (although the difference is statistically significant). Additionally, interested readers are referred to Appendix \ref{app-subsubsec-roadmapping} where the influence of catheter segmentation on the accuracy of DCR is investigated.

Table \ref{exp-tab-roadmapping-error-distribution} shows how the frame mean point distances of the 409 evaluation frames are distributed. The DCR with the proposed method has similar error distribution as the one with manual tip tracking: they both have about 1/3 of the distances less than 1 mm and 1/3 of the distances between 1 and 2 mm. The proposed method has slightly more distances larger than 5 mm than manual tip tracking. Both methods are more accurate than the DCR without tracking on intervals of small errors ($<$ 2 mm).

\begin{figure}
	\centering
	\includegraphics[width=0.48\textwidth]{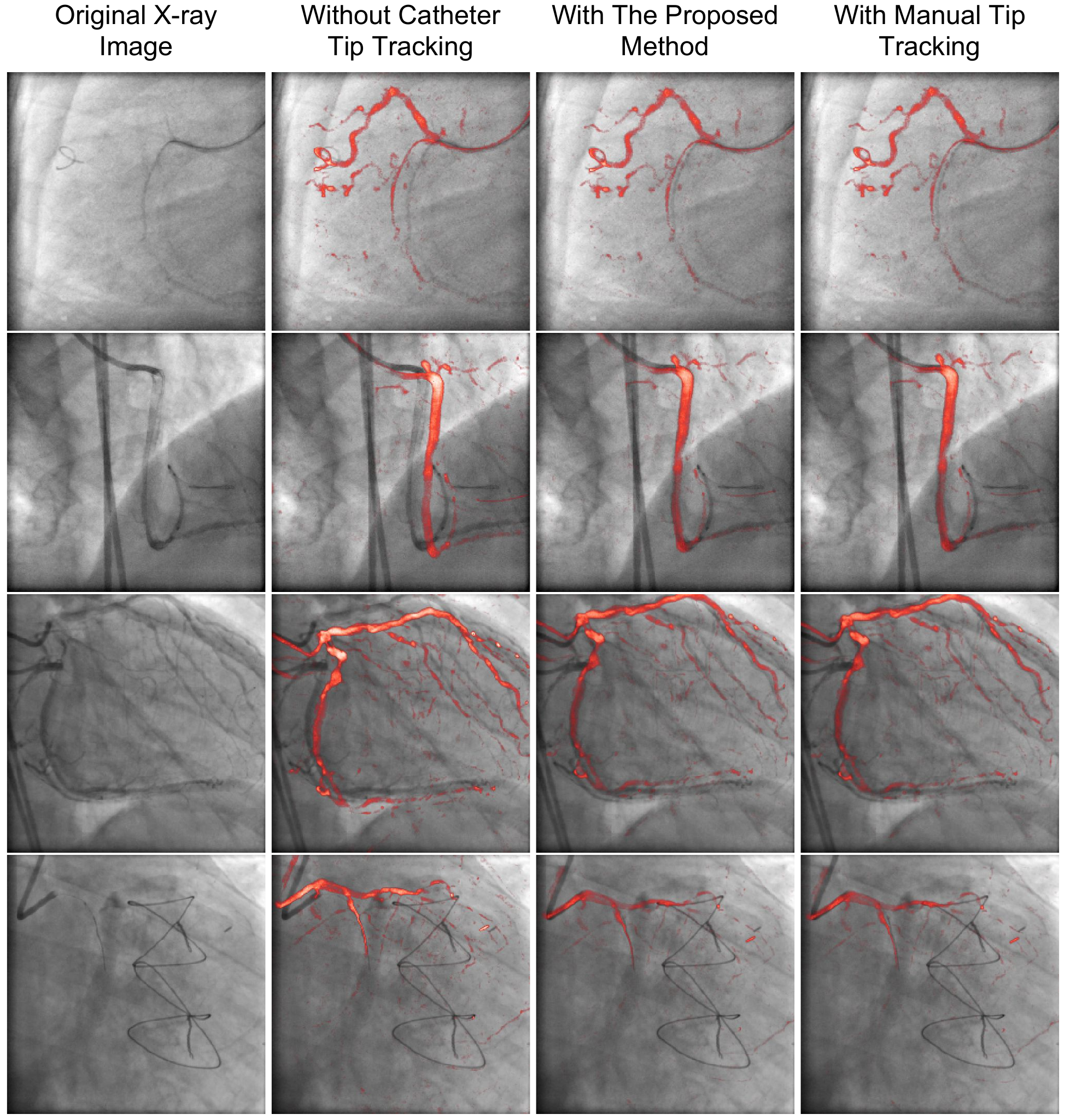}
	\caption{Examples of superimposition of selected roadmaps (red) on X-ray fluoroscopic frames. (Best viewed in color)}
	\label{exp-figure-roadmapping-examples}
\end{figure}

Figure \ref{exp-figure-roadmapping-examples} shows overlays of selected roadmaps on example frames of 4 sequences with the three DCR approaches. The DCR without tracking presents mismatch of catheters, guidewires or residual of contrast agent in the images, whereas the other methods improve the alignment and show good match between the structures in the original X-ray image and the roadmaps. Compared to the DCR with manual tip tracking, the proposed method show similar visual alignment of the roadmaps to the original X-ray images. For a dynamic view of a roadmapping example, we refer readers to the supplemental material. 

\begin{table*}[h]
	\centering
	
	\caption{The statistics of DCR accuracy (mm) with three different tracking scenarios. With the two-sided Wilcoxon signed-rank test: $\dagger$ denotes that the difference between the DCR without tracking and that with the proposed tracking method is statistically highly significant ($p < 0.001$); * indicates a statistically significantly difference between the DCR using manual tip tracking and that with the proposed tracking approach ($p < 0.05$).}
	
	\begin{tabular}{l c c c}
		\toprule
		 & Without Tracking$\dagger$ & Proposed Tracking Method & Manual Tip Tracking* \\
		\midrule 
		\textbf{All point pairs} &  &  &  \\
		Maximal distance  & 27.19 & 20.24 & 13.12 \\
		Median distance  & 1.97 & 1.43 & 1.35 \\
		Mean distance  & 2.94 $\pm$ 2.83 & 2.07 $\pm$ 2.08 & 1.85 $\pm$ 1.72 \\
		\midrule
		\textbf{Frame mean distance}  &  &  &  \\
		Median distance & 2.11 & 1.42 & 1.38 \\
		Average distance & 2.76 $\pm$ 2.08 & 1.91 $\pm$ 1.52 & 1.75 $\pm$ 1.30 \\
		\bottomrule	
	\end{tabular}
	
	\label{exp-tab-roadmapping}
\end{table*}

\begin{table*}[h]
	\centering
	
	\caption{Distribution of frame mean point distances of the 409 evaluation frames.}
	
	\begin{tabular}{l c c c c c c}
		\toprule
		\multirow{2}{*}{Tracking Methods of DCR} & \multicolumn{6}{c}{Error Intervals (mm)} \\
		\cmidrule(lr){2-7}
		 & $<$ 1 & 1-2 & 2-3 & 3-4 & 4-5 & $\geq$ 5 \\
		\midrule 
		Without tracking & 81 & 115 & 69 & 47 & 31 & 66 \\
		Proposed tracking method & 131 & 145 & 61 & 32 & 17 & 23 \\
		Manual tip tracking & 139 & 144 & 61 & 35 & 20 & 10 \\
		\bottomrule	
	\end{tabular}
	
	\label{exp-tab-roadmapping-error-distribution}
\end{table*}

\subsection{Processing Time}
\label{exp-subsec-time}


The processing time of all steps in the proposed DCR method was measured with the hardware and software setup in Section \ref{setup-setup-implementation}. The ECG matching method for roadmap selection was running in Python on the CPU of the Linux machine; the deep neural network and the optical flow component of the tracking method were running on the GPU.

\begin{figure}[t]
	\centering
	\includegraphics[width=0.48\textwidth]{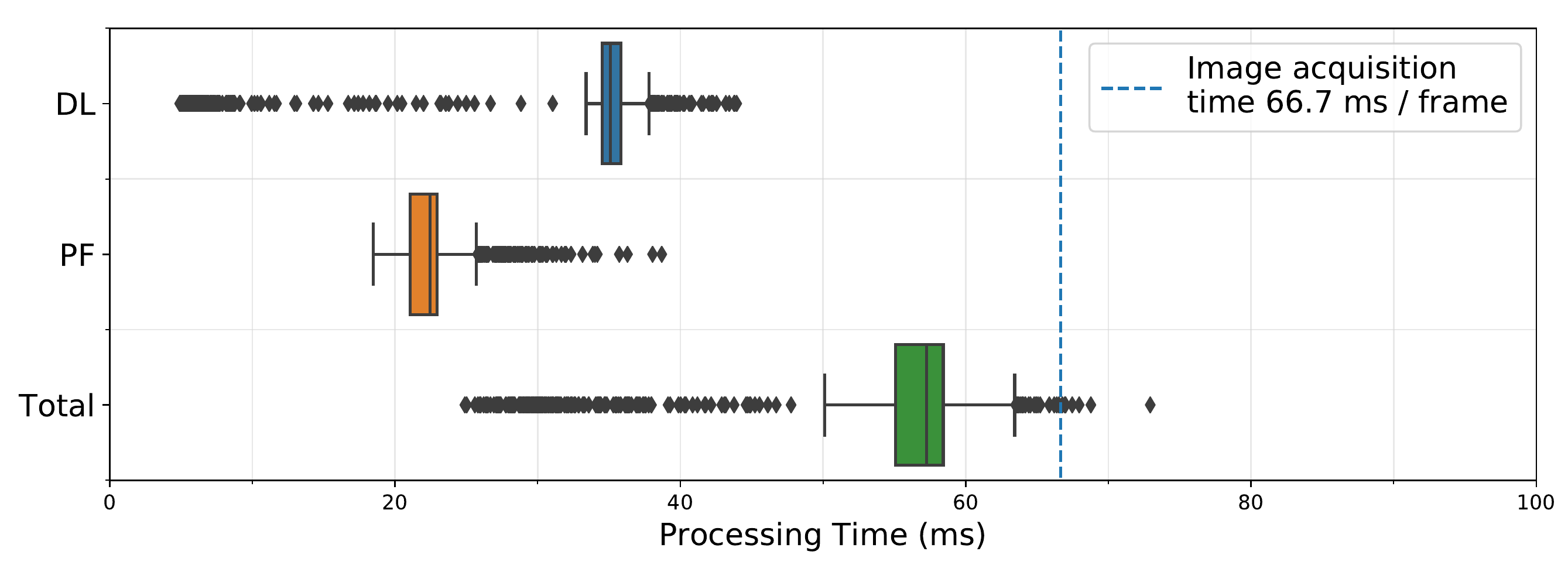}
	\caption{Runtime of catheter tip tracking (ms / frame) on all test frames.}
	\label{exp-figure-tracking-time}
\end{figure}

In the experiments, the runtimes for roadmap selection (step 1) and roadmap transformation (step 3) in Figure \ref{method-figure-overview} were negligible ($<$ 1 ms / frame). The runtime of the proposed catheter tip tracking method is shown in Table \ref{exp-tab-tracking-time} and Figure \ref{exp-figure-tracking-time}. The average time to compute the likelihood with the deep learning setup (DL) is 31.5 ms / frame. The particle filtering (PF) step, which consists of the optical flow estimation, sample propagation, sample weight update and normalization, prediction and resampling, takes on average 23 ms / frame. Therefore, the average tracking time in total is 54.5 ms / frame. The total average time of the proposed DCR including roadmap selection, catheter tip tracking and roadmap transformation is still less than the acquisition time of our data (66.7 ms / frame, 15 fps), indicating that the proposed DCR method would run in real-time with our setup.

\begin{table*}[h]
	\centering
	
	\caption{Statistics of the runtime of catheter tip tracking (ms / frame) on the test (tracking) dataset.}
	
	\begin{tabular}{c c c c}
		\toprule
		 & Deep Learning & Particle Filtering & Total Tracking Time \\
		\midrule 
		Mean & 31.5 $\pm$ 10.3 & 23.0 $\pm$ 8.7 & 54.5 $\pm$ 12.3 \\
		Median & 35.1 & 22.8 & 57.7 \\
		\bottomrule	
	\end{tabular}
	
	\label{exp-tab-tracking-time}
\end{table*}

%% file: sections/discussion/discussion.tex

\section{Discussion}

We have presented a new approach to perform online dynamic coronary roadmapping on X-ray fluoroscopic sequences for PCI procedures. The approach compensates the cardiac-induced vessel motion via selecting offline-stored roadmaps with appropriate cardiac phase using ECG matching, and corrects the respiratory motion of vessels by online tracking of guiding catheter tip in X-ray fluoroscopy using a proposed deep learning based Bayesian filtering. The proposed tracking method represents and tracks the posterior of catheter tip via a particle filter, for which a likelihood probability map is computed for updating the particle weights using a convolutional neural network. In the experiments, the proposed DCR approach has been trained and evaluated on clinical X-ray sequences for both tracking and roadmapping tasks.


One prerequisite of accurate tracking with the proposed approach is to obtain a reasonably good likelihood estimation, which requires to train the deep neural network to detect catheter tip well. In this work, we have investigated the influence of three network hyperparameters on the performance of the detection network (Section \ref{exp-subsec-detection}): the basic channel number and network depth level are model capacity parameters, the dropout adds regularization to the model. The experiment showed that the detection accuracy improves when the basic channel number and the network depth level increase (Table \ref{exp-tab-training-network}). This observation matches the expectation that a more complex model has higher capacity to model the variation in the data, hence results in better accuracy. When the complexity reaches a certain level, e.g. 64 basic channels and 5 level of depth, the network performance does not increase much compared to those with simpler settings, implying that the model starts overfitting on our dataset. 

In addition to the deep neural network, the other important component of the proposed tracking approach is the sampling in the particle filter that yields the samples for representing the prior and the posterior of catheter tip position. First, a sufficient number of samples in the whole sample space are required to well characterize the probability distributions (see Appendix \ref{exp-subsubsec-pf}). Second, the sample dynamics plays an important role in tracking, in particular, as indicated by Eq.(\ref{method-eq-state-transition-flow}), the process noise and the sample motion. The process noise influences the tracking accuracy, according to Table \ref{exp-tab-tune-particle} in Appendix \ref{exp-subsubsec-pf}. Additionally, sample motion is another key aspect of sample dynamics. Motion estimation has previouly been incorporated in a motion-based particle filter, such as adaptive block matching (\cite{bouaynaya2009optimality}). In our work, optical flow was chosen for motion estimation, as its non-parametric nature allows to characterize the complexity of motion in X-ray fluoroscopy well. In addition, the advantage of such approach from a theoretical point of view is that it takes into account of the current observation, leading to a more optimal importance density (\cite{arulampalam2002tutorial}) compared to random motion.

The tracking results in Section \ref{exp-subsubsec-tracking} show that the proposed tracking approach is able to track the catheter tip in X-ray fluoroscopy accurately with an average tracking error of about 1.3 mm. It also shows advantages over methods based only on optical flow or the detection network. The OF (pre) method relies heavily on tracking in the previous frame, hence the error could accumulate. The OF (first) method may suffer from large motion from the first frame to the current frame. The detection method uses information only from the current frame, no temporal relation between frames is utilized; therefore, it results in spikes in the longitudinal view, as shown in Figure \ref{exp-figure-line-plot}. The proposed tracking method has a CNN to provide an accurate observation on the current frame which improves the accuracy of optical flow tracking within the framework of Bayesian filtering. In the meantime, the optical flow based particle filter maintains and propagates the prior knowledge from the initial tip position to provide a constraint on searching for the potentially correct positions, which is useful especially when the CNN detector fails to find the correct target area. The association of knowledge from two sources together improves the tracking accuracy compared to each single source.

The initial state is a also key component of tracking approaches. In the context of Bayesian filtering, the initial state provides the prior knowledge of the tracking target. Most tracking algorithms assume a known initial state from which the target is tracked, e.g. our proposed method with manual initialization in Section \ref{exp-subsubsec-initialization}. In this case, the prior knowledge is provided by human. In Section \ref{exp-subsubsec-initialization}, we also investigated a scenario where the initial state is given by the detection network, so that the complete tracking process is fully automated. The results indicate that, the proposed tracking method with automatic initialization works reasonably well on most sequences even when the initialization is sometimes incorrect (Figure \ref{exp-subfigure-auto-init-examples-1}). This is because (1) the true position is covered by a few samples, and (2) the correct detection in later frames corrects the initial mistake in the first frame. The automatic initialization fails when (1) a wrong position is covered by a few samples and (2) the wrong detection in subsequent frames confirms the mistake in the initial frame (Figure \ref{exp-subfigure-auto-init-examples-2}). This happens when there is contrast agent remaining in the image or there are multiple catheters, which are the major sources causing ambiguity in detection. In practice, the automatic initialization would work well when contrast agent is washed out and only one catheter is present in the field of view, otherwise manual initialization would be needed which requires only one click to initiate tracking.

Dynamic coronary roadmapping is the direct application of the catheter tip tracking results. In our experiments, the DCR was performed with manual tip initialization to show the potential of the proposed tracking method, and was compared with the DCR without tracking and with manual tracking. The results indicate that using catheter tip tracking can improve DCR accuracy, as the respiratory-induced vessel motion is corrected by the displacement of catheter tip in addtion to cardiac motion correction. The results also show that the proposed DCR reaches a good accuracy (mean error is about 2 mm) and performs only slightly worse than its best case, the DCR with manual tip tracking which is not applicable for intraoperative use. Additionally, according to a previous study by \cite{dodge1992lumen}, the average lumen diameters of human coronary arteries are between 1.9 mm (distal left anterior descending artery) and 4.5 mm (left main artery). This means that the accuracy achieved with the proposed approach is comparable with the size of coronary arteries.

Apart from catheter tip tracking, several other possible factors in different steps of the experiments may influence the final DCR accuracy. First, in the offline phase, the signal of contrast agent may become too strong and completely cover the catheter tip, complicating the tip visibility in some cases. In this situation, the uncertainty in the manual tip annotation may result in errors in roadmap transformation. Second, in the roadmap selection step, the offline-stored roadmaps are only discrete samples of complete cardiac cycles which might not fully characterize every possible change in the cardiac motion. This problem could possibly be addressed in the future by interpolating frames between the existing frames in the data. Additionally, variation exists between different cardiac cycles (\cite{mcclelland2013respiratory}), therefore, choosing a roadmap from another cycle may cause inaccuracy for cardiac motion compensation. Finally, the way of DCR evaluation in Section \ref{setup-subsec-roadmapping} might also introduce inaccuracies in the error measurement. Since guidewires often attach to the inner curves of a vessel to take the shortest path, the small difference between the annotation of guidewire and vessel centerlines was ignored in the evaluation.

In addition to accuracy, processing speed is also critical for intraoperative applications. The results in Section \ref{exp-subsec-time} indicate that the total processing time of the proposed DCR approach is less than the image acquisition time, meaning that it runs in real-time on our setup. To build a real-time system for PCI in practice, the overall latency of the complete system needs to be considered. It is also worth noticing that the DL and PF steps of the proposed tracking method are independent from each other. In practice, in case more than one GPU are available, the proposed DCR approach can be further accelerated by paralleling the DL and PF steps, making them running on different GPUs.

Compared to the previous works on DCR, the proposed approach in this paper shows advancement in several aspects. First, our systems works on non-cardiac-gated sequences which does not require additional setups for cardiac motion gating that were needed for some methods (\cite{zhu2010image, manhart2011self}). Second, our approach compensates both respiratory- and cardiac-induced vessel motion, which is more accurate than systems that correct only cardiac motion (\cite{elion1989dynamic}). In addition, the proposed DCR approach follows a data-driven paradigm that learns target feature from sequences acquired from different patients and various view angles, making it more robust than the method that relies on traditional vesselness filtering (\cite{kim2018registration}) or methods that require specific tissue being present (\cite{zhu2010image, manhart2011self}). These are the major advantages of the proposed DCR over the existing direct roadmapping systems. Compared to model-based motion compensation, our approach does not require the extraction of motion surrogate signals and train a motion model for each new patient, but can be directly run with a trained model. 

The proposed deep learning based Bayesian filtering method has several advantages over the existing instrument tracking methods. First, the deep learning component enables a more general framework to detect instruments in medical images than methods tailored for specific tools (\cite{ma2012clinical, ma2013real}). Compared to the existing detection methods based on deep learning (\cite{baur2016cathnets, laina2017concurrent, du2018articulated}), our approach takes into account of the information between frames; the Bayesian filtering framework allows interaction between temporal information and the detection of a convolutional neural network, making the tracking more robust. Bayesian frameworks have been used in many previous temporal instrument tracking methods. Particularly, the likelihood term in some works was designed based on registration or segmentation outcomes (\cite{ambrosini2017hidden, speidel2006tracking}) or traditional machine learning approaches with handcrafted features (\cite{wang2009robust, wu2012fast, pauly2010machine}). In our method, we approximated the likelihood with a deep neural network learned from the clinical data which exempts the need of feature engineering but yet possesses more discriminative power; the network directly outputs the probability map, making it more straightforward to use. Finally, compared to the existing instrument tracking approaches based on Bayesian filtering (\cite{ambrosini2017hidden, speidel2006tracking, speidel2014visual}), the state transition in our method was based on the motion estimated from two adjacent frames, which is more reliable than totally random motion or artificially-designed state transition models.

From a practical point of view, the proposed DCR approach could potentially fit into the clinical workflow of PCI. The offline phase of the method can be done efficiently by a person who assists the procedures: selecting and creating roadmaps from an angiography acquisition, annotating the catheter tip (one point) in the images. This phase is typically done before a fluoroscopy acquisition during which the guidewire advancement and stent placement are performed. In the online phase, when a fluoroscopic image is acquired, the proposed system selects the most suitable roadmap, tracks the catheter tip and transforms the roadmap to prospectively show a vessel overlay on the fluorosocpic image. The online updated coronary roadmap can provide real-time visual guidance to cardiologists to manipulate interventional tools during the procedure without the need of administering extra contrast agent.

In the future, it may be worth investigating the following directions related to this work. As the data used in this study was acquired from one hospital using a machine from a single vendor, it would be interesting to evaluate the proposed approach on multi-center data acquired with machines from different vendors. Next, since the ECG signals of our data appear to be regular, it may be necessary in a future study to acquire data with irregular ECG that could be obtained in practice, and validate the proposed approach on those data. Besides, it would be also interesting to validate our approach during PCI procedures in an environment simulating the real clinical settings. Additionally, from a methodological point of view, although the proposed tracking method is invariant under different view angles, the whole DCR approach works only when the offline and online phase have the same view angle, i.e. it is a 2D roadmapping system. Therefore, one future direction would be to develop a 3D DCR system that would work with various view angles in the online phase.

\section{Conclusion}

We have developed and validated a novel approach to perform dynamic coronary roadmapping for PCI image guidance. The approach compensates cardiac motion through ECG alignment and respiratory motion by guiding catheter tip tracking during fluoroscopy with a deep learning based Bayesian filtering method. The proposed tracking and roadmapping approaches were trained and evaluated on clinical X-ray image datasets and were proved to perform accurately on both catheter tip tracking and dynamic coronary roadmapping tasks. Our approach also runs in real-time on a setup with a modern GPU and thus has the potential to be integrated into routine PCI procedures, assisting the operator with real-time visual image guidance.

%% file: sections/acknowledgment/acknowledgment.tex

\section*{Acknowledgment}

This work was supported by NWO-domain TTW (The Division of Applied and Engineering Sciences in The Netherlands Organisation for Scientific Research), IMAGIC project under the iMIT program (grant number 12703). Ries and Simon van Walsum are acknowledged for their contribution in the manual annotations.

%% file: sections/references/references.tex




\input{sections/references/references.bbl}

%% file: sections/appendix/appendix.tex
\section*{Appendix}
\renewcommand{\thesubsection}{\Alph{subsection}}

\subsection{Details of the Training Setup}
\label{app-subsec-training-setup}

\subsubsection{Data Augmentation}
\label{app-subsubsec-data-augmentation}

To increase the number of training samples and their diversity, data augmentation was used. The augmentation includes geometric transformation such as flipping (left-right, up-down), rotation of multiple of 90 degrees, random affine transformation (translation -10 to 10 px, scaling 0.9 to 1.1, rotation -5 to 5 degrees, shear -5 to 5 px), random elastic deformation (deformation range -4 to 4 px, grid size of control points 64 px). A training sample has 0.5 probability of being processed with one of the transformations. The probability for applying each transformation is: flipping left-right (1/24), flipping up-down (1/24), rotation of multiple of 90 degrees (1/12), affine transformation (1/6), elastic deformation (1/6), no transformation (1/2). To make the trained model robust to noise, in addition to the geometric transformations, we also augmented data by adding Gaussian noise to the pixel value with a zero mean and a standard deviation between 0.01 and 0.03. The probability of adding the noise is 0.5.

\subsubsection{Training Settings}
\label{app-subsubsec-training-settings}

The $\lambda$ value in the training loss Eq. (\ref{method-eq-total-loss}) was set to 10 to make the scale of the two terms similar. Adam optimizer was used to minimize the loss function with a learning rate 0.0001. The number of training samples in a batch is 4. The network was trained with 100 epochs to ensure convergence.

\subsection{Parameter Tuning for Catheter Tip Tracking}
\label{app-subsec-parameter-tuning}

This section describes the details of tuning the parameters of optical flow and particle filter for catheter tip tracking.

\subsubsection{Tuning Optical Flow Parameters}
\label{exp-subsubsec-of}

The approach of \cite{farneback2003two} was used as the optical flow implementation in Algorithm \ref{method-alg-dl-bayes}. A grid search to find the optimal parameter setting was done on the following parameters of the method: (1) the image scale to build the pyramids, (2) the number pyramid levels, (3) the averaging window size, (4) the number of iterations, (5) the size of the pixel neighborhood used to find polynomial expansion in each pixel, and finally (6) the standard deviation of the Gaussian that is used to smooth derivatives used as a basis for the polynomial expansion.

The above parameters were tuned independently of the deep neural network, as optical flow directly estimates the catheter tip motion between two frames. To tune the parameters, we tracked the catheter tip in X-ray fluoroscopy starting from the ground truth tip position in the first frame using the motion field between two adjacent frames estimated with optical flow. The average and median distance between the tracked tip position and the ground truth were used as the evaluation criteria for the tuning. 

The method of \cite{farneback2003two} was implemented by using the OpenCV function \texttt{calcOpticalFlowFarneback}. With consideration of the suggested parameter values from the documentation, the parameter setting chosen for optical flow from the grid search is $\texttt{pyr\_scale}=0.5$, $\texttt{levels}=3$, $\texttt{winsize}=10$, $\texttt{iterations}=30$, $\texttt{poly\_n}=5$, $\texttt{poly\_sigma}=1.1$. Details of the parameters can be found on the function documentation page\footnote{\url{https://docs.opencv.org/2.4/modules/video/doc/motion_analysis_and_object_tracking.html?}}.

\subsubsection{Tuning Particle Filter Parameters}
\label{exp-subsubsec-pf}
The parameters to tune for the particle filter are the number of samples $N_s$ and the variance of process noise $\sigma_v^2$. When tuning them, we fixed the parameters of the trained network and the optical flow method, and used their optimal parameter settings during this experiment. Following Algorithm \ref{method-alg-dl-bayes}, we tracked the catheter tip from the ground truth position (probability map) in the first frame, and used the mean and median distance between the tracked and the true position as the validation metric.

The tracking results on the validation (tracking) set are shown in Table \ref{exp-tab-tune-particle}. The table shows that 100 samples are suboptimal, while 1000 samples seem sufficient, as 10000 samples result in tracking accuracies similar to 1000 samples. It also shows that the optimal choices of the standard deviation of the process noise are 4 or 5 px for the downsampled images. One possible reason for such choices may be that they are similar to the size of guiding catheters. In general, good choices for $N_s$ are 1000 and 10000, for $\sigma_v$ are 4 and 5. By considering the mean, the standard deviation and the median of tracking errors, the parameter setting $\sigma_v=5$, $N_s=1000$ was chosen.

\begin{table*}[th!]
	\centering
	
	\caption{Catheter tip tracking errors (mm) on the validation (tracking) dataset of different parameter settings for particle filter. The tracked tip point was rounded to the pixel center. The error of all images (mean $\pm$ std / median) are presented. Red cells show the good choices of parameters; \textit{bold} number indicates the chosen setting.}
	
	\begin{tabular}{c c c c}
		\toprule
		$\sigma_v$ & \multicolumn{3}{c}{$N_s$} \\
		\cmidrule(lr){2-4}
		(px) & 100 & 1000 & 10000 \\
		\midrule 
		 3 & 1.52 $\pm$ 2.19 / 0.79 & 1.49 $\pm$ 2.18 / 0.79 & 1.48 $\pm$ 2.18 / 0.79 \\
		 4 & 1.50 $\pm$ 2.17 / 0.79 & \cellcolor{red!30}1.46 $\pm$ 2.17 / 0.79 & \cellcolor{red!30}1.47 $\pm$ 2.18 / 0.79 \\
		 5 & 1.52 $\pm$ 2.21 / 0.79 & \cellcolor{red!30}\textbf{1.47 $\pm$ 2.17 / 0.74} & \cellcolor{red!30}1.47 $\pm$ 2.19 / 0.74 \\
		 6 & 1.53 $\pm$ 2.39 / 0.79 & 1.49 $\pm$ 2.33 / 0.79 & 1.48 $\pm$ 2.29 / 0.74 \\
		 7 & 1.56 $\pm$ 2.42 / 0.79 & 1.50 $\pm$ 2.29 / 0.74 & 1.50 $\pm$ 2.39 / 0.74 \\
		 8 & 1.58 $\pm$ 2.41 / 0.79 & 1.51 $\pm$ 2.40 / 0.74 & 1.51 $\pm$ 2.42 / 0.74 \\
		 9 & 1.56 $\pm$ 2.22 / 0.79 & 1.53 $\pm$ 2.43 / 0.79 & 1.52 $\pm$ 2.45 / 0.61 \\
		10 & 2.25 $\pm$ 6.18 / 0.79 & 1.54 $\pm$ 2.46 / 0.79 & 1.53 $\pm$ 2.47 / 0.61 \\
		\bottomrule	
	\end{tabular}
	
	\label{exp-tab-tune-particle}
\end{table*}

\subsection{Detection, tracking and roadmapping without catheter segmentation}
\label{app-subsec-detection-only}

Training of the network in Figure \ref{method-figure-network} requires catheter labels for detection and segmentation. As the segmentation labels are often more expensive to acquire than the detection label in practice, we also investigated the performance of catheter tip detection, tracking and dynamic coronary roadmapping without segmenting the catheter. To this end, we used a similar encoder-decoder network architecture as Figure \ref{method-figure-network} which computes only the detection output directly after the last \textit{up} block of the decoder with a $1\times1$ convolution followed by a spatial softmax layer, instead of having two outputs. We then followed the same way as the approach using the network with segmentation to search for (hyper-)parameters for the approach without segmentation. The following parameter setting was chosen for the experiments in this section: for deep learning, the basic channel number is 64, the depth is 5, the dropout rate is zero; for particle filtering, $\sigma_v=3$, $N_s=10000$. With this setup, we compared the performance of the approach without catheter segmentation to the proposed approach with segmentation on catheter tip detection and tracking (Appendix \ref{app-subsubsec-detection-tracking}) and dynamic coronary roadmapping (Appendix \ref{app-subsubsec-roadmapping}) on the test set from Table \ref{setup-tab-tip-experiment}.

\subsubsection{Catheter tip detection and tracking}
\label{app-subsubsec-detection-tracking}

\begin{figure}[t]
	\centering
	\subfloat[Overall view of tracking errors]{
		\includegraphics[width=0.23\textwidth]{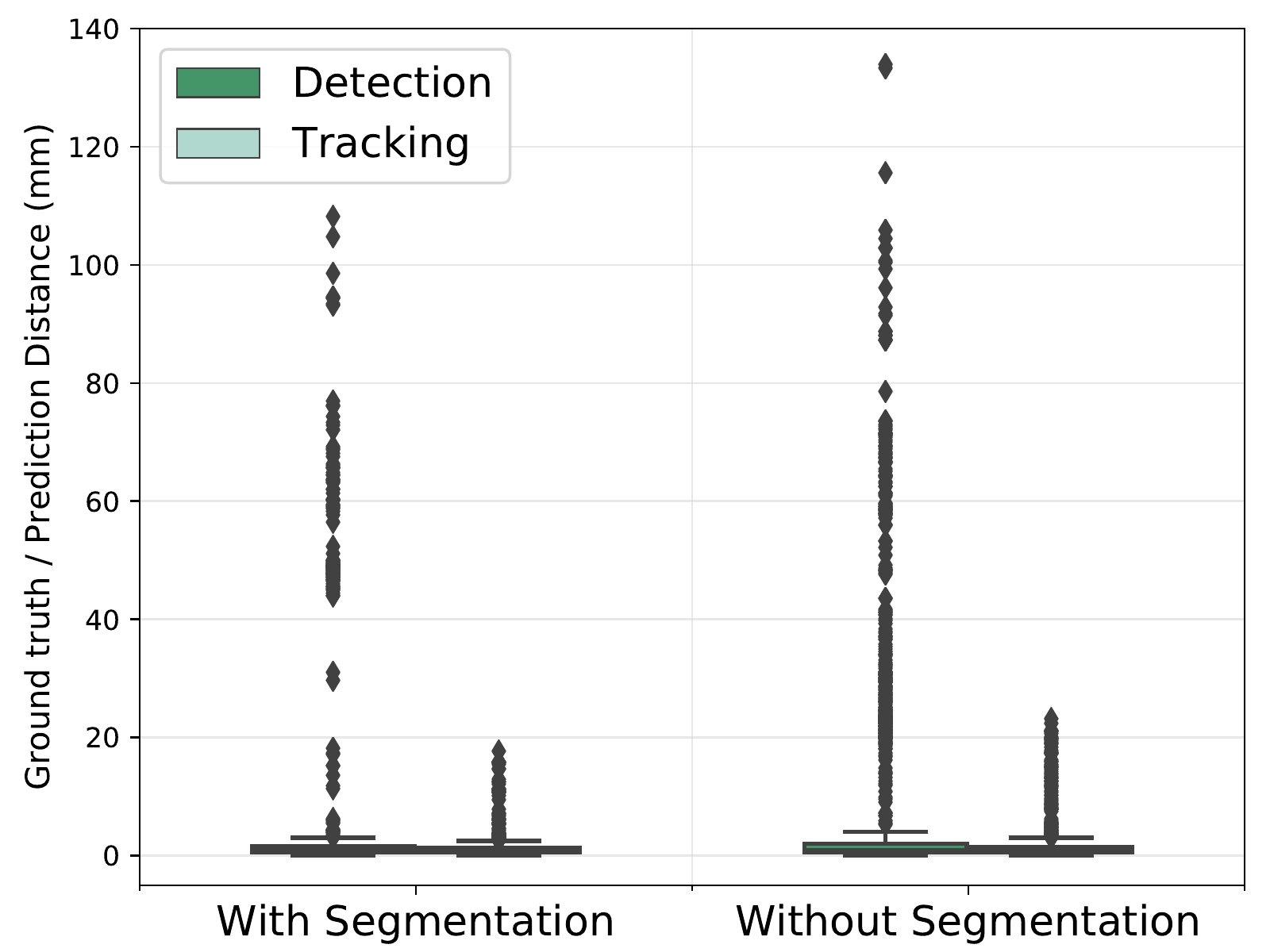}
		\label{app-subfigure-boxplot-all-dist}
	}
	\subfloat[A zoom-in view of (a)]{
		\includegraphics[width=0.23\textwidth]{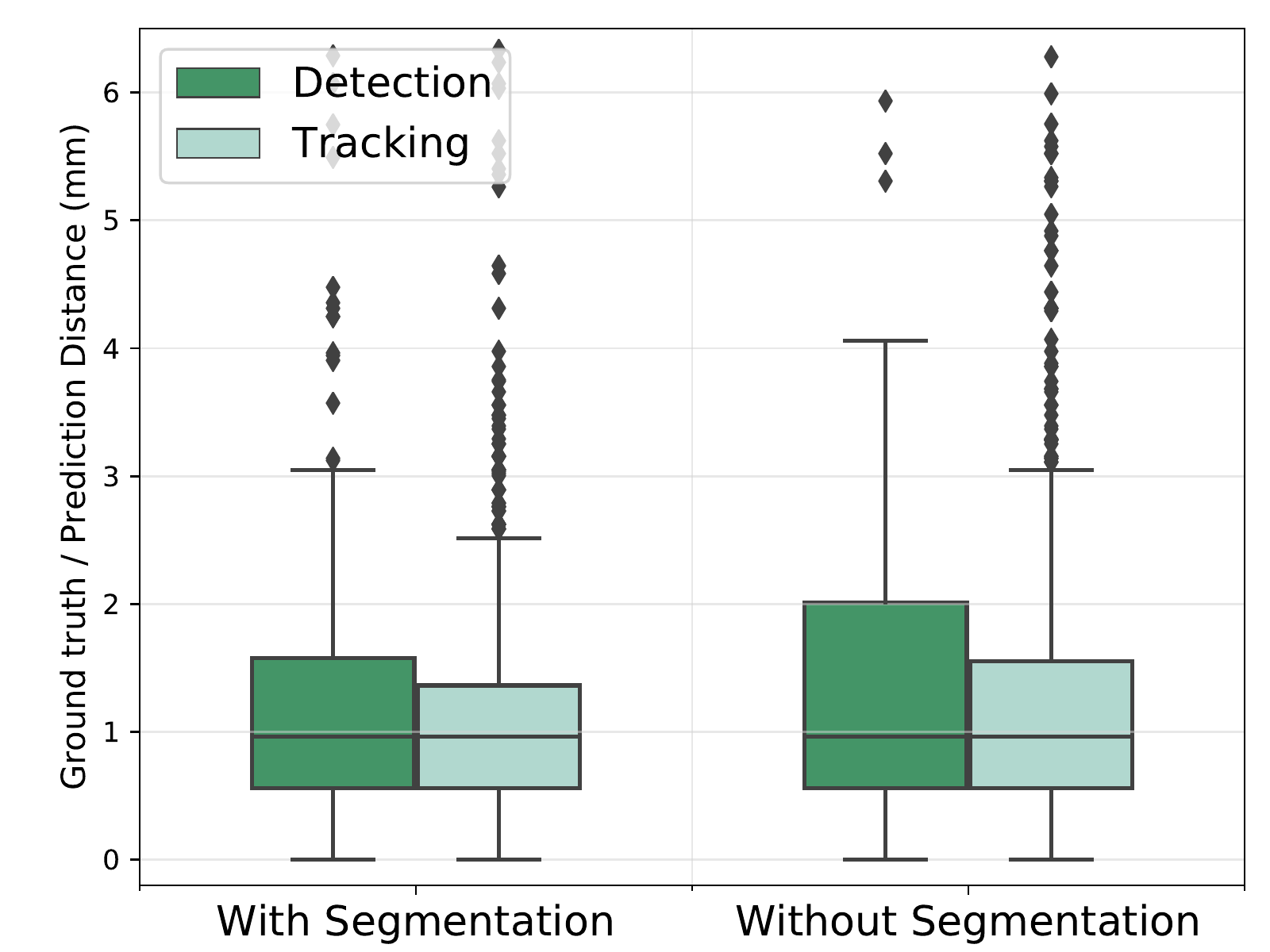}
		\label{app-subfigure-boxplot-all-dist-zoom-in}
	}
	\caption{Tracking errors on all test images with and without catheter segmentation.}
	\label{app-figure-boxplot-all_dist}
\end{figure}

The same metrics in Table \ref{exp-tab-tracking-test} are used to report the accuracy of catheter tip detection and tracking without catheter segmentation. Table \ref{app-tab-tracking-test} and Figure \ref{app-figure-boxplot-all_dist} both manifest that the segmentation sub-task improves the accuracy of catheter tip detection and tracking. Although the improvement on the tracking task is marginal and not statistically significant ($p=0.06$), the segmentation helps to reduce the magnitude and amount of outliers (large errors).

\begin{table*}[h]
	\centering
	
	\caption{Catheter tip tracking errors (mm) with and without catheter segmentation on the test (tracking) dataset. $\dagger$ indicates that the difference between the detection with and without segmentation is statistically highly significant with the two-sided Wilcoxon signed-rank test ($p < 0.001$). No statistically significantly difference is observed between the tracking with and without segmentation using the two-sided Wilcoxon signed-rank test ($p = 0.06$).}
	
	\begin{tabular}{l c c c c}
		\toprule
		\multirow{2}{*}{Evaluation Metrics} & \multicolumn{2}{c}{With Segmentation} & \multicolumn{2}{c}{Without Segmentation} \\
		\cmidrule(lr){2-3} 
		\cmidrule(lr){4-5} 
		& Detection$\dagger$ & Tracking & Detection & Tracking  \\
		\midrule 
		Maximal error of all images & 108.20 & \textbf{17.72} & 133.94 & 23.2 \\
		Median error of all images & \textbf{0.96} & \textbf{0.96} & \textbf{0.96} & \textbf{0.96}  \\
		Mean error of all images & 5.62 $\pm$ 15.91 & \textbf{1.29 $\pm$ 1.76} & 9.32 $\pm$ 19.73 & 1.75 $\pm$ 3.17 \\
		\midrule
		Average of sequence median error & 6.26 $\pm$ 17.11 & \textbf{1.03 $\pm$ 0.49} & 9.34 $\pm$ 18.64 & 1.42 $\pm$ 2.14  \\
		Average of sequence mean error & 6.83 $\pm$ 13.88 & \textbf{1.29 $\pm$ 0.94} & 10.41 $\pm$ 15.94 & 1.69 $\pm$ 1.97  \\
		\bottomrule	
	\end{tabular}
	
	\label{app-tab-tracking-test}
\end{table*}

\subsubsection{Dynamic coronary roadmapping}
\label{app-subsubsec-roadmapping}

\begin{figure}
	\centering
	\includegraphics[width=0.48\textwidth]{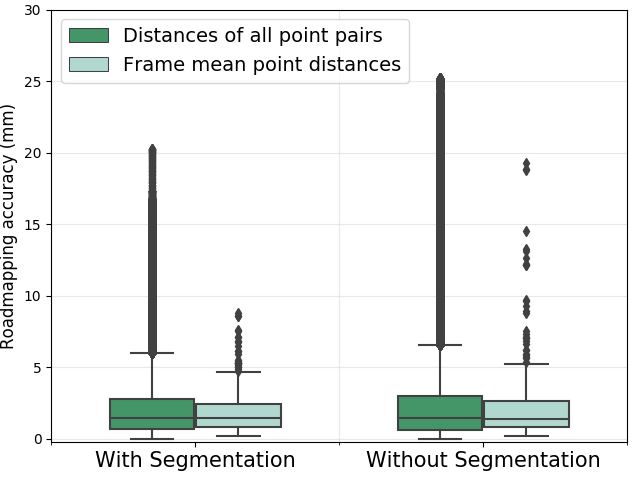}
	\caption{Accuracy (mm) of DCR via catheter tip tracking with and without catheter segmentation.}
	\label{app-figure-roadmapping-boxplot}
\end{figure}

In this experiment, the same setup in Section \ref{exp-subsec-roadmapping} was used to assess the accuracy of DCR using catheter tip tracking without segmenting the catheter. Table \ref{app-tab-roadmapping} indicate that tracking the catheter tip with catheter segmentation improves the DCR accuracy compared to that without catheter segmentation. Although the improvement is not statistically significant ($p=0.43$), the approach with segmentation is more robust by making less large roadmapping errors (Figure \ref{app-figure-roadmapping-boxplot}).

\begin{table*}[h]
	\centering
	
	\caption{The statistics of DCR accuracy (mm) via catheter tip tracking with and without catheter segmentation. With the two-sided Wilcoxon signed-rank test, no statistically significantly difference is observed between the DCR with and without segmentation ($p = 0.43$).}
	
	\begin{tabular}{l c c}
		\toprule
		 & With Segmentation & Without Segmentation \\
		\midrule 
		\textbf{All point pairs} &  &   \\
		Maximal distance  & 20.24 & 25.20  \\
		Median distance  & 1.43 & 1.43  \\
		Mean distance  & 2.07 $\pm$ 2.08 & 2.44 $\pm$ 3.15  \\
		\midrule
		\textbf{Frame mean distance}  &  &    \\
		Median distance & 1.42 & 1.40  \\
		Average distance & 1.91 $\pm$ 1.52 & 2.23 $\pm$ 2.59  \\
		\bottomrule	
	\end{tabular}
	
	\label{app-tab-roadmapping}
\end{table*}

